\documentclass[11pt]{article}
\pdfoutput=1
\usepackage{appendix}
\usepackage{tikz}
\usetikzlibrary{matrix,arrows,decorations.pathmorphing,decorations.pathreplacing}
\usepackage{jheppub}
\usepackage{amsmath,amssymb,euscript,array,mathrsfs,appendix,ctable,marvosym}
\usepackage{arydshln}
\usepackage{todonotes}
\usepackage{graphicx}
\usepackage[normalem]{ulem}
\usepackage{cleveref}

\input{tcilatex}

\numberwithin{equation}{section}

\title{A generalized 4d Chern-Simons theory}
\author{David M. Schmidtt\footnote{david@df.ufscar.br}} 

\affiliation{Departamento de F\'\i sica, Universidade Federal de S\~ao Carlos, \\
Caixa Postal 676, CEP 13565-905, S\~ao Carlos-SP, Brasil} 

\abstract{A generalization of the 4d Chern-Simons theory action introduced by Costello and Yamazaki is presented. We apply general arguments from symplectic geometry concerning the Hamiltonian action of a symmetry group on the space of gauge connections defined on a 4d manifold and construct an action functional that is quadratic in the moment map associated to the group action. The generalization relies on the use of contact 1-forms defined on non-trivial circle bundles over Riemann surfaces and mimics closely the approach used by Beasley and Witten to reformulate conventional 3d Chern-Simons theories on Seifert manifolds. 
We also show that the path integral of the generalized theory associated to integrable field theories of the PCM type, takes the canonical form of a symplectic integral over a subspace of the space of gauge connections, turning it a potential candidate for using the method of non-Abelian localization. Alternatively, this new quadratic completion of the 4d Chern-Simons theory can also be deduced in an intuitive way from manipulations similar to those used in T-duality. Further details on how to recover the original 4d Chern-Simons theory data, from the point of view of the Hamiltonian formalism applied to the generalized theory, are included as well.
\begin{flushleft}
\end{flushleft}
}

\begin{document}

\maketitle

%\newpage

\section{Introduction}\label{1}

A relatively new approach to integrable lattice models and 2-dimensional integrable field theories that, as a main tool, uses a gauge theory of the Chern-Simons (CS) type, has attracted a great deal of attention due to its potential in offering novel insights into the quantum integrable structure of these systems and their general properties. Such a gauge theory, known as the 4-dimensional semi-holomorphic Chern-Simons theory, or 4d CS theory for short, introduced in \cite{C1,C2} and studied in detail in \cite{CWY1,CWY2,CY} has, in the last few years, triggered several new interesting results in the fields of integrable systems and string theory. 

The 4d CS theory under consideration is defined by an action functional of the form
\begin{equation}
S=ic \dint\nolimits_{\mathbb{M}}\omega \wedge CS\left( \mathbb{A}\right), \label{1.1}
\end{equation}
where $c$ is a constant\footnote{Usually taken to be $1/2\pi$ or $1/4\pi$.}, $\mathbb{M}=\Sigma\times C$ is a 4-dimensional manifold constructed out of a cylinder $\Sigma=\mathbb{R}\times S^{1}$ with Minkowskian signature and a Riemann surface $C$, $\omega$ is a meromorphic 1-form defined on $C$ and $CS(\mathbb{A})$ is the usual Chern-Simons 3-form for a connection $\mathbb{A}$ on $\mathbb{M}$. The theory is topological along $\Sigma$ and holomorphic along $C$, hence the name semi-holomorphic. For an introduction to the theory and some of its properties, see the nice review \cite{Lacroix}.

What is important for the narrative of this work, is the fact that if we consider an atlas $\mathcal{T}$ covering the Riemann surface $C$, we can interpret the 4-dimensional manifold $\mathbb{M}$ as the total space of a trivial bundle over the base space $C$, with a typical fiber $\Sigma$. This observation motivates us to relax such a condition in favor of a more general one, namely, that the manifold $\mathbb{M}$ indeed looks like the product $\Sigma\times C$ but only locally over any chart $\mathcal{U}\subset \mathcal{T}$. This causes some changes into the structure of the theory because, besides the 1-form $\omega$ defined on $\mathbb{M}$, which is now seen as the pullback of a 1-form $\omega_{C}$ defined on $C$, another interesting differential 1-form $\alpha$ (to be identified later on as a contact 1-form) defined on $\mathbb{M}$ can also be introduced via the pullback from $C$ to $\mathbb{M}$ of a certain symplectic 2-form $\sigma_{C}$ defined on $C$, if the fiber bundle is taken to be inherently non-trivial. This allows to generalize the 4d CS theory action \eqref{1.1} into a new one depending on the former dynamical field $\mathbb{A}$, two new non-dynamical 1-forms $\Omega$, $\kappa$ (to be specified below) constructed out of $\alpha$, $\omega$ and some parameters.

The time direction in $\mathbb{M}$ is considered, from physical grounds, as globally defined and oriented. This means we can restrict the manifold $\mathbb{M}$ a bit more and consider instead a 4-dimensional space of the form $\mathbb{M}=\mathbb{R}\times \text{M}$, hence the \textit{assumed} non-triviality of the fiber bundle now resides entirely on the manifold M. In this work we will consider a Minkowskian cylinder $\Sigma$ and identify the time direction with the $\mathbb{R}$ factor in the decomposition $\Sigma=\mathbb{R}\times S^{1}$, thus the 3-dimensional manifold M now becomes the total space of a non-trivial circle bundle over the base space $C$. Fortunately, the structure of circle bundles of this sort was exploited heavily in the seminal paper \cite{NA loc CS}, devoted to a new formulation of the conventional 3-dimensional Chern-Simons theories on Seifert manifolds which are, roughly speaking, total spaces with a circle bundle structure plus a certain technical condition over the action of an Abelian group on the total space \cite{Orlik}. This ultimately allowed to compute the partition function of some CS theories on these manifolds by using the method of non-Abelian localization of symplectic integrals \cite{Witten revisited}. See also \cite{Wilson NA loc}, for the inclusion of Wilson loops within the formalism. 

It is then natural to ask if the partition function of the theory defined by the action \eqref{1.1} is a candidate for using the non-Abelian localization method as well, due to the structure of the 4-dimensional manifold $\mathbb{M}=\mathbb{R}\times \text{M}$ that we are considering. After all, the only difference between $\mathbb{M}$ and M is a trivial $\mathbb{R}$ factor. To be more precise \cite{NA loc CS,Witten revisited}, if we want to show that this is the case, one first need to put the action \eqref{1.1} in the quadratic form
\begin{equation}
S=ic (\mu,\mu) \label{1.2}
\end{equation}
and second, one has to show that the partition function of the theory is canonical, in the sense that it can be written as a symplectic integral of the form
\begin{equation}
Z(\epsilon)\sim \dint\nolimits_{X}\text{exp} \left[ \hat{\underline{\Omega}}-\frac{1}{2\epsilon}(\mu,\mu)  \right ]. \label{1.3}
\end{equation}
Here, $X$ is a symplectic manifold with symplectic form $\hat{\underline{\Omega}}$ constructed out of the space $\mathcal{A}$ of gauge connections $\mathbb{A}$ defined on $\mathbb{M}$. 
We assume that a
Lie group $\mathcal{H}$ acts on $X$ in a Hamiltonian fashion, with moment map $\mu: X\rightarrow \mathfrak{h}^{*}$, where $\mathfrak{h}^{*}$
is the dual of the Lie algebra $\mathfrak{h}$ of $\mathcal{H}$. Also, $(\ast,\ast)$ is an invariant quadratic form on $\mathfrak{h}$ which,
by the duality induced by it between $\mathfrak{h}$ and $\mathfrak{h}^{*}$, allows to define the action $S$ in terms of $\mu\in \mathfrak{h}$. The coupling constant of the theory is $\epsilon$.

The purpose of this work is to initiate a study of the relation between the original theory \eqref{1.1}, the quadratic action \eqref{1.2} and the symplectic integral \eqref{1.3}. We will show below that such a relation exist but it is not a direct one and involves instead a generalization of the action functional \eqref{1.1}. To do this, we follow the strategy used in \cite{NA loc CS} for computing the moment map $\mu$ in the case of the conventional 3-dimensional CS theories and make the necessary modifications in order to locate the 1-form $\omega$ and the time direction $\mathbb{R}$ within the general construction. As a complementary and more direct approach, we shall recover the quadratic action \eqref{1.2} from a different perspective using known field-theoretic manipulations that are usually employed to dualize a theory, e.g. like in T-duality. The outcome is that the partition function of the theory takes the desired form \eqref{1.3}, where $X=\mathcal{A}/\mathcal{S}$ is a symplectic quotient space,
\begin{equation}
\left( \mu ,\mu \right) =\dint\nolimits_{\mathbb{M}}\Omega \wedge CS\left( 
\mathbb{A}\right) -\dint\nolimits_{\mathbb{M}}\Omega \wedge \kappa \wedge
d_{\mathbb{M}}\kappa \text{Tr}\left( \Phi ^{2}%
\right)+\dint\nolimits_{\mathbb{M}}d_{\mathbb{M}}\Omega \wedge \kappa\wedge \text{Tr}\left( \mathbb{A}i_{\mathcal{R}}\mathbb{A}  \right) \label{main 1}
\end{equation}
is a quadratic action functional\footnote{The reader interested in a fast deduction of \eqref{main 1}, may take a quick glance to section \eqref{5} first. } generalizing \eqref{1.1} and
\begin{equation}
\text{exp}\; \hat{\underline{\Omega}}=\mathcal{D}\mathbb{A}|_{X},\text{ \ \ }
\hat{\underline{\Omega}}=-\frac{i}{2}\dint\nolimits_{\mathbb{M}}\Omega \wedge
\kappa \wedge \text{Tr}\left(  \hat{\delta }\mathbb{A}\wedge \hat{\delta }\mathbb{A}\right) \label{main 2}
\end{equation}
is the path integral symplectic measure, where $\hat{\underline{\Omega}}$ is a symplectic form defined on the quotient $X$. The Lie algebra valued function $\Phi$ and the vector field $\mathcal{R}$ on $\mathbb{M}$ will be introduced below. In this work, we will construct \eqref{1.3} primarily for integrable field theories of the principal chiral model (PCM) type, as the main illustrative example. 

The paper is organized as follows. In section \eqref{2} we construct, from general results in symplectic geometry, an action functional that is, by definition, of the quadratic form \eqref{1.2}. The construction is rather general and relies on a variant of the symplectic form originally used in \cite{NA loc CS}. In section \eqref{3}, we show how this new action functional is related to the original 4d CS theory action \eqref{1.1}. The quadratic action generalizing \eqref{1.1} depends on the usual 1-form $\omega$, a contact 1-form $\alpha$ naturally linked to the non-triviality of the circle bundle M over $C$, and a pair of parameters. We also comment on a strategy devised to recover the theory \eqref{1.1}, which is based on a partial gauge fixing and involves a degenerate limit in one of the parameters. In section \eqref{4}, we specialize the general construction to a particular example concerning one of the simplest known Seifert manifolds $\text{M}=S^{3}$, i.e. the 3-sphere. This case covers integrable field theories of the principal chiral model type, where the underlying circle bundle structure is that of the Hopf fibration of $S^{3}$ over $C=\mathbb{CP}^{1}$. We study, respectively, the Riemannian and K\"ahler metrics on the space of gauge connections $\mathcal{A}$ and on an important quotient space $\overline{\mathcal{A}}$, which is identified with the symplectic manifold $X$ in the path integral expression \eqref{1.3}. 
Then, we work out the Hamiltonian approach in order to implement the necessary partial gauge fixing and degenerate limit mentioned in section \eqref{3}. After recovering the 4d CS theory for these theories, as an example, we end up by re-deriving the Lax connection for the $\lambda$-deformed PCM. In section \eqref{5}, we deduce the generalized 4d CS theory quadratic action \eqref{1.2} from simple duality manipulations. In section \eqref{6} we address, at the formal level, the path integral measure and its relation to the symplectic form $\hat{\underline{\Omega}}$ defined on $\overline{\mathcal{A}}$, hence completing the construction of the canonical integral expression \eqref{1.3}. Finally, in section \eqref{7} we make some comments and provide further explanations concerning the results presented along the text.

\section{Moment map and the quadratic action}\label{2}

The goal of this relatively long section is to introduce an action functional $S$, on a 4-dimensional manifold $\mathbb{M}$, that is proportional to the square of a moment map $\mu$, associated to the Hamiltonian action of a symmetry group $\mathcal{H}$ on a symplectic space $\overline{\mathcal{A}}$, constructed out of the space $\mathcal{A}$ of Chern-Simons gauge connections defined on $\mathbb{M}$. In order to do this, we use known facts from symplectic geometry to correctly identify $\mathcal{H}$ and $\overline{\mathcal{A}}$, to construct $\mu$ and study its symmetry properties and to introduce the main result of this section, which is materialized in the expression \eqref{mu mu 2} (or equivalently \eqref{main 1}) defining the generalization of the 4d CS theory \eqref{1.1}. We follow \cite{NA loc CS} closely and proceed formally. 

\textit{Warning.} We are not specifying reality conditions over some quantities. As a consequence, some objects that are usually expected to be real by definition could emerge as purely imaginary.

\subsection{Moment map and its properties}

Consider a principal $G$-bundle $P$, where $G$ is a Lie group with Lie algebra $\mathfrak{g}$, over a 4-dimensional orientable manifold $\mathbb{M}$, such that $\partial \mathbb{M}=0$. Denote by $\mathcal{A}$ the space of connections $\mathbb{A}$ on P and identify it with the vector space $\Omega_{\mathbb{M}}^{1}\otimes \mathfrak{g}$ of 1-forms on $\mathbb{M}$ taking values in the Lie algebra $\mathfrak{g}$. Denote by $\mathcal{G}$, the group of gauge transformations acting on $\mathcal{A}$. The Lie algebra of $\mathcal{G}$, denoted by $\mathcal{G}_{\text{Lie}}$, consists of elements in the vector space $\Omega_{\mathbb{M}}^{0}\otimes \mathfrak{g}$ of Lie algebra valued functions on $\mathbb{M}$.

Introduce a pre-symplectic form $\hat{\Omega}\in\Omega_{\mathcal{A}}^{2}$ on the space of connections $\mathcal{A}$, defined by
\begin{equation}
\hat{\Omega}=-\frac{1}{2}\dint\nolimits_{\mathbb{M}}\Omega \wedge
\kappa \wedge \text{Tr}\left( \hat{\delta }\mathbb{A}\wedge\hat{\delta }\mathbb{A}\right), \label{pre-symplectic}
\end{equation}%
where $\kappa,\Omega \in \Omega_{\mathbb{M}}^{1}$ are globally defined and everywhere non-zero 1-forms on $\mathbb{M}$ that satisfy, respectively, the following set of conditions:
\begin{equation}
i_{\mathcal{R}}\kappa =1,\text{ \ \ }i_{\mathcal{R}}\left( d_{\mathbb{M}%
}\kappa \right) =0 \label{Condition on kappa}
\end{equation}
and
\begin{equation}
i_{\mathcal{R}}\Omega =0,\text{ \ \ }\pounds_{\mathcal{R}}\Omega =0. \label{Condition on omega}
\end{equation}%

The symbol $\hat{\delta}$ denotes the exterior derivative on $\mathcal{A}$, $d_{\mathbb{M}}$ denotes the exterior derivative on $\mathbb{M}$, $\mathcal{R}\in \mathfrak{X}_{\mathbb{M}}$ is a vector field on $\mathbb{M}$ defined by the first condition (normalization condition) in \eqref{Condition on kappa} once $\kappa$ is specified, $i_{\mathcal{R}}$ is the interior product or contraction with $\mathcal{R}$ and
$\pounds_{\mathcal{R}}=d_{\mathbb{M}}\circ i_{\mathcal{R}}+i_{\mathcal{R}}\circ d_{\mathbb{M}}$ is the Lie derivative along $\mathcal{R}$. The symbol Tr, denotes an invariant quadratic form defined on the Lie algebra $\mathfrak{g}$. As a consequence of \eqref{Condition on kappa} and \eqref{Condition on omega}, we also have that%
\begin{equation}
\pounds _{\mathcal{R}}\kappa =0,\text{ \ \ }i_{\mathcal{R}}\left( d_{\mathbb{M}%
}\Omega \right) =0. \label{consequences}
\end{equation}

The 2-form $\hat{\Omega}$ is closed and invariant under the action of $\mathcal{G}$ and, in particular, under the action of the group $\mathcal{S}=U(1)\times U(1)$ generated by the following set of independent shifts 
\begin{equation}
^{\kappa}\mathbb{A}=\mathbb{A}+s\kappa ,\text{ \ \ }^{\Omega}\mathbb{A}=%
\mathbb{A}+s' \Omega ,
\end{equation}
where $s,s'\in \Omega_{\mathbb{M}}^{0}\otimes \mathfrak{g}$ are arbitrary and, as a consequence of this, the 2-form $\hat{\Omega}$ is degenerate along elements $\mathbb{A}$ of the form $\mathbb{A}=s \kappa$, $\mathbb{A}=s' \Omega$. Thus, we take the quotient of $\mathcal{A}$ by the action of the group $\mathcal{S}$ and define the symplectic space $\overline{\mathcal{A}}=\mathcal{A}/\mathcal{S}$. Under the quotient, the pre-symplectic form $\hat{\Omega}$ on $\mathcal{A}$ descends to a symplectic form on $\overline{\mathcal{A}}$, which becomes a symplectic space naturally associated to a generalization of the 4d CS theory. The action of the gauge group $\mathcal{G}$ on $\mathcal{A}$ also descends, under the quotient, to a well-defined action on $\overline{\mathcal{A}}$ and the 2-form $\hat{\Omega}$ on $\overline{\mathcal{A}}$ is invariant under the action of $\mathcal{G}$. The action of $\mathcal{G}$ is non-linear, contrary to the linear action of $\mathcal{S}$.

The gauge group $\mathcal{G}$ and the group $\mathcal{H}$ acting on $\overline{\mathcal{A}}$ in a Hamiltonian way are related, but are not the same. Thus, in order to identify $\mathcal{H}$ properly, we first consider the action of $\mathcal{G}$ on $\overline{\mathcal{A}}$ and subsequently extend the algebraic structure of $\mathcal{G}_{\text{Lie}}$, until the defining condition of a Hamiltonian action on $\overline{\mathcal{A}}$ is fulfilled. Because of we are interested in computing the square of the moment map $\mu$ associated to the action of $\mathcal{H}$ on $\overline{\mathcal{A}}$, the Lie algebra of $\mathcal{H}$, denoted by $\mathfrak{h}$, must be endowed with a well-defined invariant inner product $(*,*)$. If such an inner product exists, the action functional we are seeking is defined to be proportional to the square $(\mu,\mu)$.
However, before we continue, let us review some facts from symplectic geometry that are necessary for accomplishing this task.

Consider a Lie group $\mathcal{H}$ with Lie algebra $\mathfrak{h}$, a symplectic manifold $X$ and assume that the action of $\mathcal{H}$ on $X$ preserves the symplectic form $\hat{\Omega}$ on $X$. The action $\mathcal{H}\circlearrowright X$ of $\mathcal{H}$ on $X$  is said to be Hamiltonian, when there exists an algebra homomorphism from $\mathfrak{h}$ to the algebra of functions on $X$ under the Poisson bracket. In terms of the moment map $\mu : X\rightarrow \mathfrak{h}^{*}$ and the elements $\eta, \lambda \in \mathfrak{h}$, this statement is equivalent to the condition that $\mu$ satisfy
\begin{equation}
\left\{ \left\langle \mu ,\eta \right\rangle ,\left\langle \mu ,\lambda
\right\rangle \right\} =\left\langle \mu ,\left[ \eta ,\lambda \right] \label{Homo}
\right\rangle ,
\end{equation}
where $\left\langle \ast ,\ast \right\rangle $ is the dual pairing between $\mathfrak{h}$ and $\mathfrak{h}^{*}$, $\{*,*\}$ is the Poisson bracket and $[*,*]$ is the Lie bracket. The homomorphism from $\mathfrak{h}$ to the algebra of functions on $X$ is given by $f=\left\langle \mu,\eta \right\rangle$ and the moment map $\mu$, by definition, satisfy the relation
\begin{equation}
d\left\langle \mu, \eta \right\rangle=-i_{V(\eta)}\hat{\Omega}, \label{HVF}
\end{equation} 
where $V(\eta)$ is the induced vector field on $X$ generated by the action of $\eta$. The symbol $d$ denotes the exterior derivative on $X$. The equation \eqref{Homo} also reflects, infinitesimally, the condition that the map $\mu$ commute with the action of $\mathcal{H}$ on $X$ and the co-adjoint action of $\mathcal{H}$ on $\mathfrak{h}^{*}$. The Poisson bracket of two functions on $X$, is given by the expression
\begin{equation}
\left\{ f_{1},f_{2}\right\} =\hat{\delta }f_{2}(V_{f_{1}})=-\hat{%
\delta }f_{1}(V_{f_{2}}), \label{PB}
\end{equation}%
where $V_{f_{1}}, V_{f_{2}}\in \mathfrak{X}_{X}$ are the Hamiltonian vector fields
associated to the functions $f_{1}, f_{2}$ on $X$, respectively. The relation between $V_{f}$ and $f$ is of the form \eqref{HVF}, i.e. we have that $df=-i_{V_{f}}\hat{\Omega}$. 
 
Let us return to the case of interest and construct $\mu$ in two stages. The first one dealing with the action of the gauge group and the second one with the action of the $\mathcal{R}$ vector field.

Consider the action of the gauge group $\mathcal{G}$ on the elements of $\mathcal{A}$, given by the known relation
\begin{equation}
^{g}\mathbb{A}=g^{-1}\mathbb{A}g+g^{-1}d_{\mathbb{M}}g. \label{gauge transf}
\end{equation}%
Infinitesimally, we write $g=\exp \eta$ with $\eta\in \Omega_{\mathbb{M}}^{0}\otimes \mathfrak{g}$. This action induces, on the space $\mathcal{A}$, the following vector field%
\begin{equation}
V(0,\eta,0 )=d_{\mathbb{A}}\eta , \label{gauge vector field}
\end{equation}%
where $d_{\mathbb{A}}=d_{\mathbb{M}}+\left[ \mathbb{A}, \cdot \; \right] $. The notation $(*,*,*)$ with three entries will become clearer as we proceed. 

Now, we compute the moment map $\mu$ associated to the action of \eqref{gauge vector field}. Start by calculating the contraction on the left hand side (lhs) of the defining expression
\begin{equation}
-i_{V(0,\eta,0 )}\hat{\Omega }=\hat{\delta }\left\langle \mu ,(0,\eta,0) \right\rangle .
\end{equation}%
We find that
\begin{equation}
-i_{V(0,\eta,0 )}\hat{\Omega }=\dint\nolimits_{\mathbb{M}}\Omega \wedge
\kappa \wedge \text{Tr}\left( d_{\mathbb{A}}\eta \wedge \hat{\delta }%
\mathbb{A}\right) .
\end{equation}%
After integrating by parts, we obtain
\begin{equation}
\left\langle \mu ,(0,\eta,0) \right\rangle =-\dint\nolimits_{\mathbb{M}}\Omega
\wedge \kappa \wedge \text{Tr}\big( \eta F_{\mathbb{A}}\big)
-\dint\nolimits_{\mathbb{M}}d_{\mathbb{M}}\left( \Omega \wedge \kappa
\right) \wedge \text{Tr}\Big( \eta (\mathbb{A}-\mathbb{A}_{0})\Bigl), \label{moment for gauge}
\end{equation}%
where $F_{\mathbb{A}}=d_{\mathbb{M}}\mathbb{A}+\mathbb{A}\wedge \mathbb{A}$ is the curvature of the connection $\mathbb{A}$ and $\mathbb{A}_{0}$ is a constant connection with respect to $\hat{\delta}$. In what follows we take a vanishing basepoint, i.e. $\mathbb{A}_{0}=0$. 

Using the general identity
\begin{equation}
F_{\mathbb{A}-\theta s}=F_{\mathbb{A}}+\theta \wedge d_{\mathbb{A}}s-d_{%
\mathbb{M}}\theta s, \label{identity-1}
\end{equation}%
where $\theta\in \Omega_{\mathbb{M}}^{1}$ and $s\in \Omega_{\mathbb{M}}^{0}\otimes\mathfrak{g}$, it is not difficult to show that \eqref{moment for gauge} descends to a functional on $\overline{\mathcal{A}}$, as it is invariant under both transformations generated by the shift group $\mathcal{S}$.

In order to find the Poisson algebra for two functionals of the form \eqref{moment for gauge}, we use the general result \eqref{PB}.
We find that
\begin{equation}
\big\{\left\langle \mu ,(0,\eta,0) \right\rangle ,\left\langle \mu ,(0,\lambda,0)
\right\rangle \big\} =\hat{\delta }\left\langle \mu ,(0,\lambda,0)
\right\rangle \left( V(0,\eta,0 )\right) =-\dint\nolimits_{\mathbb{M}}\Omega
\wedge \kappa \wedge \text{Tr}\left( d_{\mathbb{A}}\eta \wedge d_{\mathbb{A%
}}\lambda \right) .
\end{equation}%
Using the identity%
\begin{equation}
\text{Tr}\left( d_{\mathbb{A}}\epsilon \wedge d_{\mathbb{A}}\epsilon
^{\prime }\right) =\text{Tr}\left( \left[ \epsilon ,\epsilon ^{\prime }%
\right] F_{\mathbb{A}}\right) +d_{\mathbb{M}}\text{Tr}\left( \epsilon d_{%
\mathbb{A}}\epsilon ^{\prime }\right) ,
\end{equation}%
for arbitrary $\epsilon,\epsilon'\in \Omega_{\mathbb{M}}^{0}\otimes \mathfrak{g}$, we obtain%
\begin{equation}
\big\{ \left\langle \mu ,(0,\eta,0) \right\rangle ,\left\langle \mu ,(0,\lambda,0)
\right\rangle \big\} =\left\langle \mu ,(0,\left[ \eta ,\lambda \right],0)
\right\rangle -c\left( \eta ,\lambda \right) , \label{gauge moment algebra}
\end{equation}%
where we notice the presence of a Lie algebra cocycle defined by the integral
\begin{equation}
c\left( \eta ,\lambda \right) =-\dint\nolimits_{\mathbb{M}}d_{\mathbb{M}%
}\left( \Omega \wedge \kappa \right) \wedge \text{Tr}\left( \eta d_{\mathbb{%
M}}\lambda \right) . \label{co-cycle}
\end{equation}

If the cohomology class of this cocycle is not zero, something we assume from now on, the action of the gauge group $\mathcal{G}$ on the symplectic space $\overline{\mathcal{A}}$ is not Hamiltonian, as it violates the homomorphism condition \eqref{Homo}. The cocycle \eqref{co-cycle} then determines a central extension\footnote{There are two contributions in \eqref{co-cycle} but, as we shall see, one of them is required to vanish.} $\mathbb{R}$ of the gauge algebra $\mathcal{G}_{\text{Lie}}$ which, as a vector space, is given by $\tilde{\mathcal{G}}_{\text{Lie}}=\mathcal{G}_{\text{Lie}}\oplus \mathbb{R}$ and comes equipped with the bracket
\begin{equation}
\Big[ (0,\eta ,a),(0,\lambda ,b)\Big] =\Big(0, \left[ \eta ,\lambda \right]
,c(\eta ,\lambda )\Big), \label{ext Lie 1}
\end{equation}
where $\eta,\lambda \in \mathcal{G}_{\text{Lie}}$ and $a,b\in \mathbb{R}$. The pairing between an element $\mu\in \tilde{\mathcal{G}}_{\text{Lie}}^{*}$ and an element $(0,\eta,a)\in\tilde{\mathcal{G}}_{\text{Lie}}$ is defined by
\begin{equation}
\left\langle \mu,(0,\eta,a) \right\rangle=\left \langle \mu,(0,\eta,0) \right\rangle-a.
\end{equation}

We assume that the central subgroup of $\tilde{\mathcal{G}}$ acts trivially on $\overline{\mathcal{A}}$, so that the moment
map for the central generator $(0,0,a)$ of $\tilde{\mathcal{G}}_{\text{Lie}}$ is constant. Then, by construction,
we realize that the new moment map for the action of $\tilde{\mathcal{G}}$ on $\overline{\mathcal{A}}$, which is now given by
\begin{equation}
\left\langle \mu ,(0,\eta,a) \right\rangle =-\dint\nolimits_{\mathbb{M}}\Omega
\wedge \kappa \wedge \text{Tr}\big( \eta F_{\mathbb{A}}\big)
-\dint\nolimits_{\mathbb{M}}d_{\mathbb{M}}\left( \Omega \wedge \kappa
\right) \wedge \text{Tr}\big( \eta \mathbb{A}\big)-a, 
\end{equation}%  
satisfies the Hamiltonian action condition, i.e.
\begin{equation}
\big\{ \left\langle \mu ,(0,\eta,a) \right\rangle ,\left\langle \mu ,(0,\lambda,b)
\right\rangle \big\} =\left\langle \mu ,\left[ (0,\eta ,a),(0,\lambda ,b)\right]
\right\rangle . 
\end{equation}% 

Once the moment map is properly identified, the next step is to introduce a non-degenerate, invariant inner product $(*,*)$ on $\tilde{\mathcal{G}}_{\text{Lie}}$ in order to dualize $\tilde{\mathcal{G}}_{\text{Lie}}^{*}$ and subsequently compute the square $(\mu,\mu)$. However, the Lie algebra $\tilde{\mathcal{G}}_{\text{Lie}}$ does not provide the algebraic structure we are looking for and this is because we still have not taken into account the induced action of the vector field $\mathcal{R}\in \mathfrak{X}_{\mathbb{M}}$ on the spaces $\mathcal{G}_{\text{Lie}}$ and $\mathcal{A}$.  

In order to show how the vector field $\mathcal{R}$ acts on the gauge algebra, let us write the cocycle \eqref{co-cycle} in a slightly different form. Take
\begin{equation}
c\left( \eta ,\lambda \right) =-\dint\nolimits_{\mathbb{M}}d_{\mathbb{M}%
}\Omega \wedge \kappa \wedge \text{Tr}\left( \eta d_{\mathbb{M}}\lambda
\right) +\dint\nolimits_{\mathbb{M}}\Omega \wedge d_{\mathbb{M}}\kappa
\wedge \text{Tr}\left( \eta d_{\mathbb{M}}\lambda \right) 
\end{equation}%
and consider the contracted 5-form%
\begin{equation}
0 =i_{\mathcal{R}}\Big( \Omega \wedge \kappa \wedge d_{\mathbb{M}}\kappa
\wedge \text{Tr}\left( \eta d_{\mathbb{M}}\lambda \right) \Big)=-\Omega \wedge d_{\mathbb{M}}\kappa \wedge \text{Tr}\left( \eta d_{%
\mathbb{M}}\lambda \right) +\Omega \wedge \kappa \wedge d_{\mathbb{M}%
}\kappa \text{Tr}\Big( \eta i_{\mathcal{R}}\left( d_{\mathbb{M}%
}\lambda \right) \Big) .
\end{equation}%
Using this result above gives
\begin{equation}
c\left( \eta ,\lambda \right) = \dint\nolimits_{\mathbb{M}}\Omega \wedge \kappa \wedge d_{\mathbb{%
M}}\kappa \text{Tr}\left( \eta \pounds _{\mathcal{R}}\lambda \right)+d\left( \eta ,\lambda \right) , \label{cocycle expanded}
\end{equation}%
where, for further reference, we have defined
\begin{equation}
d\left( \eta ,\lambda \right) =-\dint\nolimits_{\mathbb{M}}d_{\mathbb{M}%
}\Omega \wedge \kappa \wedge \text{Tr}\left( \eta d_{\mathbb{M}}\lambda
\right) . \label{d}
\end{equation}
The first contribution in \eqref{cocycle expanded} is antisymmetric and exhibits a $U(1)$ action generated by the vector field $\mathcal{R}$ on the Lie algebra elements of $\mathcal{G}_{\text{Lie}}$. Hence, the rigid action of $\mathcal{R}$ on $\mathbb{M}$ induces a natural $U(1)$ group action on the gauge algebra $\mathcal{G}_{\text{Lie}}$, as well as on the space $\mathcal{A}$, something we shall see right below. Notice that \eqref{d} is antisymmetric as well if $d_{\mathbb{M}}\Omega\wedge d_{\mathbb{M}}\kappa=0$ holds.

Let us understand now, from the symplectic geometry point of view, what are the implications of the action of the vector field $\mathcal{R}$ on the space of gauge connections $\mathcal{A}$. 

The action of $\mathcal{R}$ on $\mathbb{M}$, induces the following vector field on $\mathcal{A}$, given by
\begin{equation}
V(p,0,0)=p\pounds _{\mathcal{R}}\mathbb{A}, \label{R vector field}
\end{equation}
where $p\in \mathbb{R}$. Notice that, because $\kappa, \Omega$ are invariant, i.e. $\pounds_{\mathcal{R}}\kappa=\pounds_{\mathcal{R}}\Omega=0$, the $U(1)$ action of $\mathcal{R}$ on $\mathcal{A}$, also descends to a corresponding action on the quotient space $\overline{\mathcal{A}}$. 

As done before with the action of the gauge group, now we compute the moment map $\mu$ associated to the action of the vector field \eqref{R vector field}. Compute then, the contraction on the lhs of the defining relation
\begin{equation}
-i_{V(p,0,0)}\hat{\Omega }=\hat{\delta }\left\langle \mu ,(p,0,0)
\right\rangle .
\end{equation}%
Consider the expression
\begin{equation}
-i_{V(p,0,0)}\hat{\Omega }=\dint\nolimits_{\mathbb{M}}\Omega \wedge \kappa
\wedge \text{Tr}\left( p\pounds _{\mathcal{R}}\mathbb{A}\wedge \hat{%
\delta }\mathbb{A}\right) 
\end{equation}
and notice the following result%
\begin{equation}
\pounds _{\mathcal{R}}\left( \Omega \wedge \kappa \wedge \text{Tr}\left( 
\mathbb{A}\wedge \hat{\delta }\mathbb{A}\right) \right) =\Omega \wedge
\kappa \wedge \text{Tr}\left( \pounds _{\mathcal{R}}\mathbb{A}\wedge 
\hat{\delta }\mathbb{A+A}\wedge \pounds _{\mathcal{R}}\left( \hat{%
\delta }\mathbb{A}\right) \right) .
\end{equation}%
Then, inside integrals, we can replace%
\begin{equation}
\Omega \wedge \kappa \wedge \text{Tr}\left( \pounds _{\mathcal{R}}\mathbb{A}%
\wedge \hat{\delta }\mathbb{A}\right) =\Omega \wedge \kappa \wedge 
\text{Tr}\left( \pounds _{\mathcal{R}}\left( \hat{\delta }\mathbb{A}%
\right) \wedge \mathbb{A}\right) .
\end{equation}
Using this fact above, we obtain that%
\begin{equation}
\left\langle \mu ,(p,0,0)\right\rangle =\frac{p}{2}\dint\nolimits_{\mathbb{M}%
}\Omega \wedge \kappa \wedge \text{Tr}\left( \pounds _{\mathcal{R}}\mathbb{A%
}\wedge \mathbb{A}\right) . \label{moment for R}
\end{equation}
The latter expression is manifestly invariant under the action of $\mathcal{S}$ and also descends to $\overline{\mathcal{A}}$.

The actions of $\mathcal{R}$ on $\mathcal{G}_{\text{Lie}}$ and $\mathcal{A}$ are intertwined. To see this, we compute the Poisson
algebra for two functionals of the form \eqref{moment for gauge} and \eqref{moment for R}. Thus,%
\begin{equation}
\big\{ \left\langle \mu ,(p,0,0)\right\rangle ,\left\langle \mu ,(0,\lambda,0)
\right\rangle \big\} =\hat{\delta }\left\langle \mu ,(0,\lambda,0)
\right\rangle \left( V(p,0,0)\right) =-p\dint\nolimits_{\mathbb{M}}\Omega
\wedge \kappa \wedge \text{Tr}\left( \pounds _{\mathcal{R}}\mathbb{A}\wedge
d_{\mathbb{A}}\lambda \right) .
\end{equation}
To simplify this expression, consider the following result%
\begin{equation}
d_{\mathbb{M}}\Big( \Omega \wedge \kappa \wedge \text{Tr}\left( \mathbb{A}%
\pounds _{\mathcal{R}}\lambda \right) \Big) =d_{\mathbb{M}}\left( \Omega
\wedge \kappa \right) \wedge \text{Tr}\left( \mathbb{A}\pounds _{\mathcal{R}%
}\lambda \right) +\Omega \wedge \kappa \wedge \text{Tr}\Big( d_{\mathbb{M}}%
\mathbb{A}\pounds _{\mathcal{R}}\lambda -\mathbb{A}\wedge d_{\mathbb{M}%
}\left( \pounds _{\mathcal{R}}\lambda \right) \Big).
\end{equation}
Then, inside the integral, we can write
\begin{equation}
\Omega \wedge \kappa \wedge \text{Tr}\Big( \mathbb{A}\wedge d_{\mathbb{M}%
}\left( \pounds _{\mathcal{R}}\lambda \right) \Big) =d_{\mathbb{M}}\left(
\Omega \wedge \kappa \right) \wedge \text{Tr}\left( \mathbb{A}\pounds _{%
\mathcal{R}}\lambda \right) +\Omega \wedge \kappa \wedge \text{Tr}\left( d_{%
\mathbb{M}}\mathbb{A}\pounds _{\mathcal{R}}\lambda \right) .
\end{equation}%
After acting with  $\pounds _{\mathcal{R}}$ on $d_{\mathbb{A}}\lambda $
and\ simplifying, we find that
\begin{equation}
\big\{ \left\langle \mu ,(p,0,0)\right\rangle ,\left\langle \mu ,(0,\lambda,0)
\right\rangle \big\} =\left\langle \mu ,\left(0, -p\pounds _{\mathcal{R}%
}\lambda,0 \right) \right\rangle . \label{R with F}
\end{equation}%
The last result follows because $\partial \mathbb{M}=0$, imply that
\begin{equation}
\dint\nolimits_{\mathbb{M}}\Omega \wedge \kappa \wedge \text{Tr}\Big( 
\mathbb{A}\wedge \mathbb{A}\pounds _{\mathcal{R}}\lambda -\mathbb{A\wedge 
}\pounds _{\mathcal{R}}\left[ \mathbb{A},\lambda \right] \Big)
=-\dint\nolimits_{\mathbb{M}}\pounds _{\mathcal{R}}\Big( \Omega \wedge
\kappa \wedge \text{Tr}\big( \mathbb{A\wedge A\lambda }\big) \Big)=0. 
\end{equation}

From these results, it turns out that the algebraic structure extending \eqref{ext Lie 1}, is given by the semi-direct product $U(1)\ltimes\tilde{\mathcal{G}}$ of the $U(1)$ group generated by the rigid action of $\mathcal{R}$ and the central extension of the gauge group $\tilde{\mathcal{G}}$ . As a vector space, the Lie algebra of $U(1)\ltimes\tilde{\mathcal{G}}$ is identified with $\mathbb{R}\oplus \tilde{\mathcal{G}}_{\text{Lie}}=\mathbb{R}\oplus\mathcal{G}_{\text{Lie}}\oplus \mathbb{R}$ and comes equipped with the following bracket \cite{Loop}
\begin{equation}
\Big[ \left( p,\eta ,a\right) ,\left( q,\lambda ,b\right) \Big] =\Big( 0,%
\left[ \eta ,\lambda \right] -p\pounds _{\mathcal{R}}\lambda +q\pounds _{%
\mathcal{R}}\eta ,c(\eta ,\lambda )\Big) . \label{total bracket}
\end{equation}%

Summarizing, the complete moment map associated to the action of the vector field
\begin{equation}
V(p,\eta ,a)=d_{\mathbb{A}}\eta +p\pounds _{\mathcal{R}}\mathbb{A} \label{total vector field}
\end{equation}
on the quotient space $\overline{\mathcal{A}}$, takes the form%
\begin{equation}
\left\langle \mu ,\left( p,\eta ,a\right) \right\rangle =\frac{p}{2}%
\dint\nolimits_{\mathbb{M}}\Omega \wedge \kappa \wedge \text{Tr}\left( 
\pounds _{\mathcal{R}}\mathbb{A}\wedge \mathbb{A}\right) -\dint\nolimits_{%
\mathbb{M}}\Omega \wedge \kappa \wedge \text{Tr}\left( \eta F_{\mathbb{A}%
}\right) -\dint\nolimits_{\mathbb{M}}d_{\mathbb{M}}\left( \Omega \wedge
\kappa \right) \wedge \text{Tr}\left( \eta \mathbb{A}\right) -a. \label{total moment}
\end{equation}%

The Poisson algebra for two functionals \eqref{total moment} is given by the desired expression, see \eqref{Homo},
\begin{equation}
\big\{ \left\langle \mu ,\left( p,\eta ,a\right) \right\rangle,\left\langle \mu ,\left(
q,\lambda ,b\right) \right\rangle  \big\} =\left\langle \mu ,
\left[ \left( p,\eta ,a\right) ,\left( q,\lambda ,b\right) \right]
\right\rangle .
\end{equation}%
Thus, the action of the group $U(1)\ltimes\tilde{\mathcal{G}}$ on $\overline{\mathcal{A}}$, is Hamiltonian. Its Lie algebra $\mathbb{R}\oplus \tilde{\mathcal{G}}_{\text{Lie}}$ admits \cite{Loop} a hyperbolic, non-degenerate and invariant inner product defined by
\begin{equation}
\Big( \left( p,\eta ,a\right) ,\left( q,\lambda ,b\right) \Big)
=-\dint\nolimits_{\mathbb{M}}\Omega \wedge \kappa \wedge d_{\mathbb{M}%
}\kappa \text{Tr}\left( \eta \lambda \right) -pb-qa, \label{inner product}
\end{equation}%
provided \eqref{d} vanish and
where the top-form given by\footnote{Below, we will see from duality arguments, why we have defined \eqref{inner product} in terms of the 4-form \eqref{Top}.}
\begin{equation}
\Omega \wedge \kappa \wedge d_{\mathbb{M}}\kappa\in\Omega_{\mathbb{M}}^{4} \label{Top}
\end{equation}
is globally defined and everywhere non-zero over $\mathbb{M}$. Notice that $\mathcal{H}=U(1)\ltimes\tilde{\mathcal{G}}$ and $\mathfrak{h}=\mathbb{R}\oplus \tilde{\mathcal{G}}_{\text{Lie}}$ because of $\mathcal{H}$ acts on the quotient space $\overline{\mathcal{A}}$ in a Hamiltonian fashion and $\mathfrak{h}$ is endowed with a well-defined invariant inner product. 

A comment is in order. The condition that \eqref{d} vanishes is absent in \cite{NA loc CS,Loop}, because of the cocycle used there is, roughly speaking, recovered from \eqref{cocycle expanded} by setting $\Omega\rightarrow 1$ and $\mathbb{M}\rightarrow \text{M}$. To show how such a condition emerges, we verify the invariance property of the inner product \eqref{inner product} under the adjoint action of $\mathfrak{h}$, which is given by
\begin{equation}
\Big( \left[ \left( p,\eta ,a\right) ,\left( q,\lambda ,b\right) \right]
,\left( r,\phi ,c\right) \Big) =\Big( \left( p,\eta ,a\right) ,\left[
\left( q,\lambda ,b\right) ,\left( r,\phi ,c\right) \right] \Big) . \label{inv}
\end{equation}
Then, \eqref{inv} boils down to
\begin{equation}
rd(\eta,\lambda)=pd(\lambda,\phi) \label{Inv 1}
\end{equation}
and because $r,p, \eta,\lambda, \phi$ are all arbitrary, we end up by enforcing that $d(\ast,\ast)=0$. We will interpret this as a condition to be imposed over the gauge parameters $\eta\in \Omega_{\mathbb{M}}^{0}\otimes\mathfrak{g}$. Although an expression generalizing \eqref{Inv 1} will be considered below in \eqref{Inv 2}, once we notice the existence of a second vector field $\mathcal{R}'$, besides $\mathcal{R}$, acting on $\mathcal{G}_{\text{Lie}}$ and $\mathcal{A}$ as well. More on this below.  

Once $\mathfrak{h}$ is equipped with an appropriate inner product, one is encouraged to use the definition \eqref{inner product} in order to dualize the moment map $\mu\in \mathfrak{h}^{*}$ and by this we mean solving the following equation%
\begin{equation}
\left\langle \mu ,\left( q,\lambda ,b\right) \right\rangle =\left( \mu
,\left( q,\lambda ,b\right) \right) ,
\end{equation}%
for an element $\mu =\left( p,\eta ,a\right)\in\mathfrak{h} $ on the right hand side (rhs). We quickly find that
\begin{equation}
\mu =\left( 1,\frac{\Omega \wedge \kappa \wedge F_{%
\mathbb{A}}+d_{\mathbb{M}}\left( \Omega \wedge \kappa \right) \wedge 
\mathbb{A}}{\Omega \wedge \kappa \wedge d_{\mathbb{M}}\kappa },-\frac{1}{2}%
\dint\nolimits_{\mathbb{M}}\Omega \wedge \kappa \wedge \text{Tr}\left( 
\pounds _{\mathcal{R}}\mathbb{A}\wedge \mathbb{A}\right) \right) . \label{dual total moment}
\end{equation}
A few words concerning the latter result. Any 4-form $\gamma\in\Omega_{\mathbb{M}}^{4}\otimes\mathfrak{g}$ is proportional to \eqref{Top} and can be written as
\begin{equation}
\gamma=\phi \, \Omega \wedge \kappa \wedge d_{\mathbb{M}}\kappa,
\end{equation}
for some $\phi\in\Omega_{\mathbb{M}}^{0}\otimes \mathfrak{g}$. Thus, `dividing' by \eqref{Top} actually means picking the term $\phi$, so that
\begin{equation}
\frac{\gamma}{\Omega \wedge \kappa \wedge d_{\mathbb{M}}\kappa}=\phi.
\end{equation}
This unusual notation, originally introduced in \cite{NA loc CS}, turns out to be very useful for performing algebraic calculations. 

Let us simplify the expression \eqref{dual total moment} further. Consider now the contracted 5-form%
\begin{equation}
0=i_{\mathcal{R}}\Big( \Omega \wedge \kappa \wedge d_{\mathbb{M}}\kappa
\wedge \mathbb{A}\Big) =-\Omega \wedge d_{\mathbb{M}}\kappa \wedge 
\mathbb{A+}\Omega \wedge \kappa \wedge d_{\mathbb{M}}\kappa i_{\mathcal{R}}%
\mathbb{A}\text{.}
\end{equation}%
From this follows that we can express the contraction $i_{\mathcal{R}}\mathbb{A}$ in an equivalent way%
\begin{equation}
i_{\mathcal{R}}\mathbb{A=}\frac{\Omega \wedge d_{\mathbb{M}}\kappa \wedge 
\mathbb{A}}{\Omega \wedge \kappa \wedge d_{\mathbb{M}}\kappa }. \label{cont}
\end{equation}%
Also, introduce the quantities
\begin{equation}
\Phi =\frac{\Omega \wedge \kappa \wedge F_{\mathbb{A}}+d_{\mathbb{M}%
}\Omega \wedge \kappa \wedge \mathbb{A}}{\Omega \wedge \kappa \wedge d_{%
\mathbb{M}}\kappa },\text{ \ \ }B=\frac{d_{\mathbb{M}}\Omega \wedge \kappa
\wedge \mathbb{A}}{\Omega \wedge \kappa \wedge d_{\mathbb{M}}\kappa }. \label{Phi and B}
\end{equation}%
In this way, we get a more compact expression for the moment map \eqref{dual total moment}, i.e.
\begin{equation}
\mu =\left( 1,\Phi -i_{\mathcal{R}}\mathbb{A},a\right) , \label{simpler}
\end{equation}%
where%
\begin{equation}
a=-\frac{1}{2}\dint\nolimits_{\mathbb{M}}\Omega \wedge \kappa \wedge \text{%
Tr}\left( \pounds _{\mathcal{R}}\mathbb{A}\wedge \mathbb{A}\right) . \label{a}
\end{equation}

Under the shift $^{\kappa }\mathbb{A}=\mathbb{A+\kappa }s$, we
have that
\begin{equation}
^{\kappa }\Phi =\Phi +s,\text{ \ \ }i_{\mathcal{R}}\left(^{\kappa }\mathbb{A}\right) =i_{%
\mathcal{R}}\mathbb{A}+s,\text{ \ \ }^{\kappa }a=a, \label{1 rule}
\end{equation}%
while under $^{\Omega }\mathbb{A}=\mathbb{A}+\Omega s'$,
we get
\begin{equation}
^{\Omega }\Phi =\Phi ,\text{ \ \ }i_{\mathcal{R}}(^{\Omega }\mathbb{A})=i_{%
\mathcal{R}}\mathbb{A},\text{ \ \ }^{\Omega }a=a. \label{2 rule}
\end{equation}%
To show this, we use the identity \eqref{identity-1}. Thus, the moment map $\mu\in \mathfrak{h}$ is, as expected, invariant under the action of the shift group $\mathcal{S}$
\begin{equation}
\mu (^{\kappa }\mathbb{A})=\mu (\mathbb{A}),\text{ \ \ }\mu (^{\Omega }\mathbb{A}%
)=\mu (\mathbb{A})
\end{equation}%
and it is a well-defined functional on the quotient space $\overline{\mathcal{A}}$. In particular, any object defined in terms of it is invariant too. 

Now we verify the equivariance property of $\mu$ under the action of $\mathfrak{h}$ on $\overline{\mathcal{A}}$. The co-adjoint action of an element $(q,\lambda,b)\in\mathfrak{h}$ on the moment map $\mu\in\mathfrak{h}^{*}$, is defined by
\begin{equation}
ad^{*}_{(p,\eta,a)}\mu\left(\mathbb{A}\right)=\mu\big(V(p,\eta,a) \big), \label{def}
\end{equation}
where we have emphasized that $\mu$ depends on $\mathbb{A}$. Recall that $V(p,\eta,a)$ represents the action of $(p,\eta,a)\in\mathfrak{h}$ on $\mathbb{A}$. By pairing this expression against an element $(q,\lambda,b)\in\mathfrak{h}$ and working out \eqref{total moment} to first order we find, after some algebra, that
\begin{equation}
\left \langle ad^{*}_{(p,\eta,a)}\mu\left(\mathbb{A}\right),(q,\lambda,b)\right \rangle=\left \langle \mu(\mathbb{A}), ad_{(p,\eta,a)}(q,\lambda,b)  \right \rangle,
\end{equation}
where
\begin{equation}
ad_{(p,\eta,a)}(q,\lambda,b)=[(p,\eta,a),(q,\lambda,b)]
\end{equation}
is the adjoint action of $(p,\eta,a)$ on $(q,\lambda,b)$, which is given by the bracket \eqref{total bracket}. 

For completeness, it is important to study how $\mu$ behaves under the action of $\tilde{\mathcal{G}}\subset \mathcal{H}$, i.e. under finite gauge transformations. After some algebra, we obtain cf. \cite{Loop}
\begin{equation}
\begin{aligned}
&\left\langle  \mu \left( ^{g}\mathbb{A}\right) ,\left( q,\lambda ,b\right)
\right\rangle =\left \langle Ad^{\ast}_{(0,g,a)}\mu(\mathbb{A}),(q,\lambda, b)   \right \rangle=\left\langle \mu(\mathbb{A}),Ad_{(0,g,a)}(q,\lambda, b)  \right\rangle \\
&=\left\langle \mu \left( \mathbb{A}\right) ,\left( q,g\lambda
g^{-1}+q\pounds _{\mathcal{R}}gg^{-1},b+\dint\nolimits_{\mathbb{M}}d_{%
\mathbb{M}}\left( \Omega \wedge \kappa \right) \wedge \text{Tr}\left( 
\mathbb{I\lambda }\right) +\frac{q}{2}\dint\nolimits_{\mathbb{M}}\Omega
\wedge \kappa \wedge \text{Tr}\left( \mathbb{I\wedge }\pounds _{\mathcal{R}}%
\mathbb{I}\right) \right) \right\rangle ,\label{Adjoint}
\end{aligned}
\end{equation}%
where $\mathbb{I}=g^{-1}d_{\mathbb{M}}g$. To first order in $\eta$ with $g=e^{\eta}$, we have%
\begin{equation}
\left\langle \mu \left( ^{g}\mathbb{A}\right) ,\left( q,\lambda ,b\right)
\right\rangle =\left\langle \mu \left( \mathbb{A}\right) ,\left( q,\lambda
,b\right) \right\rangle +\left\langle \mu \left( \mathbb{A}\right) ,\left(
0,\left[ \eta ,\lambda \right] +q\pounds _{\mathcal{R}}\eta ,c(\eta ,\lambda
)\right) \right\rangle ,
\end{equation}%
or%
\begin{equation}
\left\langle ad_{(0,\eta ,a)}^{\ast }\mu \left( \mathbb{A}\right) ,\left(
q,\lambda ,b\right) \right\rangle =\left\langle \mu \left( \mathbb{A}%
\right) ,\left[ \left( 0,\eta ,a\right) ,\left( q,\lambda ,b\right) \right]
\right\rangle ,
\end{equation}%
where%
\begin{equation}
ad_{(0,\eta ,a)}^{\ast }\mu \left( \mathbb{A}\right) =\mu \left( ^{g}%
\mathbb{A}\right) -\mu \left( \mathbb{A}\right), 
\end{equation}%
in agreement with \eqref{def} for $p=0$. 

Furthermore, by setting $\nu'=(q',\lambda ',b')\equiv Ad_{(0,g,a)}(q,\lambda,b)$ we may ask if the norm of $\nu=(q,\lambda ,b)$ is preserved under the Adjoint action of $\tilde{\mathcal{G}}$. From the inner product \eqref{inner product}, we find 
\begin{equation}
\begin{aligned}
\big( \nu',\nu' \big)&=-\dint\nolimits_{\mathbb{M}}\Omega \wedge \kappa \wedge d_{\mathbb{M}}\kappa \text{Tr}\big(\lambda'^{2}\big)-2q'b' \\
&=\big( \nu,\nu \big)-2q\dint\nolimits_{\mathbb{M}}d_{%
\mathbb{M}}\Omega \wedge \kappa  \wedge \text{Tr}\left( \mathbb{I} \big(\lambda+\frac{q}{2}g^{-1}\pounds_{\mathcal{R}}g \big) \right)-q^{2}\dint\nolimits_{\mathbb{M}}\Omega \wedge \kappa \wedge \text{Tr} \left(g^{-1}\pounds_{\mathcal{R}}g \mathbb{I \wedge I}    \right),
\end{aligned}
\end{equation}
where we have used $\pounds_{\mathcal{R}}\mathbb{I}=g^{-1}d_{\mathbb{M}}\left( \pounds_{\mathcal{R}}gg^{-1}  \right)g$ and integrated by parts with $\partial \mathbb{M}=0$. 
Now, using the contracted 5-form
\begin{equation}
0=i_{\mathcal{R}}\Big( \Omega \wedge \kappa \wedge \text{Tr}\big(\mathbb{I \wedge I \wedge I}    \big)   \Big)
=-\Omega \wedge \text{Tr}\big( \mathbb{I \wedge I \wedge I} \big)+3\Omega \wedge \kappa \wedge \text{Tr} \left(g^{-1}\pounds_{\mathcal{R}}g \mathbb{I \wedge I}    \right),
\end{equation}
we get
\begin{equation}
\Omega \wedge \kappa \wedge \text{Tr} \left(g^{-1}\pounds_{\mathcal{R}}g \mathbb{I \wedge I}    \right)=-\Omega \wedge \chi(g),\label{gen WZ term}
\end{equation}
where we have introduced a 3-form $\chi(g) \in \Omega_{\mathbb{M}}^{3}$, defined by
\begin{equation}
\chi(g)=-\frac{1}{3}\text{Tr}\left( \mathbb{I \wedge I \wedge I}  \right). \label{chi}
\end{equation}
Then, we obtain the equivalent form 
\begin{equation}
\big( \nu',\nu' \big)=\big( \nu,\nu \big)-2q\dint\nolimits_{\mathbb{M}}d_{%
\mathbb{M}}\Omega \wedge \kappa  \wedge \text{Tr}\left( \mathbb{I} \big(\lambda+\frac{q}{2}g^{-1}\pounds_{\mathcal{R}}g \big) \right)+q^{2}\dint\nolimits_{\mathbb{M}}\Omega \wedge \chi(g). \label{gauge inner}
\end{equation}
The second term on the rhs right above is related to the `obstruction' \eqref{d}, while the third term resembles the behavior of a Chern-Simons theory under the action of finite gauge transformations. It is a modified Wess-Zumino (WZ) term. This will be verified later on when we consider the quadratic expression $(\mu, \mu)$ for the dualized moment map $\mu$. To first order in $\eta$ the inner product is then invariant if \eqref{d} vanishes.

After this digression on symmetry properties, we now proceed to simplify the quantity $a$ defined in \eqref{a}. Let us write
\begin{equation}
\begin{aligned}
-2a =\dint\nolimits_{\mathbb{M}}\Omega \wedge \kappa \wedge \text{Tr}%
\left( \pounds _{\mathcal{R}}\mathbb{A}\wedge \mathbb{A}\right)=\dint\nolimits_{\mathbb{M}}\Omega \wedge \kappa \wedge \text{Tr}\Big(
d_{\mathbb{M}}\left( i_{\mathcal{R}}\mathbb{A}\right) \wedge \mathbb{A+}%
i_{\mathcal{R}}\left( d_{\mathbb{M}}\mathbb{A}\right) \wedge \mathbb{A}%
\Big) .
\end{aligned}
\end{equation}%
Using%
\begin{equation}
\begin{aligned}
d_{\mathbb{M}}\Big( \Omega \wedge \kappa \wedge \text{Tr}\left( i_{%
\mathcal{R}}\mathbb{AA}\right) \Big) 
=d_{\mathbb{M}}\left( \Omega \wedge
\kappa \right) \wedge \text{Tr}\left( i_{\mathcal{R}}\mathbb{AA}\right)
+\Omega \wedge \kappa \wedge \text{Tr}\Big( d_{\mathbb{M}}\left( i_{%
\mathcal{R}}\mathbb{A}\right) \wedge \mathbb{A+}i_{\mathcal{R}}\mathbb{A}%
d_{\mathbb{M}}\mathbb{A}\Big) ,
\end{aligned}
\end{equation}
we can write inside the integral%
\begin{equation}
\Omega \wedge \kappa \wedge \text{Tr}\Big( d_{\mathbb{M}}\left( i_{%
\mathcal{R}}\mathbb{A}\right) \wedge \mathbb{A}\Big) =-d_{\mathbb{M}%
}\left( \Omega \wedge \kappa \right) \wedge \text{Tr}\left( i_{\mathcal{R}}%
\mathbb{AA}\right) -\Omega \wedge \kappa \wedge \text{Tr}\left( i_{\mathcal{%
R}}\mathbb{A}d_{\mathbb{M}}\mathbb{A}\right) .
\end{equation}%
Now, consider the contracted 5-form%
\begin{equation}
\begin{aligned}
0=i_{\mathcal{R}}\Big( \Omega \wedge \kappa \wedge \text{Tr}\big( d_{%
\mathbb{M}}\mathbb{A}\wedge \mathbb{A}\big) \Big) 
=-\Omega \wedge 
\text{Tr}\left( d_{\mathbb{M}}\mathbb{A}\wedge \mathbb{A}\right) +\Omega
\wedge \kappa \wedge \text{Tr}\Big( i_{\mathcal{R}}\left( d_{\mathbb{M}}%
\mathbb{A}\right) \wedge \mathbb{A}+d_{\mathbb{M}}\mathbb{A}i_{\mathcal{R%
}}\mathbb{A}\Big) .
\end{aligned}
\end{equation}
From this follows that%
\begin{equation}
\Omega \wedge \kappa \wedge \text{Tr}\Big( i_{\mathcal{R}}\left( d_{%
\mathbb{M}}\mathbb{A}\right) \wedge \mathbb{A}\Big) =\Omega \wedge 
\text{Tr}\left( d_{\mathbb{M}}\mathbb{A}\wedge \mathbb{A}\right) -\Omega
\wedge \kappa \wedge \text{Tr}\left( d_{\mathbb{M}}\mathbb{A}i_{\mathcal{R}%
}\mathbb{A}\right) .
\end{equation}
Now, inserting both results above into their respective positions, gives%
\begin{equation}
\begin{aligned}
-2a=\dint\nolimits_{\mathbb{M}}\Bigl\{ \Omega \wedge \text{Tr}\left( 
\mathbb{A\wedge }d_{\mathbb{M}}\mathbb{A}\right) +2\Omega \wedge
\kappa \wedge &\text{Tr}\left( i_{\mathcal{R}}\mathbb{AA}\wedge \mathbb{A}%
\right)\\
& -2\Omega \wedge \kappa \wedge \text{Tr}\left( i_{\mathcal{R}}%
\mathbb{A}F_{\mathbb{A}}\right) -d_{\mathbb{M}}\left( \Omega \wedge
\kappa \right) \wedge \text{Tr}\left( i_{\mathcal{R}}\mathbb{AA}\right)\Bigr\} .
\end{aligned}
\end{equation}
We can simplify this expression even further by considering the contracted
5-form%
\begin{equation}
\begin{aligned}
0=i_{\mathcal{R}}\Big( \Omega \wedge \kappa \wedge \text{Tr}\left( \mathbb{%
A\wedge A}\wedge \mathbb{A}\right) \Big) 
=-\Omega \wedge \text{Tr}\left( 
\mathbb{A\wedge A}\wedge \mathbb{A}\right) +3\Omega \wedge \kappa \wedge 
\text{Tr}\left( i_{\mathcal{R}}\mathbb{AA}\wedge \mathbb{A}\right) .
\end{aligned}
\end{equation}%
Using this result and the definitions introduced above, allows to write
\begin{equation}
-2a=\dint\nolimits_{\mathbb{M}}\Omega \wedge CS\left( \mathbb{A}\right)
-\dint\nolimits_{\mathbb{M}}\Omega \wedge \kappa \wedge d_{\mathbb{M}%
}\kappa \text{Tr}\Big( \left( 2\Phi-B-i_{\mathcal{R}}\mathbb{A}\right) i_{%
\mathcal{R}}\mathbb{A}\Big) , \label{-2a}
\end{equation}
where
\begin{equation}
CS(\mathbb{A})=\text{Tr}\left( \mathbb{A}\wedge d_{\mathbb{M}}\mathbb{A+}%
\frac{2}{3}\mathbb{A\wedge A\wedge A}\right),
\end{equation}
being an element of $\Omega_{\mathbb{M}}^{3}$, is the well known Chern-Simons 3-form. 

\subsection{Generalized 4d CS theory action functional}

Armed with the previous results, now we are able to compute the square of the moment map \eqref{dual total moment} with respect to the inner product \eqref{inner product}. As an element of $\mathfrak{h}$, the moment map is of the form $\mu=(p,\eta,a)$, thus
\begin{equation}
\left( \mu ,\mu \right) =-\dint\nolimits_{\mathbb{M}}\Omega \wedge \kappa
\wedge d_{\mathbb{M}}\kappa \text{Tr}\left( \eta ^{2}\right) -2a,
\end{equation}%
equals
\begin{equation}
\left( \mu ,\mu \right) =\dint\nolimits_{\mathbb{M}}\Omega \wedge CS\left( 
\mathbb{A}\right) -\dint\nolimits_{\mathbb{M}}\Omega \wedge \kappa \wedge
d_{\mathbb{M}}\kappa \text{Tr}\left( \Phi ^{2}-Bi_{\mathcal{R}}\mathbb{A}%
\right) , \label{mu mu}
\end{equation}
by virtue of \eqref{simpler} and \eqref{-2a}. Equivalently, we have the final form (see \eqref{main 1})
\begin{equation}
\left( \mu ,\mu \right) =\dint\nolimits_{\mathbb{M}}\Omega \wedge CS\left( 
\mathbb{A}\right) -\dint\nolimits_{\mathbb{M}}\Omega \wedge \kappa \wedge
d_{\mathbb{M}}\kappa \text{Tr}\left( \Phi ^{2}%
\right)+\dint\nolimits_{\mathbb{M}}d_{\mathbb{M}}\Omega \wedge \kappa\wedge \text{Tr}\left( \mathbb{A}i_{\mathcal{R}}\mathbb{A}  \right) . \label{mu mu 2}
\end{equation}%

At this point, we see that the 1-form $\Omega$ corresponds to a generalization of the `twist' form $\omega$ entering the definition of the conventional 4d CS theory \eqref{1.1}. The action \eqref{mu mu 2} is also invariant under the following changes (rescalings)
\begin{equation}
\kappa \rightarrow t\kappa, \text{ \ \ }\Omega \rightarrow \Omega, \label{rescalings}
\end{equation}
for any non-zero function $t\in \Omega_{\mathbb{M}}^{0}$. The normalization condition $i_{\mathcal{R}}\kappa=1$, requires $\mathcal{R}\rightarrow t^{-1}\mathcal{R}$. In particular, we have that $\Phi \rightarrow t^{-1} \Phi$ and $\Omega \wedge \kappa \wedge d_{\mathbb{M}}\kappa \rightarrow t^{2}\Omega \wedge \kappa \wedge d_{\mathbb{M}}\kappa$ and the sign of \eqref{Top} is preserved under the arbitrary rescalings \eqref{rescalings}, reflecting the orientability of $\mathbb{M}$. %Furthermore, the local equivalence class of 1-forms $\kappa \sim t\kappa$ define a 1-dimensional subbundle of the cotangent bundle $T^{\ast}\mathbb{M}$ of $\mathbb{M}$. 

The $\mathcal{S}$-invariant action functional on the quotient space $\overline{\mathcal{A}}$, is then defined by
\begin{equation}
S=ic\left( \mu ,\mu \right). \label{Generalized 4CS action}
\end{equation}
The action \eqref{Generalized 4CS action} can be interpreted as a quadratic completion of the 4d CS theory defined in terms of the Lagrangian density $L\sim\omega\wedge CS(\mathbb{A})$. To find its functional variation, we use the following results
\begin{equation}
\begin{aligned}
\Omega \wedge \delta CS\left( \mathbb{A}\right)  &=2\Omega \wedge \text{Tr}\left(
\delta \mathbb{A}\wedge F_{\mathbb{A}}\right) +d_{\mathbb{M}}\Omega
\wedge \text{Tr}\left( \delta \mathbb{A}\wedge \mathbb{A}\right) , \\
\Omega \wedge \kappa \wedge d_{\mathbb{M}}\kappa \delta \text{Tr}\left(
\Phi ^{2}\right)  &=-2\Omega \wedge \text{Tr}\Big( \delta \mathbb{A}%
\wedge \left( \kappa \wedge d_{\mathbb{A}}\Phi -d_{\mathbb{M}}\kappa \Phi
\right) \Big) .
\end{aligned}
\end{equation}%
Then, 
\begin{equation}
\delta (\mu,\mu)=2\dint\nolimits_{\mathbb{M}}\Omega \wedge \text{Tr}\left( \delta 
\mathbb{A\wedge }F_{\mathbb{A }-\kappa \Phi}\right) +\dint\nolimits_{%
\mathbb{M}}d_{\mathbb{M}}\Omega \wedge \text{Tr}\left( \delta \mathbb{%
A\wedge A}\right) +\dint\nolimits_{\mathbb{M}}d_{\mathbb{M}}\Omega
\wedge \kappa \wedge \delta \text{Tr} \left( \mathbb{A}i_{\mathcal{R}}%
\mathbb{A}\right) ,
\end{equation}
where we have used \eqref{identity-1}. To simplify this expression further, consider the contracted 5-form%
\begin{equation}
0=i_{\mathcal{R}}\Big(\ d_{\mathbb{M}}\Omega \wedge \kappa \wedge \text{Tr}\left( \mathbb{%
A\wedge \delta A}\right)\Big)=d_{\mathbb{M}}\Omega\wedge \text{Tr}( \mathbb{A}\wedge \delta \mathbb{A})-
d_{\mathbb{M}}\Omega\wedge\kappa\wedge \text{Tr}(i_{\mathcal{R}}\mathbb{A} \delta \mathbb{A}-\mathbb{A}i_{\mathcal{R}}\delta\mathbb{A}). 
\end{equation}%
From this result, it follows that
\begin{equation}
d_{\mathbb{M}}\Omega \wedge \kappa \wedge \text{Tr}\left( \mathbb{A}i_{%
\mathcal{R}}\delta \mathbb{A}\right) =d_{\mathbb{M}}\Omega \wedge \text{Tr}%
\Big( \delta \mathbb{A\wedge }\left( \mathbb{A-\kappa }i_{\mathcal{R}}%
\mathbb{A}\right) \Big) .
\end{equation}%
Then, the general variation takes the form
\begin{equation}
\delta (\mu,\mu)=2\dint\nolimits_{\mathbb{M}}\Omega \wedge \text{Tr}\left( \delta 
\mathbb{A\wedge }F_{\mathbb{A}-\kappa \Phi}\right) +2\dint\nolimits_{%
\mathbb{M}}d_{\mathbb{M}}\Omega \wedge \text{Tr}%
\Big( \delta \mathbb{A\wedge }\left( \mathbb{A-\kappa }i_{\mathcal{R}}%
\mathbb{A}\right) \Big)
\end{equation}
and from this we get the equations of motion (eom) of the theory, which are given by
\begin{equation}
\Omega\wedge F_{\mathbb{A}-\kappa \Phi}=d_{\mathbb{M}}\Omega \wedge\left( \mathbb{A-\kappa }i_{\mathcal{R}}%
\mathbb{A}\right).\label{full eom}
\end{equation}

To understand better the r\^ole played by the object $\Phi $ introduced before in \eqref{Phi and B}, we rewrite such a definition in the equivalent form
\begin{equation}
\kappa \wedge \Big( \Omega \wedge F_{\mathbb{A}-\kappa \Phi }-d_{\mathbb{M%
}}\Omega \wedge \mathbb{A}\Big) =0. \label{def of Phi}
\end{equation}%
This means that $\Phi $ is actually determined by wedging the eom of the theory against the 1-form $\kappa $. Later on, we shall see how \eqref{def of Phi} raises as a genuine eom from a well-defined variational problem applied to an extended action functional $S(\mathbb{A},\Phi)$.

Now, let us we analyze the behavior of the quadratic expression \eqref{mu mu} under the action of the gauge group $\mathcal{G}$. Using \eqref{gauge transf}, we get the transformation properties of the CS 3-form under the action of finite gauge transformations, namely,
\begin{equation}
CS\left(^{g}\mathbb{A} \right)=CS(\mathbb{A})+d_{\mathbb{M}}\Big(\text{Tr}\left( \mathbb{A }\wedge\mathbb{J}\right)\Big) +\chi (g), \label{CS gauge change}
\end{equation}
where $\mathbb{J}=d_{%
\mathbb{M}}gg^{-1}$. Using the contracted 5-form
\begin{equation}
0=i_{\mathcal{R}}\Big( d_{\mathbb{M}}\Omega \wedge \kappa \wedge \text{Tr}%
\left( \mathbb{A\wedge J}\right) \Big) =d_{\mathbb{M}}\Omega \wedge
\text{Tr}\left( \mathbb{A\wedge J}\right) -d_{\mathbb{M}}\Omega \wedge \kappa
\wedge \text{Tr}\left( i_{\mathcal{R}}\mathbb{AJ-A}i_{\mathcal{R}}\mathbb{J%
}\right) ,\label{relation}
\end{equation}%
we find that%
\begin{equation}
\begin{aligned}
\Omega \wedge \kappa \wedge d_{\mathbb{M}}\kappa \text{Tr}\left(^{g}X^{2}\right) =\Omega \wedge \kappa  \wedge d_{%
\mathbb{M}}\kappa \text{Tr}\left( X^{2}
\right) +d_{\mathbb{M}}\Omega \wedge \text{Tr}\left( \mathbb{A\wedge J}%
\right) +d_{\mathbb{M}}\Omega \wedge \kappa \wedge \text{Tr}\big( \mathbb{%
J}Y \big) ,
\end{aligned}
\end{equation}
where 
\begin{equation}
\begin{aligned}
X^{2}=\Phi^{2}-Bi_{\mathcal{R}}\mathbb{A},\text{ \ \ } Y=-2\Phi +2i_{\mathcal{R}}\mathbb{A}-B(\mathbb{J})+i_{%
\mathcal{R}}\mathbb{J} \label{Y}
\end{aligned},
\end{equation}
with $B(\mathbb{J})$ given by the second relation of \eqref{Phi and B} with $\mathbb{J}$ in the place of $\mathbb{A}$. Putting all together gives, cf. \eqref{gauge inner}
\begin{equation}
^{g}(\mu,\mu)=(\mu,\mu)+\int_{\mathbb{M}}\Omega \wedge \chi (g)+\dint\nolimits_{\mathbb{M}}d_{%
\mathbb{M}}\Omega \wedge \kappa \wedge \text{Tr}\big( Y\mathbb{J }% 
 \big).\label{gauge change}
\end{equation}
The last contribution on the rhs right above is again related to the obstruction \eqref{d}, while the second term on the rhs is a generalization of the usual WZ shift $\chi(g)$ common to CS theories. Thus, for a path integral formulation of the theory it is reasonable to consider that further analytic constraints over the gauge group elements $g$ must, in principle, be imposed in order to obtain an honest gauge theory. In this paper we will adopt this point of view. 

The eom \eqref{full eom} also reflects a lack of gauge covariance. If we demand that the eom \eqref{full eom} are preserved under gauge transformations, i.e. if 
\begin{equation}
\Omega\wedge F_{^{g}\mathbb{A}-\kappa ^{g}\Phi}=d_{\mathbb{M}}\Omega \wedge\left( ^{g}\mathbb{A}-\kappa i_{\mathcal{R}}
(^{g}\mathbb{A})\right), \label{gauge full eom}
\end{equation}
then an equation for $g$ emerge. Indeed, using
\begin{equation}
^{g}\Phi=g^{-1}\left (\Phi+B(\mathbb{J})    \right )g \label{on-shell Phi under gauge}
\end{equation}
and \eqref{full eom}, we find that \eqref{gauge full eom} holds, provided $g\in \mathcal{G}$ satisfy the following equation
\begin{equation}
\Omega\wedge \kappa \wedge d_{\mathbb{A}}B(\mathbb{J})=0. \label{boring eq}
\end{equation} 
This equation can be solved $\forall \mathbb{A}$, if we choose $g$ to be such that $B(\mathbb{J})=0$. Notice that \eqref{d} can be written alternatively as
\begin{equation}
d(\eta,\lambda)=-\int_{\mathbb{M}}\Omega \wedge \kappa \wedge d_{\mathbb{M}}\kappa
\text{Tr}\left ( \eta B(d_{\mathbb{M}}\lambda)  \right ). \label{d2}
\end{equation}
Then, if $B(j)=0$ for $j=\mathbb{I}=\mathbb{J}=0$, we have that \eqref{d2} vanishes as well. In this case, \eqref{on-shell Phi under gauge} reduces to
\begin{equation}
^{g}\Phi=g^{-1}\Phi g
\end{equation}
and the field $\Phi$ then transforms in the adjoint. Below, we will see how \eqref{d} extends the usual `boundary' condition imposed over the gauge parameters $\eta$ in conventional 4d CS theories, that they must vanish at the set of poles of the twist 1-form $\omega$. 

We now proceed to verify if the quadratic action \eqref{Generalized 4CS action} is a real number. To show it, we invoke the same reality conditions used in \cite{unifying}, when applied to the conventional 4d CS theories. In the present case, these conditions impose suitable properties to be obeyed by the 1-forms $\Omega, \kappa$ and the gauge connection $\mathbb{A}$. Notice that all manipulations done above have been at the formal level and, as a consequence of this, it should not come as a surprise to find that some objects are purely imaginary instead of real. This is because we have not specified any reality conditions on the Lie algebra $\mathfrak{g}$, the gauge connection $\mathbb{A}$, the central extensions, the manifold $\mathbb{M}$ and so on. 

Let us first declare that the Lie algebra $\mathfrak{g}$ is actually a complex Lie algebra. Let $\tau:\mathfrak{g}\rightarrow \mathfrak{g} $ be an anti-linear involutive automorphism. It provides $\mathfrak{g}$ with an action of the cyclic group $\mathbb{Z}_{2}$. Its fixed point
subset is a real Lie subalgebra $\mathfrak{g}_{\mathbb{R}}$ of $\mathfrak{g}$, regarded itself as a real Lie algebra. The
anti-linear involution $\tau$ is compatible with the bilinear form on $\mathfrak{g}$, in the sense that
\begin{equation}
\overline{\text{Tr}(ab)}=\text{Tr}(\tau (a)\tau (b)),
\end{equation}
for any $a,b$ in the Lie algebra $\mathfrak{g}$ or by extension in $ \mathcal{G}_{\text{Lie}}$. 
Denote by $x$ a set of local coordinates on the manifold $\mathbb{M}$ and endow it with a complex structure. The complex conjugation $x\rightarrow \overline{x}$ defines an involution $\nu : \mathbb{M}\rightarrow \mathbb{M}$, which also provides $\mathbb{M}$ with a $\mathbb{Z}_{2}$ action. We then require $\gamma \in \Omega_{\mathbb{M}}^{p}$ and $\rho\in \Omega_{\mathbb{M}}^{p}\otimes \mathfrak{g}$ to be equivariant under this action of $\mathbb{Z}_{2}$, i.e. 
\begin{equation}
\overline{\gamma}=\nu^{\ast}\gamma, \text{ \ \ } \tau(\rho)=\nu^{*}\rho. \label{reality cond}
\end{equation}

For example, $\gamma$ represents any of the differential forms $\Omega, \kappa$ and their exterior derivatives and $\rho$ represents the gauge connection $\mathbb{A}$ the Lie algebra valued functions $\Phi$ and the contraction $i_{\mathcal{R}}\mathbb{A}$. The rest of the proof goes exactly like in \cite{unifying} (see \S 2.5) after noticing the fact that $\nu$ has the effect of
conjugating the complex structure on $\mathbb{M}$ and thus also of reversing its orientation. Let us perform a sample computation to see how this works. Consider the first term in \eqref{mu mu 2} and conjugate it, then
\begin{equation}
\begin{aligned}
\int_{\mathbb{M}}\overline{\Omega }\wedge \overline{CS(\mathbb{A})}
&=\int_{\mathbb{M}}\overline{\Omega }\wedge CS(\tau (\mathbb{A)})=\int_{%
\mathbb{M}}\nu ^{\ast }\Omega \wedge CS(\nu ^{\ast }\mathbb{A})
\\
&=\int_{\mathbb{M}}\nu ^{\ast }\Big( \Omega \wedge CS(\mathbb{A}%
)\Big) =\int_{\nu \left( \mathbb{M}\right) }\Omega \wedge CS(%
\mathbb{A})=-\int_{\mathbb{M}}\Omega \wedge CS(\mathbb{A}).
\end{aligned}
\end{equation}

After repeating for each contribution in \eqref{mu mu 2} (or in the form \eqref{S's} below), we end up with
\begin{equation}
\overline{(\mu,\mu)}=-(\mu,\mu).
\end{equation} 
As a consequence, the action \eqref{Generalized 4CS action} obeys $\overline{S}=S$ and $S\in \mathbb{R}$. The gauge transformations are also required to preserve the conditions \eqref{reality cond} for $\rho=\mathbb{A}$, which is equivalent to restricting the gauge elements to the ones satisfying the condition $\tau(g)=\nu^{\ast}g$. In the latter expression, $\tau$ also denotes the lift
of the antilinear automorphism $\tau$ acting on $\mathcal{G}_{\text{Lie}}$ to the group $\mathcal{G}$. With this, the last contribution to \eqref{gauge change} also flips sign. After repeating the same procedure for the pre-symplectic form \eqref{pre-symplectic}, we conclude that $\underline{\hat{\Omega}}=i\hat{\Omega}$ is real.

For future reference, now we introduce a second vector field $\mathcal{R}' \in \mathfrak{X}_{\mathbb{M}}$ besides $\mathcal{R}$, defined by the following set of conditions
\begin{equation}
i_{\mathcal{R}'}\kappa =0,\text{ \ \ }i_{\mathcal{R}'}\left( d_{\mathbb{M}%
}\kappa \right) =0 \label{basic rel R prime}
\end{equation}
and
\begin{equation}
i_{\mathcal{R}'}\Omega =1,\text{ \ \ }\pounds_{\mathcal{R}'}\Omega =0. \label{basic rel for R prime}
\end{equation}%
As a consequence of \eqref{basic rel R prime} and \eqref{basic rel for R prime}, we also have that%
\begin{equation}
\pounds _{\mathcal{R}'}\kappa =0,\text{ \ \ }i_{\mathcal{R}'}\left( d_{\mathbb{M}%
}\Omega \right) =0. \label{consequences 2}
\end{equation}

The obstruction \eqref{d} can be written in terms of the vector field $\mathcal{R}'$ as well. In fact, by using the contracted 5-form 
\begin{equation}
0=i_{\mathcal{R}^{\prime }}\Big( d_{\mathbb{M}}\Omega \wedge \Omega \wedge
\kappa \wedge \text{Tr}\left( \eta d_{\mathbb{M}}\lambda \right)\Big)
=d_{\mathbb{M}}\Omega \wedge \kappa \wedge \text{Tr}\left( \eta d_{\mathbb{%
M}}\lambda \right) +d_{\mathbb{M}}\Omega \wedge \Omega \wedge \kappa \text{%
Tr}\left( \eta \pounds _{\mathcal{R}^{\prime }}\lambda \right) ,
\end{equation}
we find that
\begin{equation}
d(\eta ,\lambda )=\int\nolimits_{\mathbb{M}}d_{%
\mathbb{M}}\Omega\wedge \Omega \wedge \kappa \text{Tr}\left( \eta \pounds _{\mathcal{R}^{\prime
}}\lambda \right) . \label{d in pounds}
\end{equation}
A possible solution to the condition $B(d_{\mathbb{M}}g)=0$, may be taken to be of the form
\begin{equation}
d_{\mathbb{M}}g=\kappa \pounds_{\mathcal{R}}g+\gamma(g), \label{eq over g}
\end{equation}
where\footnote{Not to be confused with the $\gamma$ used above.} $\gamma \in \text{ker}_{i_{\mathcal{R}}}\cap \text{ker}_{i_{\mathcal{R}'}}$ is such that $d_{%
\mathbb{M}}\Omega\wedge \gamma=0$. Notice that $\pounds_{\mathcal{R}'}g=0$. 

Now we have formally constructed the generalized 4d CS theory action functional and studied some of its properties, the next natural step is to explore how it is related to the conventional 4d CS theory, i.e. we would like to know what the actions \eqref{Generalized 4CS action} and \eqref{1.1} have in common. Fortunately, answering this question is precisely the content of the next section. 

\section{Relation to the 4d Chern-Simons theory}\label{3} 

We now solve explicitly the conditions \eqref{Condition on kappa} and \eqref{Condition on omega} imposed over $\kappa$ and $\Omega$ in a simple way. To do so, we take the 4-dimensional manifold $\mathbb{M}$ to be of the form $\mathbb{M}=\mathbb{R}\times \text{M}$, where the $\mathbb{R}$ factor is identified with the time direction and where $\text{M}$ is taken to be the total space of a non-trivial circle bundle of degree $n\neq 0$ over a Riemann surface $C$. The time direction $\mathbb{R}$ and the typical fiber $S^{1}$ of M being provided by the cylinder $\Sigma=\mathbb{R}\times S^{1}$. Thus, we have that M is defined by
\begin{equation}
S^{1}\overset{n}\longrightarrow \text{M}\overset{\underline{\pi}}{\longrightarrow} C. \label{bundle}
\end{equation}
This is a natural generalization of the space $\mathbb{M}=\Sigma\times C$ originally used to define the 4d CS theories, in which $\text{M}=S^{1}\times C$ is a trivial $S^{1}$ fiber bundle over the base manifold $C$. In the present case, the non-trivial bundle structure in \eqref{bundle} provides extra room for introducing some geometric structures that can be exploited in order to generalize the 4d CS theories. In particular, the circle bundle \eqref{bundle} admits the existence of invariant contact structures defined on the total space M. Notice that locally $\mathbb{R}\times \text{M}$ and $\Sigma\times C$ coincide. We follow \cite{NA loc CS} closely. 

Consider the following solutions $\kappa,\Omega\in\Omega_{\mathbb{M}}^{1}$ to the conditions \eqref{Condition on kappa} and \eqref{Condition on omega}, given by
\begin{equation}
\kappa =\kappa_{\tau}d\tau+\tilde{\kappa},\text{ \ \ }%
\Omega =\Omega_{\tau}d\tau+\tilde{\Omega},\label{solutions}
\end{equation}%
where
\begin{equation}
\kappa_{\tau}=\frac{1}{2}\alpha_{\tau},\text{ \ \ }\tilde{\kappa}=\frac{1}{2}\alpha\text{ \ \ }\Omega_{\tau}=2\zeta,
\text{ \ \ }\tilde{\Omega}=\omega-\frac{\Omega_{\tau}}{\alpha_{\tau}}\alpha.  \label{solutions 2}
\end{equation}
We have chosen $\alpha_{\tau }\neq 0 \in\mathbb{R}$ and $\zeta>0\in \mathbb{R}$ to be non-zero constants, $\alpha\in\Omega_{\text{M}}^{1} $ to be a
contact\footnote{Any compact, orientable 3-manifold possesses a contact structure \cite{Martinet}. Recall that $\mathbb{M}=\mathbb{R}\times \text{M}$ is assumed to be orientable and such that $\partial\mathbb{M}=0$.} 1-form on the 3-dimensional manifold M and $\omega =\underline{\pi} ^{\ast }\omega _{C}\in\Omega_{\text{M}}^{1}$, with $\omega _{C}\in\Omega_{C}^{1,0}$ to be a $(1,0)$-form on the Riemann surface $C$. The `twist' 1-form $\omega_{C}$ is a meromorphic differential with a set of zeroes and poles on $C$ denoted, respectively, by $\mathfrak{z}$ and $\mathfrak{p}$. Notice that, in order for the solutions \eqref{solutions} to exists, the manifold M must admit a contact structure, hence $(\text{M},\alpha)$ is required to be a contact manifold\footnote{For a comprehensive description of contact manifolds, the reader is referred to the references \cite{Etnyre,Blair,Geiges} }. 
Furthermore, over \eqref{solutions} we have imposed the conditions $\partial_{\tau}\alpha=0$ and $\partial_{\tau}\omega=0$. Concerning the rescalings \eqref{rescalings}, we notice that $\kappa \rightarrow t \kappa$, implies
\begin{equation}
\alpha_{\tau} \rightarrow t \alpha_{\tau},\text{ \ \ } \alpha \rightarrow t \alpha
\end{equation}
and from this follows that $\Omega$ is invariant, as initially assumed in \eqref{rescalings}. 

Split now the exterior derivative 
\begin{equation}
d_{\mathbb{M}}=d\tau\wedge \partial_{\tau}+d_{\text{M}}, \label{differential split}
\end{equation}
where $d_{\text{M}}$ denotes the exterior differential on M.  

The total space M admits a free $U(1)$ action arising from the rotations of the fibers $S^{1}$ and also admits a natural contact structure, which is invariant under the action of $U(1)$. Such a contact 1-form $\alpha $ is introduced via the Boothby-Wang
construction \cite{Boothby-Wang}, see also \cite{Blair,Geiges}. This means that $\alpha$ is defined by the relation
\begin{equation}
d_{\text{M}}\alpha =n\underline{\pi} ^{\ast }\sigma _{C}, \label{Boothby-Wang}
\end{equation}%
where $\sigma _{C}\in\Omega_{C}^{1,1}$ is a symplectic 2-form on $C$, normalized to one, i.e.
\begin{equation}
\dint\nolimits_{C}\sigma _{C}=1
\end{equation}%
and where $n>0$ (after a suitable choice of orientation) is the degree of the bundle M. The geometric meaning of the relation \eqref{Boothby-Wang} is that $\alpha$ defines a $U(1)$-connection on M regarded now as the total space of a $U(1)$ principal bundle over $C$, that is induced by the symplectic form $\sigma_{C}$ and has a non-trivial curvature given by $d_{\text{M}}\alpha$. The contact 1-form $\alpha$ constructed in this way is $U(1)$ invariant, globally defined on M and satisfies the contact condition\footnote{In general, if $(\text{M},\alpha)$ is a $2n+1$ dimensional contact manifold, the contact condition is $\alpha \wedge (d_{\text{M}}\alpha)^{n}\neq 0$. Compare this against its analogue on the $2n$ dimensional symplectic manifold $(\text{M},\sigma)$, given by $\sigma^{n}\neq 0$.} that the top-form $\alpha\wedge d_{\text{M}}\alpha\in \Omega_{\text{M}}^{3}$ is nowhere vanishing all over M. 

The action of $U(1)$ along the fibers is generated by the Reeb vector field $R\in\mathfrak{X}_{\text{M}}$. It is a non-vanishing vector field, globally defined over M and canonically associated to $\alpha$ by the normalization condition $\alpha (R)=i_{R}\alpha=1$. The 1-form $\alpha$ is invariant because, as a connection 1-form on $\text{M}$, it separates any tangent space $T_{p}\text{M}$ over $p\in\text{M}$ into vertical and horizontal spaces. R is vertical and $d_{\text{M}}\alpha$ is a horizontal 2-form, thus $\pounds_{R}\alpha=0$. Furthermore, the integral of $\alpha$ over any fiber $S^{1}$ is normalized to one, i.e.
\begin{equation}
\dint\nolimits_{S^{1}}\alpha =1.
\end{equation}% 
Hence, after a fiber integration, we get that
\begin{equation}
\dint\nolimits_{\text{M}}\alpha \wedge d_{\text{M}}\alpha =n\dint\nolimits_{\text{M}}\alpha\wedge\underline{\pi} ^{\ast }\sigma _{C}=n\dint\nolimits_{C}\sigma _{C}=n \label{int over M}
\end{equation}
is never zero because, by assumption, the bundle M is non-trivial.

Returning to the solutions \eqref{solutions}, for the vector field $\mathcal{R}\in\mathfrak{X}_{\mathbb{M}}$, we have taken
\begin{equation}
\mathcal{R=} \frac{1}{\alpha _{\tau }}\partial _{\tau }+R.  \label{time+Reeb}
\end{equation}% 
All conditions written in \eqref{Condition on kappa} and \eqref{Condition on omega} are satisfied after noticing that $\omega$ is a horizontal 1-form, i.e. it is obtained by pulling $\omega_{C}$ back from $C$ to $\mathbb{M}$, with vanishing $d\tau$ component. Actually, both conditions are solved in general by first considering
\begin{equation}
\kappa=\frac{1}{2}\Big(\alpha_{\tau}d\tau + \alpha    \Big),\text{ \ \ }\mathcal{R=} f_{\tau }\partial _{\tau }+R,
\end{equation}
for $\alpha_{\tau},f_{\tau}\in \Omega_{\mathbb{M}}^{0}$ arbitrary non-zero functions. The normalization condition $i_{\mathcal{R}}\kappa=1$, requires that $f_{\tau}=1/ \alpha_{\tau}$ and $i_{\mathcal{R}}\left (  d_{\mathbb{M}}\kappa \right )=0$ is equivalent to $\partial_{\tau}\alpha-d\alpha_{\tau}=0$. The condition $i_{\mathcal{R}}\Omega=0$ can be solved by taking
\begin{equation}
\Omega=\omega+\Omega_{\tau}\Big (d\tau-\frac{\alpha}{\alpha_{\tau}}     \Big ),
\end{equation} 
such that $i_{\mathcal{R}}\omega=0$ and with $\Omega_{\tau}\in \Omega_{\mathbb{M}}^{0}$ being an arbitrary non-zero function. Finally, the invariance condition $\pounds_{\mathcal{R}}\Omega=0$ implies $i_{\mathcal{R}}\left (  d_{\mathbb{M}}\Omega \right )=0$, which is equivalent to
\begin{equation}
i_{\mathcal{R}}\left (d_{\mathbb{M}}\omega+d_{\mathbb{M}}\Omega_{\tau}\wedge\Big(d\tau-\frac{\alpha}{\alpha_{\tau}}   \Big)-\Omega_{\tau}\Big(\frac{d_{\mathbb{M}}\alpha}{\alpha_{\tau}}-\frac{d_{\mathbb{M}}\alpha_{\tau}\wedge \alpha}{\alpha_{\tau}^{2}}   \Big)   \right )=0.
\end{equation} 
Taking $\alpha$ and $\omega$ as defined above and restricting $\Omega_{\tau}$ to be a constant, requires $\alpha_{\tau}$ to be a constant too. Thus, in \eqref{solutions 2} we have chosen a simple non-trivial solution. The explicit form of the vector field $\mathcal{R}'\in\mathfrak{X}_{\mathbb{M}}$ introduced above in the last section, will be given in due course.   

Now, we consider a key result concerning the 4-form \eqref{Top} introduced above. Notice that%
\begin{equation}
\omega \wedge d_{\text{M}}\alpha =n\underline{\pi} ^{\ast }\omega _{C}\wedge \underline{\pi} ^{\ast }\sigma
_{C}=n\underline{\pi} ^{\ast }\left( \omega _{C}\wedge \sigma _{C}\right) =0, \label{Clever}
\end{equation}%
because of $\omega _{C}\wedge \sigma _{C}\in\Omega_{C}^{2,1}$ is a $(2,1)$-form on $C$, hence it vanishes by dimensionality reasons. Then, the result \eqref{Clever} imply that the 4-dimensional top-form \eqref{Top} takes the more explicit form
\begin{equation}
\Omega \wedge \kappa \wedge d_{\mathbb{M}}\kappa =\zeta d\tau \wedge \alpha
\wedge d_{\text{M}}\alpha . \label{Top explicit}
\end{equation}
This expression is globally defined on $\mathbb{M}$ and nowhere vanishing. It is basically proportional to the volume form $d\text{Vol}_{\mathbb{M}}$ of the manifold $\mathbb{M}$. 
If we integrate \eqref{Top explicit} over $\mathbb{M}$, we get
\begin{equation}
\dint\nolimits_{\mathbb{M}}\Omega \wedge \kappa \wedge d_{\mathbb{M}}\kappa=n\zeta \Delta\tau, \label{undo}
\end{equation} 
where we have integrated the $\tau$ direction over a finite interval of size $\Delta\tau$. At this point, it is interesting to compare \eqref{Top explicit} with the top-form obtained from the symplectisation \cite{Geiges} $\left (\mathbb{R}\times \text{M},\sigma=d_{\mathbb{M}}\left (e^{\zeta \tau /2}\alpha   \right )\right )$ of the contact manifold $(\text{M}, \alpha)$, which is given by
\begin{equation}
\sigma \wedge \sigma =\zeta e^{\zeta \tau}d\tau \wedge \alpha \wedge d_{\text{M}}\alpha.
\end{equation}
The factor $e^{\zeta\tau}$ can be absorbed into $\sigma$ but this spoils the closedness of the symplectic form $\sigma$. \\
In what follows, we denote $d_{\text{M}}=d$ in order to avoid clutter.

To see more explicitly how the action \eqref{Generalized 4CS action} generalize the conventional 4d Chern-Simons theories, we proceed by interpreting $\zeta$
as a deformation parameter. Exhibiting $\zeta$ in the expressions%
\begin{equation}
\begin{aligned}
\Omega \wedge \kappa  &=\omega \wedge \kappa +2\zeta d\tau \wedge \alpha ,
\\
d_{\mathbb{M}}\Omega \wedge \kappa  &=d\omega \wedge \kappa -\frac{2\zeta 
}{\alpha _{\tau }}\kappa \wedge d\alpha , \\
\Omega \wedge d_{\mathbb{M}}\kappa  &=\zeta 
\Big( d\tau -\frac{\alpha}{\alpha_{\tau}} \Big) \wedge d\alpha , \label{defining results}
\end{aligned}
\end{equation}
we find that
\begin{equation}
\Phi =\frac{\Phi _{-1}}{\zeta }+\Phi _{0},\text{ \ \ }i_{\mathcal{R}}\mathbb{A=}\left( i_{\mathcal{R}}\mathbb{%
A}\right) _{0}.
\end{equation}%
Right above, we get
\begin{equation}
\Phi _{-1}=\frac{\omega \wedge \kappa \wedge F_{\mathbb{A}}+d\omega \wedge
\kappa \wedge \mathbb{A}}{d\tau \wedge \alpha \wedge d\alpha },\text{ \ \ }%
\Phi _{0}=2\frac{d\tau \wedge \alpha \wedge F_{\mathbb{A}}-\alpha _{\tau
}^{-1}\kappa \wedge d\alpha \wedge \mathbb{A}}{d\tau \wedge \alpha \wedge
d\alpha },
\end{equation}%
and%
\begin{equation}
i_{\mathcal{R}}\mathbb{A=}\frac{\left( d\tau -\alpha / \alpha_{\tau} \right) \wedge d\alpha \wedge \mathbb{A}}{d\tau \wedge
\alpha \wedge d\alpha }.
\end{equation}

Now, inserting these results into the action \eqref{mu mu 2}, we obtain the following $\zeta$ expansion
\begin{equation}
S=S_{0}+\zeta S_{1}+\zeta ^{-1}S_{-1}, \label{expansion zeta}
\end{equation}%
where%
\begin{equation}
\begin{aligned}
S_{0} &=ic\dint\nolimits_{\mathbb{M}}\omega \wedge CS\left( \mathbb{A}%
\right) -2ic\dint\nolimits_{\mathbb{M}}d\tau \wedge \alpha \wedge d\alpha 
\text{Tr}\left( \Phi _{0}\Phi _{-1} \right)+ic\dint\nolimits_{\mathbb{M}}d\omega\wedge \kappa\wedge \text{Tr}\left( \mathbb{A}i_{\mathcal{R}}\mathbb{A} \right)
, \\
S_{1} &=2ic\dint\nolimits_{\mathbb{M}}\left( d\tau -\frac{\alpha}{\alpha_{\tau}} \right) \wedge CS\left( \mathbb{A}\right) -ic\dint\nolimits_{%
\mathbb{M}}d\tau \wedge \alpha \wedge d\alpha \text{Tr}\left( \Phi
_{0}^{2}\right)-2i\dint\nolimits_{\mathbb{M}}\frac{\kappa\wedge d\alpha}{\alpha_{\tau}} \wedge \text{Tr}\left( \mathbb{A}i_{\mathcal{R}}\mathbb{A} \right), \\
S_{-1} &=-ic\dint\nolimits_{\mathbb{M}}d\tau \wedge \alpha \wedge d\alpha 
\text{Tr}\left( \Phi _{-1}^{2}\right) . \label{S's}
\end{aligned}
\end{equation}%
The first term in the rhs of the first line above matches perfectly with the 4d Chern-Simons theory \eqref{1.1}. Thus, if we are interested in recovering the 4d CS theories it is desirable to find a way to do so. Fortunately, inspired by \eqref{expansion zeta} we can implement the following two-step strategy:
\begin{itemize}
\item Step I, we gauge fix the $\kappa$-shift
symmetry by imposing the gauge fixing condition $\Phi \approx 0$. This step simplifies drastically the expressions \eqref{S's} and we end up with an expansion \eqref{expansion zeta} involving only the powers $\zeta^{0}=1$ and $\zeta^{1}=\zeta$ in the deformation parameter $\zeta$. Because of $^{\kappa}\Phi=\Phi+s$, this gauge fixing condition is accessible. Yet,
we still need to verify if it is a good gauge fixing condition. This requires running the Dirac algorithm. \\
The partially gauged fixed theory is still invariant under the $\Omega$-shifts. 
\item Step II, we take $\zeta \rightarrow 0$ at the end. This is a degenerate limit rendering several expressions ill-defined. For instance, from \eqref{defining results} we realize that in this limit, the 2-form $\Omega\wedge\kappa$ vanishes at the set of zeroes $\mathfrak{z}^{\prime}$ of the twist 1-form $\omega$ and the term $d_{\mathbb{M}}\Omega \wedge \kappa$ localizes at the set of poles $\mathfrak{p}^{\prime}$ of $\omega$. In the first case, the pre-symplectic form \eqref{pre-symplectic} vanishes at the set $\mathfrak{z}^{\prime}$ and from \eqref{Top explicit}, we have that the first contribution to the inner product \eqref{inner product} is absent. Thus, $\zeta$ can be interpreted also as a regularizing parameter. 
\end{itemize}

Let us notice that these steps make no sense if performed in reverse order. However, by assuming everything is fine we get, after setting $\Phi=\zeta= 0$, the partially gauge fixed action%
\begin{equation}
S=ic\dint\nolimits_{\mathbb{M}}\omega \wedge CS\left( \mathbb{A}\right)
+ic\dint\nolimits_{\mathbb{M}}d\omega \wedge \kappa \wedge \text{Tr}\left( 
\mathbb{A}i_{\mathcal{R}}\mathbb{A}\right) . \label{zeta 0 limit action}
\end{equation}%
When $\zeta \rightarrow 0$, $\Omega =\omega $ and the
$\Omega$-shift symmetry is reduced to $^{\omega }\mathbb{A}=\mathbb{A}+s\omega $. There is a new `boundary'
term in \eqref{zeta 0 limit action} that is not present in the original theory\footnote{Boundary terms are understood as those contributions to the action that localize at the set of poles $\mathfrak{p}\in C$. In the present case, this can be seen by using local bundle coordinates.}. However, by demanding that this
contribution vanish, we can fix part of the analytic structure of the connection $\mathbb{A}$ that later on will define the Lax connection $\mathscr{L}$ of an integrable field theory associated to the 4d CS theory. We will work out this explicitly in the next section. Then, if the field $\mathbb{A}$ satisfies the condition
\begin{equation}
\dint\nolimits_{\mathbb{M}}d\omega \wedge \kappa \wedge \text{Tr}\left( 
\mathbb{A}i_{\mathcal{R}}\mathbb{A}\right) =0, \label{new bdry condition}
\end{equation}
the action functional \eqref{zeta 0 limit action} formally reduces, to that of the 4d Chern-Simons theory \eqref{1.1}
\begin{equation}
S=ic\dint\nolimits_{\mathbb{M}}\omega \wedge CS\left( \mathbb{A}\right). \label{hol CS action}
\end{equation}
The only trace of the non-triviality of the circle bundle M over $C$ lies in the integration domain $\mathbb{M}$ and locally, the generalized and the original theory \eqref{1.1} coincide. Notice the important r\^ole played by the deformation parameter $\zeta$ in achieving the final expression \eqref{hol CS action}.

Now we have clarified how the generalized and the usual 4d CS theory are related, we proceed now to make a choice for the 3-dimensional contact manifold $(\text{M},\alpha)$ in order to work out explicitly the generalized 4d CS theory in a well known situation. In the next section, we choose $\text{M}=S^{3}$. Within our construction, this manifold is naturally associated to integrable field theories of the PCM type.   

\section{Principal Chiral Model type theories}\label{4}

In this section, we specialize the construction introduced above to a particular case and consider an example corresponding to the description of integrable field theories of the Principal Chiral Model (PCM) type, i.e. we take $\text{M}=S^{3}$, $C=S^{2}$. We first gather some basic results concerning the Hopf fibration of $S^{3}$. Then, following \cite{NA loc CS} we show that the induced metric on the space $\overline{\mathcal{A}}$ is K\"ahler with respect to the symplectic form $\hat{\Omega}$ and a complex structure $J$ to be defined below. This is an essential result required to prove \eqref{main 2}. We also perform the Hamiltonian analysis in this case, where we implement a partial gauge fixing for the action of the shift group $\mathcal{S}$ by means of the condition $\Phi\approx 0$ and subsequently, take the limit $\zeta\rightarrow 0$, where we recover the known 4d CS theory action, making contact with the discussion made in section \eqref{3}. Finally, as an example of a solution to the condition \eqref{new bdry condition}, we re-derive the Lax connection for the lambda deformed PCM.  

\subsection{A contact form and the Hopf fibration}

This case corresponds to the Riemann surface $C=\mathbb{CP}^{1}$, which is the spectral space associated to integrable field theories of the principal chiral model type \cite{CY}. Thus, we have a $S^{1}$ bundle over $S^{2}$ and
\begin{equation}
S^{1}\longrightarrow S^{3}\overset{\underline{\pi}}\longrightarrow S^{2}
\end{equation} 
is the Hopf fibration. The degree of this bundle is $n=1$ \cite{Bott-Tu}. Furthermore, it is known that $S^{3}$ is one of the simplest Seifert manifolds \cite{Orlik}.

In what follows we gather some basic facts concerning the Hopf fibration of $S^{3}$ and then move to the explicit construction of the contact 1-form $\alpha$ and its associated Reeb vector field $R$. Here, for sake of completeness, we try to be as self-contained as possible.

Consider $S^{3}$ as the unit sphere in $\mathbb{C}^{2}$ with coordinates $\left( z_{0},z_{1}\right) $ and $\mathbb{CP}^{1}$ as the quotient space of $S^{3}$ under the equivalence relation $%
\left( z_{0},z_{1}\right) \sim \lambda \left( z_{0},z_{1}\right) $, for any $%
\lambda \in S^{1}$. Define the projection $\underline{\pi} :S^{3}\rightarrow \mathbb{CP}^{1}$ by the natural map%
\begin{equation}
\underline{\pi} :(z_{0},z_{1})\longrightarrow \left[ z_{0},z_{1}\right] ,
\end{equation}%
where $\left[ z_{0},z_{1}\right] $ are the homogeneous coordinates of $\mathbb{CP}^{1}$. Let 
\begin{equation}
\mathcal{U}_{i}=\left\{ \left[ z_{0},z_{1}\right] :z_{i}\neq 0\right\} ,%
\text{ \ \  }i=0,1 \label{Charts}
\end{equation}%
be the coordinate charts of $\mathbb{CP}^{1}$. From \eqref{Charts}, we have that the local coordinates on $\mathcal{U}_{0}$ and $\mathcal{U}_{1}$ are, respectively, given by $w=z_{1}/z_{0}$ and $z=z_{0}/z_{1}$, so that $w=1/z$ .  

The bundle structure is introduced via the local trivializations%
\begin{equation}
\Phi _{i}:\underline{\pi} ^{-1}\left( \mathcal{U}_{i}\right) \longrightarrow \mathcal{U}%
_{i}\times S^{1},\text{ \ \ }\Phi _{i}(z_{0},z_{1})=\left( \left[ z_{0},z_{1}%
\right] ,\frac{z_{i}}{\left\vert z_{i}\right\vert }\right) , \label{Phi}
\end{equation}%
with inverses given by%
\begin{equation}
\Phi _{i}^{-1}:\mathcal{U}_{i}\times S^{1}\longrightarrow \underline{\pi} ^{-1}\left( 
\mathcal{U}_{i}\right) ,\text{ \ \ }\Phi _{i}^{-1}\left( \left[ z_{0},z_{1}\right]
,e^{i\sigma }\right) =\frac{e^{i\sigma }\left\vert z_{i}\right\vert }{\sqrt{%
\left\vert z_{0}\right\vert ^{2}+\left\vert z_{1}\right\vert ^{2}}z_{i}}%
\left( z_{0},z_{1}\right) . \label{Phi inverse}
\end{equation}%
For $j\neq i$, one has that
\begin{equation}
\Phi _{j}\Phi _{i}^{-1}:\mathcal{U}_{i}\times S^{1}\longrightarrow \mathcal{U}%
_{j}\times S^{1},\text{ \ \ }\Phi _{j}\Phi _{i}^{-1}\left( \left[ z_{0},z_{1}%
\right] ,e^{i\sigma }\right) =\left( \left[ z_{0},z_{1}\right] ,e^{i\sigma }%
\frac{z_{j}\left\vert z_{i}\right\vert }{z_{i}\left\vert z_{j}\right\vert }%
\right) ,
\end{equation}%
so the transition map is given by%
\begin{equation}
t_{ji}:\mathcal{U}_{j}\cap \mathcal{U}_{i}\longrightarrow S^{1},\text{ \ \ }%
\left[ z_{0},z_{1}\right] \rightarrow \frac{z_{j}\left\vert z_{i}\right\vert 
}{z_{i}\left\vert z_{j}\right\vert }.
\end{equation}
 
Let us write some expressions in a more explicit way. Define $S^{2}$ and $S^{3}$ by the elements $(u_{1},u_{2},u_{3})\in 
%TCIMACRO{\U{211d} }%
%BeginExpansion
\mathbb{R}
%EndExpansion
^{3}$ and $(x_{1},x_{2},x_{3},x_{4})\in 
%TCIMACRO{\U{211d} }%
%BeginExpansion
\mathbb{R}
%EndExpansion
^{4}$ obeying $u_{1}^{2}+u_{2}^{2}+u_{3}^{2}=1$ and  $%
x_{1}^{2}+x_{2}^{2}+x_{3}^{2}+x_{4}^{2}=1,$ respectively$.$ 
In terms of the complex coordinates of $%
%TCIMACRO{\U{2102} }%
%BeginExpansion
\mathbb{C}
%EndExpansion
^{2}$ introduced above, we set $z_{0}=x_{1}+ix_{2}$ and $z_{1}=x_{3}+ix_{4}$. The latter defining the complex structures to be considered here.

Let $(x,y)$ be the stereographic projection coordinates of a point in the
southern hemisphere $\mathcal{U}_{1}$ of $S^{2}$ from the north pole. We have that%
\begin{equation}
(x,y)=\left( \frac{u_{1}}{1-u_{3}},\frac{u_{2}}{1-u_{3}}\right) ,
\end{equation}%
and 
\begin{equation}
z=x+iy=\frac{u_{1}+iu_{2}}{1-u_{3}}=\frac{x_{1}+ix_{2}}{x_{3}+ix_{4}}=\frac{%
z_{0}}{z_{1}}.
\end{equation}%
In a similar way, the stereographic coordinates $(u,v)$ of the northern
hemisphere $\mathcal{U}_{0}$ projected from the south pole are%
\begin{equation}
(u,v)=\left( \frac{u_{1}}{1+u_{3}},\frac{u_{2}}{1+u_{3}}\right). 
\end{equation}%
Then,
\begin{equation}
w=u-iv=\frac{u_{1}-iu_{2}}{1+u_{3}}=\frac{x_{3}+ix_{4}}{x_{1}+ix_{2}}=\frac{%
z_{1}}{z_{0}}.
\end{equation}
On the equator of $S^{2}$, $u_{3}=0$, $\left\vert
z_{0}\right\vert =\left\vert z_{1}\right\vert =1/\sqrt{2}$ and the transition
function $t_{01}$ become $t_{01}=z_{0}/z_{1}=u_{1}+iu_{2}$. 

From \eqref{Phi} and \eqref{Phi inverse}, we write for $i=1$, the bundle coordinate relations   
\begin{equation}
\left( z,e^{i\sigma }\right) =\left( \frac{z_{0}}{z_{1}},\frac{z_{1}}{%
\left\vert z_{1}\right\vert }\right) ,\text{ \ \ }\left( z_{0},z_{1}\right) =%
\frac{e^{i\sigma }}{\sqrt{1+\left\vert z\right\vert ^{2}}}\left( z,1\right). \label{chart U1} 
\end{equation}%
We now use the local trivialization coordinates \eqref{chart U1} over the chart $%
\mathcal{U}_{1}\subset \mathbb{CP}^{1}$ to compute the contact form $\alpha$ and mainly to run the Hamiltonian analysis of the theory below\footnote{We will be mostly working on the chart $\mathcal{U}_{1}$, covering all $\mathbb{CP}^{1}$ but the north pole. If something special occurs at the point $z=\infty$, we will properly comment on it when necessary.}. 

Notice that $\underline{\pi}:(z_{0},z_{1})\rightarrow z=z_{0}/z_{1}$. Consider the symplectic form on $\mathbb{CP}^{1}$, given by the K\"ahler form
\begin{equation}
\sigma _{C}=\frac{i}{2\pi }\frac{dz\wedge d\overline{z}}{\big( 1+\left\vert
z\right\vert ^{2}\big) ^{2}}. \label{Kahler}
\end{equation}%
To find the pull-back $\underline{\pi} ^{\ast }\sigma _{C},$ simply take $z=z_{0}/z_{1}$ in order to obtain a local expression on M, which is given by
\begin{equation}
\underline{\pi} ^{\ast }\sigma _{C}=\frac{i}{2\pi }\Big( dz_{0}\wedge d\overline{z}%
_{0}+dz_{1}\wedge d\overline{z}_{1}\Big) ,
\end{equation}%
where we have used $\left\vert z_{0}\right\vert ^{2}+\left\vert
z_{1}\right\vert ^{2}=1$ and $d\big( \left\vert z_{0}\right\vert
^{2}+\left\vert z_{1}\right\vert ^{2}\big) =0$ in order to reach the final form. The contact 1-form is then given by the defining relation \eqref{Boothby-Wang}, implying in
\begin{equation}
\alpha =\frac{i}{4\pi }\Big( z_{0}d\overline{z}_{0}-\overline{z}%
_{0}dz_{0}+z_{1}d\overline{z}_{1}-\overline{z}_{1}dz_{1}\Big) .
\end{equation}%
Now, the Reeb vector field satisfying the condition $\alpha(R)=1$, is
\begin{equation}
R=2\pi i\Big( z_{0}\partial _{z_{0}}-\overline{z}_{0}\partial _{\overline{z}%
_{0}}+z_{1}\partial _{z_{1}}-\overline{z}_{1}\partial _{\overline{z}%
_{1}}\Big) \label{Reeb bundle}
\end{equation}%
and its integral curves, given by
\begin{equation}
\big( z_{0}(t),z_{1}(t)\big) =\lambda(t) (z_{0},z_{1}),\text{ \ \ }\lambda(t)
=e^{2\pi it}\in S^{1},
\end{equation}%
with $t \in \mathbb{R}$ are, not surprisingly, the $S^{1}$ fibers of the Hopf bundle. Then, we get
\begin{equation}
 \left( \frac{z_{0}(t)}{z_{1}(t)},\frac{z_{1}(t)}{%
\left\vert z_{1}(t)\right\vert }\right)=\left( z,e^{i(\sigma+2\pi t) }\right).
\end{equation}

In real coordinates, we alternatively have
\begin{equation}
\begin{aligned}
\alpha  =\frac{1}{2\pi }\Big(
x_{1}dx_{2}-x_{2}dx_{1}+x_{3}dx_{4}-x_{4}dx_{3}\Big) , \text{ \ \ } d\alpha  =\frac{1}{\pi }\Big( dx_{1}\wedge dx_{2}+dx_{3}\wedge
dx_{4}\Big) 
\end{aligned}
\end{equation}%
and from this follows that%
\begin{equation}
\alpha \wedge d\alpha =\frac{1}{2\pi ^{2}}d\text{Vol}_{S^{3}},
\end{equation}%
where%
\begin{equation}
d\text{Vol}_{S^{3}}=i^{\ast }\left( i_{u}d\text{Vol}_{%
%TCIMACRO{\U{211d} }%
%BeginExpansion
\mathbb{R}
%EndExpansion
^{4}}\right) =\sum_{i=1}^{4}(-1)^{i-1}x_{i}dx_{1}\wedge ...\wedge \widehat{%
dx_{i}}\wedge ...\wedge dx_{4}.
\end{equation}%
Right above, $u=\left( x_{1},x_{2},x_{3},x_{4}\right) $ is a unit vector
normal to $S^{3},$ $d\text{Vol}_{%
%TCIMACRO{\U{211d} }%
%BeginExpansion
\mathbb{R}
%EndExpansion
^{4}}=dx_{1}\wedge dx_{2}\wedge dx_{3}\wedge dx_{4}$ is the volume form of $%
%TCIMACRO{\U{211d} }%
%BeginExpansion
\mathbb{R}
%EndExpansion
^{4}$ and $i$ is the inclusion map $S^{3}\hookrightarrow 
%TCIMACRO{\U{211d} }%
%BeginExpansion
\mathbb{R}
%EndExpansion
^{4}$. The $2\pi^{2}$ is the 3-dimensional surface volume of a 3-sphere of unit radius. The integral of $\alpha \wedge d\alpha$ over M is one, cf. \eqref{int over M}. 

In the local bundle coordinates $(z,\sigma)$, which are the ones that we will use later on to run the Dirac algorithm, we have
\begin{equation}
\begin{aligned}
\alpha =\alpha_{z}dz+\alpha_{\overline{z}}d\overline{z}+\alpha_{\sigma}d\sigma =-\frac{i}{4\pi }\frac{\overline{z}dz-zd\overline{z}}{1+\left\vert
z\right\vert ^{2}}+\frac{d\sigma }{2\pi },\text{ \ \ } d\alpha =ig_{z\overline{z}}dz\wedge d\overline{z}=\frac{i}{2\pi }\frac{dz\wedge d\overline{z}}{\big( 1+\left\vert z\right\vert ^{2}\big)
^{2}}. \label{bundle form alpha e dalpha}
\end{aligned}
\end{equation}
The contraction \eqref{cont}, is equivalent to 
\begin{equation}
i_{\mathcal{R}}\mathbb{A=}\frac{A_{\tau }}{\alpha _{\tau }}+i_{R}A=\frac{%
A_{\tau }}{\alpha _{\tau }}+\frac{d\alpha \wedge A}{\alpha \wedge d\alpha }, \label{contraction Abb}
\end{equation}
where we have used the decomposition
\begin{equation}
\mathbb{A}=A_{\tau}d\tau+A, \text{ \ \ } A=A_{z}dz+A_{\overline{z}}d\overline{z}+A_{\sigma}d\sigma. \label{natural}
\end{equation}
Notice that 
\begin{equation}
i_{R}A=\frac{A_{\sigma}}{\alpha_{\sigma}},\text{ \ \ }\alpha_{\sigma}=\frac{1}{2\pi}. \label{contraction A}
\end{equation}
Then, in these coordinates the Reeb vector field takes the simple form
\begin{equation}
R=\frac{1}{\alpha_{\sigma}}\partial_{\sigma} \label{Reeb in sigma}
\end{equation}
and from \eqref{time+Reeb}, we get that
\begin{equation}
\mathcal{R}=\frac{1}{\alpha_{\tau}}\partial_{\tau}+\frac{1}{\alpha_{\sigma}}\partial_{\sigma}.
\end{equation}
Alternatively, the expression \eqref{Reeb in sigma} can be obtained from \eqref{Reeb bundle} and \eqref{chart U1}.

In the rest of this work we will set $\alpha_{\tau}=\alpha_{\sigma}$. This particular choice ensures that in local coordinates, the vector fields $\mathcal{R}$ and $\mathcal{R}'$ are proportional to the usual light-cone expressions $\partial_{\pm}$, respectively. Indeed, locally $\Sigma=\mathbb{R}\times S^{1}$ is a Minkowskian cylinder and the integral curves of the vector field $\mathcal{R}$ are spirals drawn on $\Sigma$. After introducing light-cone coordinates via the definitions $\sigma^{\pm}=\tau\pm\sigma$, $\partial_{\pm}=\frac{1}{2}(\partial_{\tau}\pm\partial_{\sigma})$, we have that $\mathcal{R}=4\pi\partial_{+}$. Then, \eqref{time+Reeb} can be understood as a global definition of the light-cone vector $\partial_{+}\in T\Sigma$. The global counterpart of $\partial_{-}$, denoted by $\mathcal{R}'$ and announced before, will be introduced later on when needed.

Now, we verify explicitly the conditions \eqref{reality cond} for $\kappa, \Omega$ with $x=(\tau,\sigma,z)$ and $\overline{x}=(\tau, \sigma, \overline{z})$. The reality conditions for $\kappa$ are trivially satisfied, while for $\Omega$ they imply that $\overline{\varphi(z)}=\varphi(\overline{z})$. The latter condition is equivalent to the statement that the zeroes and the poles in the sets $\mathfrak{z}$, $\mathfrak{p}$, are either real of coming in complex conjugate pairs \cite{unifying,Lacroix}. In what follows, we will assume that this is the case.

Let us consider the expression \eqref{d} in more detail. After writing 
\begin{equation}
d_{\mathbb{M}}\Omega =d\tilde{\Omega }=d \omega-4\pi\zeta d \alpha=\Omega _{z\overline{z}}dz\wedge d%
\overline{z},
\end{equation}%
we get that
\begin{equation}
d(\eta ,\lambda )=\frac{1}{2\pi }\dint\nolimits_{\mathbb{M}}d\tau \wedge
d\sigma \wedge dz\wedge d\overline{z}\Omega _{z\overline{z}}\text{Tr}\left(
\eta \partial _{-}\lambda \right) . \label{d-explicit}
\end{equation}%
Then, the condition for the inner product \eqref{inner product} to be
invariant under the adjoint action of $\mathfrak{h}$, requires restricting the elements $\eta \in \Omega _{\mathbb{M}%
}^{0}\otimes \mathfrak{g}$ of the gauge algebra to depend on the $\Sigma$ coordinates $\tau ,\sigma 
$ only through the light-cone coordinate $\sigma ^{+}$. In general they may be chosen to satisfy the global condition $\pounds_{\mathcal{R}'}\eta=0$, see \eqref{d in pounds}. In the degenerate limit $\zeta
\rightarrow 0$, it suffices instead to impose the condition $\eta |_{\mathfrak{p}}=0$, that the
gauge parameters vanish at the set of poles $\mathfrak{p}$ of the twist 1-form $\omega$. This is already a well known fact in the literature. We also verify explicitly that $d_{\mathbb{M}}\Omega\wedge d_{\mathbb{M}}\kappa=0$ in both situations, so the expression \eqref{d} is anti-symmetric.

\subsection{Riemannian and K\"ahler metrics on $\mathcal{A}$ and $\overline{\mathcal{A}}$} \label{4.2}

We work now in the Hopf coordinates, $0\leq \eta \leq \pi /2,$ $0\leq \xi _{1},\xi
_{2}\leq 2\pi $, which are defined by
\begin{equation}
z_{0}=e^{i\xi _{1}}\sin \eta ,\text{ \ \ }z_{1}=e^{i\xi _{2}}\cos \eta .
\end{equation}
Some useful expressions to be used in what follows are%
\begin{equation}
\alpha  =\frac{1}{2\pi }\Big( \sin ^{2}\eta d\xi _{1}+\cos ^{2}\eta d\xi
_{2}\Big) , \text{ \ \ \ }
d\alpha  =\frac{1}{\pi }\sin \eta \cos \eta d\eta \wedge \left( d\xi
_{1}-d\xi _{2}\right) . \label{useful}
\end{equation}%

In these coordinates, the round metric of $\text{M}=S^{3}$ takes the form%
\begin{equation}
g_{\text{M}}=d\eta \otimes d\eta +\sin ^{2}\eta d\xi _{1}\otimes d\xi
_{1}+\cos ^{2}\eta d\xi _{2}\otimes d\xi _{2}.
\end{equation}
We want to relate $g_{\text{M}}$ to the pull-back of the K\"ahler metric $g_{C}
$ on $C$ associated to the K\"ahler form \eqref{Kahler} given by%
\begin{equation}
g_{C}=\frac{1}{2\pi }\frac{dz\otimes d\overline{z}+d\overline{z}\otimes dz}{\big( 1+\left\vert z\right\vert ^{2}\big)
^{2}} 
\end{equation}%
and to the contact form $\alpha $. In order to compute $\underline{\pi} ^{\ast }g_{C}$,
we set $z=z_{0}/z_{1}$ as before and obtain
\begin{equation}
\pi \left( \underline{\pi} ^{\ast }g_{C}\right) =d\eta \otimes d\eta +\sin ^{2}\eta \cos
^{2}\eta \left( d\xi _{1}-d\xi _{2}\right) \otimes \left( d\xi _{1}-d\xi
_{2}\right) .
\end{equation}
The metric $g_{\text{M}}$, then takes the compact form%
\begin{equation}
g_{\text{M}}=\pi \left( \underline{\pi} ^{\ast }g_{C}\right) +\left( 2\pi \right)
^{2}\alpha \otimes \alpha .
\end{equation}%
In this guise, we see that the action of the vector field $R$ is an isometry of the metric $g_{\text{M}}$ and this follows from  
\begin{equation}
i_{R}(d\eta)=0, \text{ \ \ }i_{R}\left( d\xi _{1}-d\xi
_{2}\right)=0 \label{Hori}
\end{equation}
as can be seen from the expression for $d\alpha$ written in \eqref{useful}.

Now, we introduce a pseudo-Riemannian metric $g_{\mathbb{M}}$ on the 4-dimensional manifold $\mathbb{M}=\mathbb{R}\times \text{M}$. It is defined by $g_{%
\mathbb{M}}=\sqrt{\rho }\left( d\tau \otimes d\tau -g_{\text{M}}\right) $,
where $\rho $ is a real positive constant to be fixed below. In matrix form, we have that
\begin{equation}
\left[ g_{\mathbb{M}}\right] =\sqrt{\rho }\left( 
\begin{array}{cccc}
1 & 0 & 0 & 0 \\ 
0 & -1 & 0 & 0 \\ 
0 & 0 & -\sin ^{2}\eta  & 0 \\ 
0 & 0 & 0 & -\cos ^{2}\eta 
\end{array}%
\right) . \label{metric matrix}
\end{equation}

In order to proceed, we need an explicit expression for the second vector field $\mathcal{R}%
^{\prime }\in\mathfrak{X}_{\mathbb{M}}$, satisfying the conditions \eqref{basic rel R prime} and \eqref{basic rel for R prime}. We quickly find that
\begin{equation}
\mathcal{R}'=\frac{\kappa_{\tau}}{\Omega_{\tau}}\Big( \frac{1}{\alpha_{\tau}}\partial_{\tau}-R \Big).\label{time-Reeb}
\end{equation}
Let us write now
\begin{equation}
\Omega=\omega+\Omega',\text{ \ \ }\Omega'=\Omega_{\tau}\left( d\tau-\frac{\alpha}{\alpha_{\tau}}    \right). \label{Omega prime}
\end{equation}
Thus $i_{\mathcal{R}'}\Omega'=1$, because of $i_{\mathcal{R}'}\omega=0$. In the local bundle coordinates $(z,\sigma)$, the vector field \eqref{time-Reeb} becomes $\mathcal{R'}=(1/\Omega_{\tau})\partial_{-}$ and we see that \eqref{time-Reeb} corresponds, up to a constant, to a global definition of the tangent vector $\partial_{-}\in T\Sigma$. Its integral curves are also spirals drawn on $\Sigma$. Thus, the vectors $\partial_{\pm}$ span $T\Sigma$ locally, as expected. 

Consider now the contracted 5-form
\begin{equation}
0=i_{\mathcal{R}^{\prime }}\Big( \Omega \wedge \kappa \wedge d_{\mathbb{M}%
}\kappa \wedge \gamma \Big) =\kappa \wedge d_{\mathbb{M}}\kappa \wedge
\gamma +\Omega \wedge \kappa \wedge d_{\mathbb{M}}\kappa i_{\mathcal{R}%
^{\prime }}\gamma , 
\end{equation}%
for any $\gamma \in \Omega _{\mathbb{M}}^{1}.$ This result allows to write the contraction with $\mathcal{R}'$ as
\begin{equation}
i_{\mathcal{R}^{\prime }}\gamma =-\frac{\kappa \wedge d_{\mathbb{M}}\kappa
\wedge \gamma }{\Omega \wedge \kappa \wedge d_{\mathbb{M}}\kappa }.
\end{equation}
Using a similar result for $i_{\mathcal{R}}\gamma$, see \eqref{cont}, we get the basic contractions%
\begin{eqnarray}
\begin{aligned}
i_{\mathcal{R}}(d\tau ) &=\frac{1}{\alpha _{\tau }},\text{ \ \ \ \, }i_{\mathcal{%
R}}(d\eta )=0,\text{ \ \, }i_{\mathcal{R}}(d\xi _{1})=\frac{1}{\alpha _{\tau }%
},\text{ \ \ \ \ \ \ \ }i_{\mathcal{R}}(d\xi _{2})=\frac{1}{\alpha _{\tau }}, \\
i_{\mathcal{R}^{\prime }}(d\tau ) &=\frac{1}{2\Omega _{\tau }},\text{ \ \ }%
i_{\mathcal{R}^{\prime }}(d\eta )=0,\text{ \ \ }i_{\mathcal{R}^{\prime
}}(d\xi _{1})=-\frac{1}{2\Omega _{\tau }},\text{ \ \ }i_{\mathcal{R}^{\prime
}}(d\xi _{2})=-\frac{1}{2\Omega _{\tau }}.
\end{aligned}
\end{eqnarray}
From these results, we find that
\begin{equation}
g_{\mathbb{M}}\left( \mathcal{R},\mathcal{R}\right) =g_{\mathbb{M}}\left( 
\mathcal{R}^{\prime },\mathcal{R}^{\prime }\right) =0,\text{ \ \ }\pounds
_{\mathcal{R}}g_{\mathbb{M}}=\pounds_{\mathcal{R}^{\prime }}g_{%
\mathbb{M}}=0.
\end{equation}
Thus, $\mathcal{R}$ and $\mathcal{R}^{\prime}$ are light-like Killing vectors. They are not orthogonal under $g_{\mathbb{M}}$ and the explicit value for $g_{\mathbb{M}}\left( \mathcal{R},\mathcal{R}'\right) $ will not be important in what follows.

Once $g_{\mathbb{M}}$ has been defined, we consider now a Riemannian inner product on the space $\mathcal{A}$ of gauge connections defined by the expression
\begin{equation}
\left( \mathbb{A},\mathbb{A}'\right) _{\mathbb{M}}=-\dint\nolimits_{%
\mathbb{M}}\text{Tr}\left( \mathbb{A\wedge \ast A'}\right),  \label{metric inner}
\end{equation}%
for $\mathbb{A},\mathbb{A}'\in \mathcal{A}$ and use it to implement an orthogonal decomposition of $\mathcal{A}$ in
terms of the quotient space $\overline{\mathcal{A}}=\mathcal{A}/\mathcal{S}$
and space $\mathcal{S}$ of Abelian shifts. The Hodge duality operator $*$ to be used in what follows is defined in terms of the metric introduced in \eqref{metric matrix}, the coordinates
$x^{\mu}=(\tau,\eta,\xi_{1},\xi_{2})$, with  $\mu=1,2,3,4$ and the orientation $d\tau \wedge d\eta \wedge d\xi_{1} \wedge d\xi_{2}$. It gives
\begin{equation}
\begin{aligned}
\ast (d\tau ) &=\sqrt{\rho }\sin \eta \cos \eta d\eta \wedge d\xi
_{1}\wedge d\xi _{2},\text{ \ \ }\ast (d\eta )=\sqrt{\rho }\sin \eta \cos
\eta d\tau \wedge d\xi _{1}\wedge d\xi _{2}, \\
\ast (d\xi _{1}) &=-\sqrt{\rho }\cot \eta d\tau \wedge d\eta \wedge d\xi
_{2},\text{ \ \ \ \ \ \  }\ast (d\xi _{2})=\sqrt{\rho }\tan \eta d\tau \wedge d\eta
\wedge d\xi _{1}.
\end{aligned}
\end{equation}
We also have the useful expressions
\begin{equation}
\ast 1=2\pi^{2} \rho d\tau \wedge \alpha \wedge d\alpha, \text{ \ \ }\ast (d\tau)=2\pi^{2} \sqrt{\rho} \alpha \wedge d\alpha, 
\text{ \ \ }\ast \alpha =\frac{1}{2}\sqrt{\rho}d\tau \wedge d\alpha.
\end{equation}

First, we consider the shift group space $\mathcal{S}$.
Elements in this space are of the form $s\kappa $, $s^{\prime }\Omega $ and satisfy the relations
\begin{equation}
\left( s\kappa ,s^{\prime }\kappa \right) _{\mathbb{M}}=0,\text{ \ \ }%
\left( s\Omega ,s^{\prime }\Omega \right) _{\mathbb{M}}=0,\text{ \ \ }%
\left(
s\Omega ,s^{\prime }\kappa \right) _{\mathbb{M}}=-\frac{\sqrt{\rho }}{%
\alpha _{\tau }}\dint\nolimits_{\mathbb{M}}\Omega \wedge \kappa \wedge d_{%
\mathbb{M}}\kappa \text{Tr}\left( ss^{\prime }\right) , \label{inner S}
\end{equation}%
where we have used%
\begin{equation}
\ast \kappa =\frac{\sqrt{\rho }}{\alpha _{\tau }}\kappa \wedge d_{\mathbb{M}%
}\kappa ,\text{ \ \ }\ast \Omega =\ast \omega -\frac{\sqrt{\rho }}{\alpha
_{\tau }}\Omega \wedge d_{\mathbb{M}}\kappa ,
\end{equation}%
with%
\begin{equation}
\omega =m\Big( d\eta +i\sin \eta \cos \eta \left( d\xi _{1}-d\xi
_{2}\right) \Big) ,\text{ \ \ }\ast \omega =m\sqrt{\rho }\Big( \sin \eta
\cos \eta d\tau \wedge d\xi _{1}\wedge d\xi _{2}-2\pi id\tau \wedge d\eta
\wedge \alpha \Big) 
\end{equation}%
and $m=\varphi (z_{0}/z_{1})e^{i(\xi _{1}+\xi _{2})}/z_{1}^{2}$. The norm of any element belonging to $\mathcal{S}$ is then
\begin{equation}
\left (s\kappa+s'\Omega, s\kappa + s'\Omega    \right )_{\mathbb{M}}=2\left(
s\Omega ,s^{\prime }\kappa \right) _{\mathbb{M}}. \label{norm S}
\end{equation}
Notice that the last term in \eqref{inner S} is proportional to the first contribution on the rhs of the inner product defined before in \eqref{inner product} and we can consider \eqref{norm S} as being defined independently of any metric in $\mathbb{M}$. Some useful results used are
\begin{equation}
\kappa\wedge \ast \omega=\omega\wedge\ast\kappa=\omega\wedge\ast\omega=\Omega\wedge\ast\omega=\omega\wedge\ast\Omega=0. \label{boring}
\end{equation} 

Second, we consider the quotient space $\overline{\mathcal{A}}$. 
Elements in this space are taken to be of the form $\mathbb{A}%
^{\perp }=\mathbb{A}-p\kappa -q\Omega $, where $p,q\in \Omega _{\mathbb{M}%
}^{0}\otimes \mathfrak{g}$ are to be determined by the orthogonality conditions%
\begin{equation}
\left( \mathbb{A}^{\perp },s\kappa \right) _{\mathbb{M}}=0,%
\text{ \ \ }\left( \mathbb{A}^{\perp },s\Omega \right) _{\mathbb{M}}=0.
\end{equation}%
We find that%
\begin{equation}
p=i_{\mathcal{R}}\mathbb{A}+X,\text{ \ \ }q=i_{\mathcal{R}^{\prime }}%
\mathbb{A},
\end{equation}%
where 
\begin{equation}
X=\frac{\alpha _{\tau }}{\sqrt{\rho }}\frac{\mathbb{A}\wedge \ast \omega}{\Omega\wedge \kappa \wedge d_{\mathbb{M}}\kappa} .
\end{equation}
As a consequence, the norm of any element $\mathbb{A}$ can be decomposed in the form
\begin{equation}
\left( \mathbb{A},\mathbb{A}\right) _{\mathbb{M}}=\left( \mathbb{A}%
^{\perp },\mathbb{A}^{\perp }\right)_{\mathbb{M}}+2\left( p\kappa
,q\Omega \right) _{\mathbb{M}}. \label{norm}
\end{equation}

Now, we proceed to relate the first term on the rhs in \eqref{norm} to the pre-symplectic form \eqref{pre-symplectic} in order to define a K\"ahler metric structure on $\overline{\mathcal{A}}$. Using 
\begin{equation}
i_{\mathcal{R}^{\prime }}\mathbb{A}^{\perp }=i_{\mathcal{R}^{\prime }}%
\mathbb{A}-q=0,\text{ \ \ }i_{\mathcal{R}^{\prime }}\left( \Omega \wedge 
\mathbb{A}^{\perp }\right) =\mathbb{A}^{\perp }
\end{equation}%
and the contracted 5-form%
\begin{equation}
0=i_{\mathcal{R}^{\prime }}\left( \Omega \wedge \mathbb{A}^{\perp }\wedge
\ast \mathbb{A}^{\perp }\right) =\mathbb{A}^{\perp }\wedge \ast \mathbb{A}%
^{\perp }+\Omega \wedge \mathbb{A}^{\perp }\wedge \left( i_{\mathcal{R}%
^{\prime }}\circ \ast \right) \mathbb{A}^{\perp },
\end{equation}%
we have that%
\begin{equation}
\mathbb{A}^{\perp }\wedge \ast \mathbb{A}^{\perp }=-\Omega \wedge \mathbb{%
A}^{\perp }\wedge \left( i_{\mathcal{R}^{\prime }}\circ \ast \right) 
\mathbb{A}^{\perp }.
\end{equation}%
In a similar way, we use
\begin{equation}
i_{\mathcal{R}}\mathbb{A}^{\perp }=i_{\mathcal{R}}\mathbb{A}-p=-X,\text{ \
\ }i_{\mathcal{R}}\left( \Omega \wedge \kappa \wedge \mathbb{A}^{\perp
}\right) =-\Omega \wedge \mathbb{A}^{\perp }+\Omega \wedge \kappa i_{%
\mathcal{R}}\mathbb{A}^{\perp }
\end{equation}%
and the contracted 5-form%
\begin{equation}
\begin{aligned}
0=i_{\mathcal{R}}\Big( \Omega \wedge \kappa \wedge & \mathbb{A}^{\perp
}\wedge \left( i_{\mathcal{R}^{\prime }}\circ \ast \right) \mathbb{A}%
^{\perp }\Big) \\
&=i_{\mathcal{R}}\left( \Omega \wedge \kappa \wedge \mathbb{%
A}^{\perp }\right) \wedge \left( i_{\mathcal{R}^{\prime }}\circ \ast \right) 
\mathbb{A}^{\perp }-\Omega \wedge \kappa \wedge \mathbb{A}^{\perp }\wedge
\left( i_{\mathcal{R}}\circ i_{\mathcal{R}^{\prime }}\circ \ast \right) 
\mathbb{A}^{\perp }
\end{aligned}
\end{equation}%
to find that%
\begin{equation}
\mathbb{A}^{\perp }\wedge \ast \mathbb{A}^{\perp }=\Omega \wedge \kappa
\wedge \mathbb{A}^{\perp }\wedge \ast _{2}\mathbb{A}^{\perp }-\Omega
\wedge \kappa X\wedge \ast _{3}\mathbb{A}^{\perp },
\end{equation}%
where we have introduced two `effective' Hodge duality operators defined by 
\begin{equation}
\ast _{2}=i_{%
\mathcal{R}}\circ i_{\mathcal{R}^{\prime }}\circ \ast  \text{ \ \ } \ast _{3}=-i_{%
\mathcal{R}^{\prime }}\circ \ast. 
\end{equation}
Then, \eqref{norm} becomes
\begin{equation}
\left( \mathbb{A},\mathbb{A}\right) _{\mathbb{M}}=-\dint\nolimits_{%
\mathbb{M}}\Omega \wedge \kappa \wedge \text{Tr}\left( \mathbb{A}^{\perp }%
\mathbb{\wedge \ast }_{2}\mathbb{A}^{\perp }\right) +\dint\nolimits_{%
\mathbb{M}}\Omega \wedge \kappa \wedge \text{Tr}\left( X\mathbb{\ast }_{3}%
\mathbb{A}^{\perp }\right) +2\left( p\kappa ,q\Omega \right) _{\mathbb{M}}. \label{metric inner 2}
\end{equation}

The expression \eqref{metric inner 2} can be simplified a bit more if now we introduce the quantity
\begin{equation}
\Pi \left( \mathbb{A}\right) =\mathbb{A}-(i_{\mathcal{R}}\mathbb{%
A})\kappa -(i_{\mathcal{R}^{\prime }}\mathbb{A})\Omega \label{A in quotient}
\end{equation}%
and write $\mathbb{A}^{\perp }=\Pi \left( \mathbb{A}\right) -X\kappa .$
Notice that $i_{\mathcal{R}}\circ \Pi =i_{\mathcal{R}^{\prime }}\circ \Pi =0$, hence $\Pi \left( \mathbb{A}\right)\subset \text{ker}_{i_{\mathcal{R}}}\cap \text{ker}_{i_{\mathcal{R}'}}$. Thus, after using the fact that $\ast _{2}\kappa =$\ $\ast _{3}\kappa =0$, we get 
\begin{equation}
\begin{aligned}
\Omega \wedge \kappa \wedge \text{Tr}\left( \mathbb{A}^{\perp }\mathbb{%
\wedge \ast }_{2}\mathbb{A}^{\perp }\right)  &=\Omega \wedge \kappa \wedge 
\text{Tr}\big( \Pi \left( \mathbb{A}\right) \mathbb{\wedge \ast }_{2}\Pi
\left( \mathbb{A}\right) \big) , \\
\Omega \wedge \kappa \wedge \text{Tr}\left( X\mathbb{\ast }_{3}\mathbb{A}%
^{\perp }\right)  &=\Omega \wedge \kappa \wedge \text{Tr}\big( X\mathcal{%
\ast }_{3}\Pi \left( \mathbb{A}\right) \big) .
\end{aligned}
\end{equation}
Explicitly, we find that
\begin{equation}
\Pi \left( \mathbb{A}\right) =A_{\eta }d\eta +\Big( \frac{\alpha_{2}}{\alpha_{\tau}}
A_{1}-\frac{\alpha_{1}}{\alpha_{\tau}} A_{2}\Big) \left( d\xi _{1}-d\xi _{2}\right) -\frac{1}{%
4\zeta }\Big( A_{\tau }-A_{1}-A_{2}\Big) \omega , \label{explicit A}
\end{equation}%
where we have taken
\begin{equation}
\mathbb{A}=A_{\tau}d\tau+A_{\eta}d\eta+A_{1}d\xi_{1}+A_{2}d\xi_{2}
\end{equation}
and used \eqref{useful} with $\alpha=\alpha_{1}d\xi_{1}+\alpha_{2} d\xi_{2}$. Notice that the 1-forms $d\eta $ and $\left( d\xi _{1}-d\xi _{2}\right) 
$ span the space $\text{ker}_{i_{R}}=\Omega_{\text{M,Hor}}^{1}$ of horizontal 1-forms defined by the vector field $R$, cf. \eqref{Hori} above.

On the one hand, we obtain that 
\begin{equation}
\Omega \wedge \kappa \wedge \mathcal{\ast }_{3}(d\eta )=0,\text{ \ \ }\Omega \wedge \kappa
\wedge \mathcal{\ast }_{3}(d\xi _{1}-d\xi _{2})=0,
\end{equation}%
showing that the second contribution on the rhs in \eqref{metric inner 2} is absent. On the other hand, from the explicit expressions
\begin{equation}
\begin{aligned}
\mathcal{\ast }_{2}(d\tau ) &=0,\text{ \ \ }\mathcal{\ast }_{2}(d\eta )=-%
\frac{\pi \sqrt{\rho }}{\zeta}\sin \eta \cos \eta \left( d\xi _{1}-d\xi _{2}\right) , \\
\mathcal{\ast }_{2}(d\xi _{1}) &=\frac{\pi \sqrt{\rho }}{\zeta }\cot \eta
d\eta ,\text{ \ \ }\mathcal{\ast }_{2}(d\xi _{2})=-\frac{\pi \sqrt{\rho }}{%
\zeta }\tan \eta d\eta ,
\end{aligned}
\end{equation}%
we find that the `effective' 2-dimensional Hodge duality operator $\ast _{2}$ actually defines a complex structure on the space $\Omega _{\text{M},\text{%
Hor}}^{1}$, i.e. it obeys $\ast _{2}^{2}=-1$, provided we fix the constant $\rho$ to be 
\begin{equation}
\sqrt{\rho} =\alpha_{\tau}\Omega_{\tau}=\frac{\zeta }{\pi }.
\end{equation}
It satisfies the basic relation 
\begin{equation}
\ast_{2}1=\zeta d\alpha=\underline{\pi}^{\ast}(\zeta\sigma_{C}),
\end{equation}
where we have used the defining relation \eqref{Boothby-Wang} with $n=1$.

As a consequence of all these results, we get the desired relation
\begin{equation}
\left( \mathbb{A},\mathbb{A}\right) _{\mathbb{M}}=\hat{\Omega }%
\big( \Pi \left( \mathbb{A}\right) ,\mathbb{\ast }_{2}\Pi \left( \mathbb{%
A}\right) \big) +2\left( p\kappa ,q\Omega \right) _{\mathbb{M}}, \label{break metric}
\end{equation}%
where we have 
\begin{equation}
\hat{\Omega }\big( \Pi \left( \mathbb{A}\right) ,\mathbb{\ast }%
_{2}\Pi \left( \mathbb{A}\right) \big) =-\dint\nolimits_{\mathbb{M}%
}\Omega \wedge \kappa \wedge \text{Tr}\Big( \Pi \left( \mathbb{A}\right) 
\mathbb{\wedge \ast }_{2}\Pi \left( \mathbb{A}\right) \Big) , \label{metric inner final}
\end{equation}%
with $\hat{\Omega }$ being the pre-symplectic form defined in \eqref{pre-symplectic}. Recall that now $\hat{\Omega}|_{\overline{\mathcal{A}}}$ is symplectic. 

Let us now consider the gauge group $\mathcal{G}$ and use \eqref{eq over g} to write
\begin{equation}
\mathbb{I}=\kappa g^{-1}\pounds_{\mathcal{R}}g+g^{-1}\gamma(g) \label{I split}.
\end{equation}
Using $^{g}\mathbb{A}=g^{-1}\mathbb{A}g+\mathbb{I}$ and \eqref{I split} in \eqref{A in quotient}, gives
\begin{equation}
\Pi(^{g}\mathbb{A})=g^{-1}\Pi(\mathbb{A})g+g^{-1}\gamma(g). \label{coset gauge}
\end{equation} 
Then $\Pi \left( ^{g}\mathbb{A}\right)\subset \text{ker}_{i_{\mathcal{R}}}\cap \text{ker}_{i_{\mathcal{R}'}}$ and the quotient space $\overline{\mathcal{A}}$ is preserved by the action of a `restricted' gauge group defined by the elements $g$ satisfying \eqref{eq over g}.

Notice that $d_{\mathbb{M}}\Omega\sim d\eta \wedge (d\xi_{1}-d\xi_{2})$. In this case, in order to get $d_{\mathbb{M}}\Omega \wedge \gamma=0$, we propose a linear combination of the form
\begin{equation}
g^{-1}\gamma(g)=Xd\eta + Y (d\xi_{1}-d\xi_{2}),
\end{equation}
with 
\begin{equation}
X=g^{-1}\partial_{\eta}g,\text{ \ \ } Y=\frac{\alpha_{2}}{\alpha_{\tau}}g^{-1}\partial_{1}g-\frac{\alpha_{1}}{\alpha_{\tau}}g^{-1}\partial_{2}g.
\end{equation}
This choice in \eqref{coset gauge} provide the usual gauge field transformations
\begin{equation}
^{g}A_{i}=g^{-1}A_{i}g+g^{-1}\partial_{i}g, \text{ \ \ }i=\tau, \eta, 1,2,
\end{equation}
with $g$ obeying $\pounds_{\mathcal{R}'}g=0$, namely
\begin{equation}
\partial_{\tau}g-\partial_{1}g-\partial_{2}g=0. \label{piri}
\end{equation}
Furthermore, the WZ-type term \eqref{gen WZ term} becomes
\begin{equation}
\dint\nolimits_{\mathbb{M}%
}\Omega \wedge \chi(g)=-\frac{\zeta}{\pi}
\dint\nolimits_{\mathbb{M}%
} d\tau \wedge d\eta \wedge d\xi_{1} \wedge d\xi_{2} \text{Tr}\Big (g^{-1}\pounds_{\mathcal{R}}g [X,Y]  \Big ).\label{Hopf coo}
\end{equation}
The expression \eqref{I split} exhibits an orthogonal decomposition, under the inner product \eqref{metric inner}, of the current $\mathbb{I}\in \Omega_{\mathbb{M}}^{1}\otimes \mathfrak{g}$ along $\text{im}_{i_{\mathcal{R}}}$ and $\text{ker}_{i_{R}}\subset \text{ker}_{i_{\mathcal{R}}}$ while keeping $\mathbb{I}\in \text{ker}_{i_{\mathcal{R}'}}$. This follows from the fact that $\text{im}_{i_{\mathcal{R}}}^{\perp}=\text{ker}_{i_{R}}$, because of
\begin{equation}
\kappa \wedge \ast d\eta=0,\text{ \ \ }  \kappa \wedge \ast d(\xi_{1}-\xi_{2})=0.
\end{equation}

Alternatively, in the local bundle coordinates $(z,\sigma)$ over $\mathcal{U}_{1}$, we have that $d_{\mathbb{M}}\Omega\sim dz \wedge d\overline{z}$ and this time we propose instead
\begin{equation}
g^{-1}\gamma(g)=Xdz +Yd\overline{z},
\end{equation}
with
\begin{equation}
X=g^{-1}\partial_{z}g-\frac{\alpha_{z}}{\alpha_{\tau}}g^{-1}\partial_{\sigma}g,\text{ \ \ }Y=g^{-1}\partial_{\overline{z}}g-\frac{\alpha_{\overline{z}}}{\alpha_{\tau}}g^{-1}\partial_{\sigma}g.
\end{equation}
Using this in
\begin{equation}
\Pi(\mathbb{A})=\Big( A_{z}-\frac{\alpha_{z}}{\alpha_{\tau}} A_{\sigma}    \Big) dz+\Big( A_{\overline{z}}-\frac{\alpha_{\overline{z}}}{\alpha_{\tau}} A_{\sigma}    \Big) d\overline{z}-\frac{1}{4\zeta}\Big( A_{\tau }-A_{\sigma}\Big) \omega,   \label{Pi bundle}
\end{equation}
gives
\begin{equation}
^{g}A_{i}=g^{-1}A_{i}g+g^{-1}\partial_{i}g, \text{ \ \ }i=\tau, \sigma, z,\overline{z},
\end{equation}
for the gauge field transformations. The condition $\pounds_{\mathcal{R}'}g=0$, now takes the form
\begin{equation}
\partial_{\tau}g-\partial_{\sigma}g=0. \label{poro}
\end{equation}
This is precisely the condition $\partial_{-}g=0$ we found before. The WZ-type term takes the same form as in \eqref{Hopf coo}, but with the volume form $d\tau \wedge d\sigma \wedge dz \wedge d\overline{z}$. 

Now, we consider the WZ-type term. Using \eqref{I split} in \eqref{Hopf coo}, allows to write\footnote{This can be shown in an explicit way by using the results \eqref{piri} or \eqref{poro}.}
\begin{equation}
\dint\nolimits_{\mathbb{M}%
}\Omega \wedge \chi(g)=2\Omega_{\tau} \dint\nolimits_{\mathbb{R}\times \text{M}%
} d\tau \wedge \chi(g)', \label{simplified WZ term}
\end{equation}
where we have defined, cf. \eqref{chi},
\begin{equation}
\chi(g)'=-\frac{1}{3}\text{Tr}(I\wedge I\wedge I),
\end{equation}
with $I=g^{-1}dg$ being the contribution along M in the decomposition $\mathbb{I}=I_{\tau}d\tau+I$. In principle, the WZ-type term contribution can be set to zero if further analytic restrictions are imposed over the elements $g\in \mathcal{G}$. For example, by enforcing the holomorphicity condition $\partial_{\overline{z}}g=0$ when using the local bundle coordinates $(z,\sigma)$. Also notice the possibility of taking the limit $\zeta\rightarrow 0$ as well. It is important to emphasize that at this level of analysis it is not clear if \eqref{simplified WZ term} could, alternatively, be related to some quantization condition as occurs in conventional CS theories.

Finally, the result \eqref{metric inner final} shows that the metric induced by the Riemannian metric \eqref{metric inner} on the quotient space $\overline{\mathcal{A}}$ is K\"ahler, with respect to the symplectic form $\hat{\Omega}|_{\overline{\mathcal{A}}}$ and complex structure 
\begin{equation}
J=\ast _{2}=i_{\mathcal{R}}\circ i_{\mathcal{R}'}\circ \ast.
\end{equation}
This is an important result and we will come back to it later on when considering the symplectic measure of the generalized 4d CS theory path integral, see \eqref{main 2}.

\subsection{Hamiltonian analysis and 4d Chern-Simons theory}

In order to perform the Hamiltonian analysis and implement explicitly the steps I and II mentioned before in section \eqref{3}, we first need to isolate the time differential $d\tau$ from
all expressions. The exterior derivative $d_{\mathbb{M}}$ was already introduced in \eqref{differential split} and now we write it in terms of the local bundle coordinates $(z,\sigma)$ over $\mathcal{U}_{1}$. Thus,
\begin{equation}
d_{\mathbb{M}}=d\tau\wedge \partial_{\tau}+d,\text{ \ \ }d=dz\wedge\partial_{z}+d\overline{z}\wedge \partial_{\overline{z}}+d\sigma \wedge
\partial_{\sigma}. 
\end{equation}

Some quantities of interest to be used below are:\\
i) The curvature of the connection $\mathbb{A}$ under the decomposition \eqref{natural}, namely,
\begin{equation}
F_{\mathbb{A}}=F_{A}+d\tau \wedge F_{\tau },\text{ \ \ }F_{\tau }=\partial \label{F split}
_{\tau }A-d_{A}A_{\tau },
\end{equation}%
where $F_{A}=dA+A\wedge A$ is the curvature of $A$ and $d_{A}=d+[A,*]$.\\
ii) The Lie algebra value field $\Phi$ defined previously in \eqref{Phi and B}, which now takes the form  
\begin{equation}
\Phi =2\frac{\big( \Omega _{\tau }\tilde{\kappa }-\kappa _{\tau }%
\tilde{\Omega }\big) \wedge F_{A}+\tilde{\Omega }\wedge \tilde{%
\kappa }\wedge F_{\tau }+d\tilde{\Omega }\wedge \left( \kappa _{\tau }A-%
\tilde{\kappa }A_{\tau }\right) }{\Omega _{\tau }\alpha \wedge d\alpha }. \label{Phi time}
\end{equation}%
iii) The Chern-Simons 3-form
\begin{equation}
CS(\mathbb{A})=-d\tau \wedge \text{Tr}\left( A\wedge \partial _{\tau
}A-2A_{\tau }F_{A}\right) -d\tau \wedge d\text{Tr}\left( A_{\tau }A\right)
+CS(A).
\end{equation}%
Also, introduce the following variables
\begin{equation}
A'_{\tau}=A_{\tau}-\kappa_{\tau}\Phi,\text{ \ \ }   A'=A-\tilde{\kappa}\Phi.
\end{equation}
They are invariant under $\kappa$-shifts and will be useful for writing results coming from the Hamiltonian analysis in a more compact way.

From the expression \eqref{mu mu 2}, we quickly obtain the Lagrangian of the theory%
\begin{equation}
\begin{aligned}
L=ic\bigg\{ \dint\nolimits_{\text{M}}\tilde{\Omega }\wedge \text{Tr}%
\big( A\wedge \partial _{\tau }A&-2A_{\tau }F_{A}\big) -\frac{1}{2}\Omega _{\tau }\dint\nolimits_{%
\text{M}}\alpha \wedge d\alpha \text{Tr}\left( \Phi ^{2}\right)\\
&+\Omega _{\tau
}\dint\nolimits_{\text{M}} CS(A)+\dint\nolimits_{\text{M}}d\tilde{\Omega }\wedge \text{Tr}\Big ( A_{\tau
}A+\left( \kappa _{\tau }A-\tilde{\kappa }A_{\tau }\right) i_{\mathcal{R}%
}\mathbb{A}\Big ) \bigg\}. \label{Lagrangian}
\end{aligned}
\end{equation}
Because of the Lagrangian is quadratic in $\Phi$, we see from \eqref{F split}, that the theory is not only linear in the velocities $\partial_{\tau}A$ but also quadratic on them.

To find the canonical momenta, we compute the variation of the Lagrangian \eqref{Lagrangian} with respect to all field components in $\delta(\partial_{\tau}\mathbb{A})$. Using the following result
\begin{equation}
\Omega _{\tau }\alpha \wedge d\alpha \delta _{\partial _{\tau }A}\Phi =2%
\tilde{\Omega }\wedge \tilde{\kappa }\wedge \delta (\partial _{\tau
}A),
\end{equation}%
we find that
\begin{equation}
\delta _{\partial _{\tau }A}L=\dint\nolimits_{\text{M}}\text{Tr}\Big(
\delta (\partial _{\tau }A)\wedge P\Big) ,
\end{equation}%
where the 2-form $P\in \Omega_{\text{M}}^{2}\otimes \mathfrak{g}$ is given by%
\begin{equation}
P=ic\left( \tilde{\Omega }\wedge A-2\tilde{\Omega }\wedge \tilde{%
\kappa }\Phi \right) . \label{momentum 2-form}
\end{equation}%
The components in the expansion
\begin{equation}
P=P_{z}d\overline{z}\wedge d\sigma +P_{\overline{z}}d\sigma \wedge
dz+P_{\sigma }dz\wedge d\overline{z}
\end{equation}%
are, actually, the usual canonical momenta defined by 
\begin{equation}
\begin{aligned}
P_{z} &=\frac{\delta L}{\delta \left( \partial _{\tau }A_{z}\right) }=ic\Big( \Omega _{\overline{z}}\left( A_{\sigma }-2\kappa _{\sigma
}\Phi \right) -\Omega _{\sigma }\left( A_{\overline{z}}-2\kappa _{\overline{z%
}}\Phi \right) \Big) , \\
P_{\overline{z}} &=\frac{\delta L}{\delta \left( \partial _{\tau }A_{\overline{z}}\right) }=ic\Big( \Omega _{\sigma }\left( A_{z}-2\kappa _{z}\Phi
\right) -\Omega _{z}\left( A_{\sigma }-2\kappa _{\sigma }\Phi \right) \Big)
, \\
P_{\sigma } &=\frac{\delta L}{\delta \left( \partial _{\tau }A_{\sigma}\right) }=ic\Big( \Omega _{z}\left( A_{\overline{z}}-2\kappa _{%
\overline{z}}\Phi \right) -\Omega _{\overline{z}}\left( A_{z}-2\kappa
_{z}\Phi \right) \Big) , \label{canonical momenta}
\end{aligned}
\end{equation}%
where we have used the local decompositions
\begin{equation}
\tilde{\Omega } =\Omega _{z}dz+\Omega _{\overline{z}}d\overline{z}%
+\Omega _{\sigma }d\sigma , \text{ \ \ } \tilde{\kappa } =\kappa _{z}dz+\kappa _{\overline{z}}d\overline{z}%
+\kappa _{\sigma }d\sigma , \text{ \ \ } A =A_{z}dz+A_{\overline{z}}d\overline{z}+A_{\sigma }d\sigma .
\end{equation}
Also, notice that
\begin{equation}
P_{\tau} =\frac{\delta L}{\delta \left( \partial _{\tau }A_{\tau}\right) }=0. \label{momentum tau}
\end{equation}

From the absence of the term $\partial _{\tau }A_{\tau }$ in the Lagrangian
and the very form of $P$, we easily detect the presence of three primary constraints given by
\begin{equation}
P_{\tau }\approx 0,\text{ \ \ }\tilde{\kappa }\wedge P+ic\tilde{%
\Omega }\wedge \tilde{\kappa }\wedge A\approx 0,\text{ \ \ }\tilde{%
\Omega }\wedge P\approx 0.
\end{equation}%
In components, the last two constraints can be written in the form%
\begin{equation}
\tilde{\kappa }\wedge P+ic\tilde{\Omega }\wedge \tilde{\kappa }%
\wedge A =\phi _{\tilde{\kappa }}dz\wedge d\overline{z}\wedge d\sigma
\approx 0, \text{ \ \ } \tilde{\Omega }\wedge P =\phi _{\tilde{\Omega }}dz\wedge d%
\overline{z}\wedge d\sigma \approx 0,
\end{equation}%
with%
\begin{equation}
\begin{aligned}
\phi _{\tilde{\kappa }} =\kappa _{z}P_{z}+\kappa& _{\overline{z}}P_{%
\overline{z}}+\kappa _{\sigma }P_{\sigma }\\
&+ic\Big( \left( \Omega _{%
\overline{z}}\kappa _{\sigma }-\Omega _{\sigma }\kappa _{\overline{z}%
}\right) A_{z}+\left( \Omega _{\sigma }\kappa _{z}-\Omega _{z}\kappa
_{\sigma}\right) A_{\overline{z}}+\left( \Omega _{z}\kappa _{\overline{z}%
}-\Omega _{\overline{z}}\kappa _{z}\right) A_{\sigma }\Big) \approx 0,
\end{aligned}
\end{equation}
and
\begin{equation}
\phi _{\tilde{\Omega }} =\Omega _{z}P_{z}+\Omega _{\overline{z}}P_{%
\overline{z}}+\Omega _{\sigma }P_{\sigma }\approx 0.
\end{equation}
Then, the three primary constraints are
\begin{equation}
P_{\tau}\approx 0,\text{ \ \ }\phi _{\tilde{\kappa }}\approx 0, \text{ \ \ }\phi _{\tilde{\Omega }}\approx 0.
\end{equation}

Let us now introduce the compact notation%
\begin{equation}
\text{Tr}\left( A,B\right) _{\left( z,\sigma \right) }=\dint\nolimits_{\text{%
M}}d\text{Vol} \text{Tr}\left( AB\right) ,
\end{equation}%
where $d\text{Vol}=dz\wedge d\overline{z}\wedge d\sigma$.

The canonical Poisson bracket of the theory is defined by 
\begin{equation}
\left\{ f,g\right\} =\text{Tr}\left( \frac{\delta f}{\delta A_{i}(\sigma ,z)}%
,\frac{\delta g}{\delta P_{i}(\sigma ,z)}-\frac{\delta f}{\delta
P_{i}(\sigma ,z)},\frac{\delta g}{\delta A_{i}(\sigma ,z)}\right) _{\left(
z,\sigma \right) },
\end{equation}%
where $i,j=\tau ,\sigma ,z,\overline{z}$. Then, the phase space coordinates satisfy the usual relations%
\begin{equation}
\left\{ A_{i}(\sigma ,z)_{\mathbf{1}},P_{j}(\sigma ^{\prime },z^{\prime })_{%
\mathbf{2}}\right\} =C_{\mathbf{12}}\delta _{ij}\delta _{\sigma \sigma
^{\prime }}\delta _{zz^{\prime }}, \label{Can PB comp}
\end{equation}
where $C_{\mathbf{12}}=\eta^{AB}T_{A}\otimes T_{B}$ is the tensor Casimir of the Lie algebra $\mathfrak{g}$, $\delta_{\sigma \sigma^{\prime}}=\delta(\sigma-\sigma^{\prime})$ and $\delta_{zz^{\prime}}=\delta(z-z^{\prime})$ are Dirac delta distributions. The latter can be written locally in the form
\begin{equation}
\delta_{zz^{\prime}}=-\frac{1}{2\pi i}\partial_{\overline{z}}\left( \frac{1}{z-z^{\prime}}\right). \label{delta complex}
\end{equation}

Define now the quantities%
\begin{equation}
\mathcal{S}_{\kappa }(s)=\text{Tr}\left( s,\phi_{\kappa} \right) _{\left( z,\sigma \right) } ,\text{ \ \ }\mathcal{S}_{\Omega }(s')=\text{Tr}%
\left( s',\phi_{\Omega}\right) _{\left( z,\sigma \right) } ,
\end{equation}%
where $s,s´\in \Omega_{\mathbb{M}}^{0}\otimes \mathfrak{g}$ are arbitrary and
\begin{equation}
\phi_{\kappa}=\kappa_{\tau }P_{\tau }+\phi _{\tilde{\kappa }},\text{ \ \ }\phi_{\Omega}=\Omega _{\tau }P_{\tau }+\phi _{\tilde{\Omega }}.
\end{equation}
They generate both Abelian $U(1)\times U(1)$ shifts in $\mathcal{S}$
\begin{equation}
\delta _{\kappa }A_{i}=\left\{ A_{i},\mathcal{S}_{\kappa }(s)\right\} =\kappa
_{i}s,\text{ \ \ }\delta _{\Omega }A_{i}=\left\{ A_{i},\mathcal{S}%
_{\Omega }(s')\right\} =\Omega _{i}s',
\end{equation}
under the bracket \eqref{Can PB comp} and Poisson commute between.

The canonical Hamiltonian is given by the Legendre transformation
\begin{equation}
H=\dint\nolimits_{\text{M}}\text{Tr}\left( P\wedge \partial _{\tau }A\right)+\dint\nolimits_{\text{M}}d\text{Vol}\text{Tr}\left( P_{\tau }\partial _{\tau }A_{\tau }\right)-L
\end{equation}%
and the total Hamiltonian is given by%
\begin{equation}
H_{T}=H+\text{Tr}\left( u_{\tau },P_{\tau }\right) _{(z,\sigma )}+\mathcal{S}%
_{\kappa }(u)+\mathcal{S}_{\Omega }(u^{\prime }), \label{total H}
\end{equation}%
where $u_{\tau },u,u^{\prime }\in \Omega_{\mathbb{M}}^{0}\otimes \mathfrak{g}$ are arbitrary Lagrange multipliers.  

The variation $\delta H$ must depend \cite{Teitel} only on the variations $(\delta A$, $\delta P)$ and $(\delta A_{\tau}$, $\delta P_{\tau})$. In order to find it, we first compute the variation of the $P$-dependent contribution
to $H$, then use \eqref{momentum 2-form}, \eqref{momentum tau} and finally subtract $\delta L$. We find that%
\begin{equation}
\delta H=\dint\nolimits_{\text{M}}\text{Tr}\left( \delta P\wedge \partial
_{\tau }A\right) +\dint\nolimits_{\text{M}}d\text{Vol}\text{Tr}\left( \delta P_{\tau }\partial _{\tau }A_{\tau
}\right) +ic\dint\nolimits_{\text{M}}\text{Tr}\left[ \delta
(\partial _{\tau }A)\wedge \left( \tilde{\Omega }\wedge A-2\tilde{%
\Omega }\wedge \tilde{\kappa }\Phi \right) \right] -\delta L.
\end{equation}%
To compute $\delta L$ correctly, we calculate the variation of $L$ without performing
any integration by parts involving the derivative $\partial_{\tau}$ along the $\tau $ direction. Thus, by using the identity%
\begin{equation}
\begin{aligned}
\Omega _{\tau }\alpha \wedge d\alpha \text{Tr}\left( \delta \Phi X
\right)  &=2\text{Tr}\left[ \delta A\wedge \left( -(\Omega _{\tau }%
\tilde{\kappa }-\kappa _{\tau }\tilde{\Omega })\wedge d_{A}X
+\Omega _{\tau }d\tilde{\kappa }X -\tilde{\Omega }\wedge 
\tilde{\kappa }\left[ A_{\tau },X \right] \right) \right]  \\
&+2\text{Tr}\left[ \delta A_{\tau }\left( -\tilde{\Omega }\wedge d%
\tilde{\kappa }X +\tilde{\Omega }\wedge \tilde{\kappa }\wedge
d_{A}X \right) \right] +2\text{Tr}\left[ \delta (\partial _{\tau
}A)\wedge \left( \tilde{\Omega }\wedge \tilde{\kappa }X \right) %
\right] , \label{imp identity}
\end{aligned}
\end{equation}
with $X=\Phi$ being an element of $\Omega_{\mathbb{M}}^{0}\otimes \mathfrak{g}$, we find from \eqref{Lagrangian}, that
\begin{equation}
\begin{aligned}
\delta L &=ic\dint\nolimits_{\text{M}}\text{Tr}\left[ \delta A\wedge \left(
-\tilde{\Omega }\wedge \partial _{\tau }A+2\tilde{\Omega }\wedge
d_{A' }A_{\tau }'+2\Omega
_{\tau }F_{A'}\right) \right]  \\
&+ic\dint\nolimits_{\text{M}}\text{Tr}\left[ \delta A_{\tau }\left( -2%
\tilde{\Omega }\wedge F_{A' }\right) \right]
+ic\dint\nolimits_{\text{M}}\text{Tr}\left[ \delta (\partial _{\tau
}A)\wedge \left( \tilde{\Omega }\wedge A-2\tilde{\Omega }\wedge 
\tilde{\kappa }\Phi \right) \right]  \\
&+2ic\dint\nolimits_{\text{M}}d\tilde{\Omega }\wedge d\sigma \wedge \text{Tr}\Big[
\delta A_{\tau }\left( A_{\sigma}-\kappa_{\sigma}i_{\mathcal{R}}\mathbb{A}\right) -\delta A_{\sigma}\left( A_{\tau}-\kappa_{\tau} i_{\mathcal{R}}\mathbb{%
A} \right) \Big] .
\end{aligned}
\end{equation}
Then,
\begin{equation}
\begin{aligned}
\delta H &=\dint\nolimits_{\text{M}}\text{Tr}\left( \delta P\wedge \partial
_{\tau }A\right) +ic\dint\nolimits_{\text{M}}\text{Tr}\left[ \delta A\wedge
\left( \tilde{\Omega }\wedge \partial _{\tau }A-2\tilde{\Omega }%
\wedge d_{A'}A_{\tau }'-2\Omega _{\tau }F_{A'}\right) \right]  \\
&+\dint\nolimits_{\text{M}}d\text{Vol}\text{Tr}\left( \delta P_{\tau }\partial _{\tau }A_{\tau }\right)
+ic\dint\nolimits_{\text{M}}\text{Tr}\left[ \delta A_{\tau
}\left( 2\tilde{\Omega }\wedge F_{A'}\right) %
\right]  \\
&-2ic\dint\nolimits_{\text{M}}d\tilde{\Omega }\wedge d\sigma \wedge \text{Tr}\Big[
\delta A_{\tau }\left( A_{\sigma}-\kappa_{\sigma}i_{\mathcal{R}}\mathbb{A}\right) -\delta A_{\sigma}\left( A_{\tau}-\kappa_{\tau} i_{\mathcal{R}}\mathbb{%
A} \right) \Big], \label{Can H var}
\end{aligned}
\end{equation}
only depends on the phase space coordinates variations, as required.

Before we continue, as a consistency test, let us find the Lagrangian eom of the theory. After integrating by parts the term $\partial_{\tau}(\delta A)$ and simplifying, we obtain
\begin{equation}
\begin{aligned} 
\delta L &=-2ic\dint\nolimits_{\text{M}}\text{Tr}\left[ \delta A\left( \tilde{\Omega }%
\wedge F_{\tau }'-\Omega _{\tau }F_{A'}\right) +\delta A_{\tau }\left( \tilde{\Omega }\wedge F_{A'}\right) \right]   \\
&+2ic\dint\nolimits_{\text{M}}d\tilde{\Omega }\wedge d\sigma \wedge \text{Tr}\Big[
\delta A_{\tau }\left( A_{\sigma}-\kappa_{\sigma}i_{\mathcal{R}}\mathbb{A}\right) -\delta A_{\sigma}\left( A_{\tau}-\kappa_{\tau} i_{\mathcal{R}}\mathbb{%
A} \right) \Big] , \label{Lagrangian eom}
\end{aligned}
\end{equation}
where $F_{\tau}'=\partial_{\tau}A'-d_{A'}A_{\tau}'$. The eom extracted from \eqref{Lagrangian eom} also derive from \eqref{full eom} and both are perfectly equivalent.

Now, we return to the expression \eqref{Can H var} and find the functional derivatives of $H$ with respect to the phase space coordinate fields. We get that
\begin{equation}
\begin{aligned}
\frac{\delta H}{\delta A_{\tau }} &=2ic\Big( \Omega _{z}F_{\overline{z}%
\sigma }^{\prime }+\Omega _{\sigma }F_{z\overline{z}}^{\prime }+\Omega _{%
\overline{z}}F_{\sigma z}^{\prime }-\Omega _{z\overline{z}}\left( A_{\sigma
}-\kappa _{\sigma }i_{\mathcal{R}}\mathbb{A}\right) \Big) ,\\
\frac{\delta H}{\delta A_{z}}& =icQ_{\overline{z}\sigma }, \text{ \ \ }
\frac{\delta H}{\delta A_{\overline{z}}} =icQ_{\sigma z},\text{ \ \ }\frac{%
\delta H}{\delta A_{\sigma }}=icQ_{z\overline{z}}+2ic\Omega _{z\overline{z}%
}\left( A_{\tau }-\kappa _{\tau }i_{\mathcal{R}}\mathbb{A}\right) 
\end{aligned}
\end{equation}%
and%
\begin{equation}
\frac{\delta H}{\delta P_{\tau }}=\partial _{\tau }A_{\tau },\text{ \ \ }%
\frac{\delta H}{\delta P_{z}}=\partial _{\tau }A_{z},\text{ \ \ }\frac{%
\delta H}{\delta P_{\overline{z}}}=\partial _{\tau }A_{\overline{z}},\text{
\ \ }\frac{\delta H}{\delta P_{\sigma }}=\partial _{\tau }A_{\sigma },
\end{equation}%
where we have used
\begin{equation}
\begin{aligned}
Q &=\tilde{\Omega }\wedge \partial _{\tau }A-2\tilde{\Omega }\wedge
d_{A^{\prime }}A_{\tau }^{\prime }-2\Omega _{\tau }F_{A^{\prime }}, \\
&=Q_{\overline{z}\sigma }d\overline{z}\wedge d\sigma +Q_{\sigma z}d\sigma
\wedge dz+Q_{z\overline{z}}dz\wedge d\overline{z}, \\
F_{A^{\prime }} &=F_{\overline{z}\sigma }^{\prime }d\overline{z}\wedge
d\sigma +F_{\sigma z}^{\prime }d\sigma \wedge dz+F_{z\overline{z}}^{\prime
}dz\wedge d\overline{z}, \\
d\tilde{\Omega } &=\Omega _{z\overline{z}}dz\wedge d\overline{z}.
\end{aligned}
\end{equation}

Armed with these expressions, now we are able to verify the time preservation of the primary constraints. Starting
with $P_{\tau}$, we find that
\begin{equation}
\big\{ \text{Tr}\left( \eta ,P_{\tau }\right) _{(z,\sigma )},H_{T}\big\}
=\gamma(\eta)\approx 0
\end{equation}%
and from this we obtain a secondary constraint given by
\begin{equation}
\gamma(\eta)=-2ic\dint\nolimits_{\text{M}}\tilde{\Omega }\wedge \text{Tr}\left( \eta
F_{A^{\prime }}\right) +2ic\dint\nolimits_{\text{M}}d\tilde{\Omega }%
\wedge \text{Tr}\Big( \eta \left( A-\tilde{\kappa }i_{\mathcal{R}}%
\mathbb{A}\right) \Big). \label{Secondary cons}
\end{equation}  

Now, we consider $\mathcal{S}_{\kappa }(s)$ and obtain
\begin{equation}
\begin{aligned}
\left\{ H_{T},\mathcal{S}_{\kappa }(s)\right\}  =& -2ic\dint\nolimits_{\text{%
M}}\text{Tr}\left[ s\left( (\Omega _{\tau }\tilde{\kappa }-\kappa _{\tau
}\tilde{\Omega })\wedge F_{A^{\prime }}+\tilde{\Omega }\wedge 
\tilde{\kappa }\wedge F_{\tau }^{\prime }\right) \right]  \\
&+2ic\dint\nolimits_{\text{M}}d\tilde{\Omega }\wedge \text{Tr}\Big[
s\Big(\tilde{\kappa }(A_{\tau }-\kappa _{\tau }i_{\mathcal{R}}\mathbb{%
A})-\kappa _{\tau }(A-\tilde{\kappa }i_{\mathcal{R}}\mathbb{A})\Big) %
\Big] .
\end{aligned}
\end{equation}
By using \eqref{Phi time}, \eqref{solutions}, we get that%
\begin{equation}
(\Omega _{\tau }\tilde{\kappa }-\kappa _{\tau }\tilde{\Omega }%
)\wedge F_{A^{\prime }}+\tilde{\Omega }\wedge \tilde{\kappa }\wedge
F_{\tau }^{\prime }=d\tilde{\Omega }\wedge \left( \tilde{\kappa }%
A_{\tau }-\kappa _{\tau }A\right) 
\end{equation}%
and from \eqref{contraction Abb}, \eqref{contraction A}, we have 
\begin{equation}
A_{\tau }-\kappa _{\tau }i_{\mathcal{R}}\mathbb{A=}\frac{1}{2}\left(
A_{\tau }-A_{\sigma }\right) ,\text{ \ \ }A_{\sigma }-\kappa _{\sigma }i_{%
\mathcal{R}}\mathbb{A}=-\frac{1}{2}\left( A_{\tau }-A_{\sigma }\right) .
\end{equation}%
Thus,%
\begin{equation}
\left\{ H_{T},\mathcal{S}_{\kappa }(s)\right\} =0.
\end{equation}

For $\mathcal{S}_{\Omega }(s')$, we follow a similar calculation to obtain
\begin{equation}
\left\{ H_{T},\mathcal{S}_{\Omega }(s')\right\} =2ic\dint\nolimits_{\text{M}}d%
\tilde{\Omega }\wedge \text{Tr}\left[ s'\left( \tilde{\Omega }%
(A_{\tau }-\kappa _{\tau }i_{\mathcal{R}}\mathbb{A})-\Omega _{\tau }(A-%
\tilde{\kappa }i_{\mathcal{R}}\mathbb{A})\right) \right] 
\end{equation}%
and after using \eqref{solutions}, we get that
\begin{equation}
\left\{ H_{T},\mathcal{S}_{\Omega }(s')\right\} =0.
\end{equation}
Not surprisingly, the canonical Hamiltonian of the theory is invariant under the two independent shifts in $\mathcal{S}$. 

There is a single secondary constraint $\gamma(\eta)$. From the expression \eqref{Secondary cons}, it is not difficult to verify that it is invariant under both shift transformations as well, hence
\begin{equation}
\left\{ \gamma(\eta),\mathcal{S}_{\kappa }(s)\right\} =0,\text{ \ \ }\left\{ \gamma(\eta),\mathcal{S}_{\Omega }(s')\right\} =0.
\end{equation}
It remains to verify if its time preservation introduce a tertiary constraint or if instead it determines some of the Lagrange multipliers. From the expression
\begin{equation}
\left\{ H_{T},\gamma (\eta )\right\} =\left\{ H,\gamma (\eta )\right\} -%
\text{Tr}\left( u_{\tau },\frac{\delta \gamma (\eta )}{\delta A_{\tau }}%
\right) _{(z,\sigma )}\approx 0, \label{time-sec}
\end{equation}%
we realize that this is actually a condition over the Lagrange multiplier $u_{\tau }$,
provided that $\frac{\delta \gamma (\eta )}{\delta A_{\tau }}\neq 0$ and it is not
difficult to check that this is indeed the case. The explicit form for $u_{\tau }$
is not required in what follows, hence we do not need to perform the
calculation explicitly.

It is clear from \eqref{Secondary cons} that the computation of
the Poisson algebra for two $\gamma ^{\prime }s$ is rather tedious.
Fortunately, at this stage, we are mainly interested in implementing the step I mentioned above\footnote{We will consider step II below as well.}, see section \eqref{3}, in the Hamiltonian formulation. Thus, we gauge fix the first class
constraint that generates the $\kappa $-shift symmetry by choosing the
following gauge fixing condition%
\begin{equation}
\Phi \approx 0. \label{fixing kappa}
\end{equation}%
Now, the pair $\phi _{\kappa }\approx0$ $,\Phi\approx0 $ of constraints become a second class set and we impose
them strongly by means of a Dirac bracket. Notice that  
\begin{equation}
\begin{aligned}
\left\{ \Phi (\sigma ,z)_{\mathbf{1}},\phi _{\kappa }(\sigma ^{\prime
},z^{\prime })_{\mathbf{2}}\right\}  =C_{\mathbf{12}}\delta _{\sigma
\sigma ^{\prime }}\delta _{zz^{\prime }}, \text{ \ \ } \left\{ \phi _{\kappa }(\sigma ,z)_{\mathbf{1}},\Phi (\sigma ^{\prime
},z^{\prime })_{\mathbf{2}}\right\} ^{-1} =C_{\mathbf{12}}\delta _{\sigma
\sigma ^{\prime }}\delta _{zz^{\prime }}.
\end{aligned}
\end{equation}%
Then, \eqref{fixing kappa} is also a good gauge fixing condition. 

The Dirac bracket is given by%
\begin{equation}
\begin{aligned}
\big\{ f,g\big\} ^{\ast } & =\left\{ f,g\right\}\\
&-\text{Tr}\left( \left\{
f,\phi _{\kappa }(\sigma ,z)_{\mathbf{1}}\right\} ,\text{Tr}\left( \left\{
\phi _{\kappa }(\sigma ,z)_{\mathbf{1}},\Phi (\sigma ^{\prime },z^{\prime
})_{\mathbf{2}}\right\} ^{-1},\left\{ \Phi (\sigma ^{\prime },z^{\prime })_{%
\mathbf{2}},g\right\} \right) _{\mathbf{2}(z^{\prime },\sigma ^{\prime })}%
\right) _{\mathbf{1}(z,\sigma )} \\
&-\text{Tr}\left( \left\{ f,\Phi (\sigma ,z)_{\mathbf{1}}\right\} ,\text{Tr}%
\left( \left\{ \Phi (\sigma ,z)_{\mathbf{1}},\phi _{\kappa }(\sigma ^{\prime
},z^{\prime })_{\mathbf{2}}\right\} ^{-1},\left\{ \phi _{\kappa }(\sigma
^{\prime },z^{\prime })_{\mathbf{2}},g\right\} \right) _{\mathbf{2}%
(z^{\prime },\sigma ^{\prime })}\right) _{\mathbf{1}(z,\sigma )},
\end{aligned}
\end{equation}%
where the labels $\textbf{1,2}$ instruct us to take the trace along the corresponding tensor factors. It reduces to
\begin{equation}
\begin{aligned}
\left\{ f,g\right\} ^{\ast }& =\left\{ f,g\right\}\\
&-\text{Tr}\Big( \left\{
f,\phi _{\kappa }(\sigma ,z)_{\mathbf{1}}\right\} ,\left\{ \Phi (\sigma ,z)_{%
\mathbf{1}},g\right\} \Big) _{\mathbf{1}(z,\sigma )}+\text{Tr}\Big(
\left\{ f,\Phi (\sigma ,z)_{\mathbf{1}}\right\} ,\left\{ \phi _{\kappa
}(\sigma ,z)_{\mathbf{1}},g\right\} \Big) _{\mathbf{1}(z,\sigma )}
\end{aligned}
\end{equation}%
and equals the canonical Poisson bracket if we restrict to phase space functionals that are
$\kappa$-shift invariant. Step I is then accomplished. 

The partially gauge fixed theory is no longer quadratic in the velocity $\partial_{\tau}A$ but linear and this restructures completely the whole set of Hamiltonian constraints allowing to `restart' the Dirac algorithm again. To see this, consider a theory with a Lagrangian that can be decomposed in the form $L=L_{0}+L_{1}+L_{2}$, where $L_{0}$, $L_{1}$ and $L_{2}$ are the terms independent, linear and quadratic in the velocities. The Hamiltonian energy function is $H=L_{2}-L_{0}$, i.e. the quadratic term is preserved, the linear term is absent and the velocity-independent term flips sign. As the gauge fixed theory is now linear in the velocities, the Hamiltonian energy function and the canonical Hamiltonian coincide when restricted to the constraint surface defined by \eqref{momentum 2-form}, after taking $\Phi=0$. Thus, $H=-L_{0}$ and \eqref{momentum 2-form} leads now to three new `primary' constraints. 

Explicitly, the gauge fixed action is invariant under the remaining $\Omega$-shifts, it is given by  
\begin{equation}
S =ic\dint\nolimits_{\mathbb{M}}\Omega \wedge CS\left( 
\mathbb{A}\right) +ic\dint\nolimits_{\mathbb{M}}d_{\mathbb{M}}\Omega \wedge \kappa\wedge \text{Tr}\left( \mathbb{A}i_{\mathcal{R}}\mathbb{A}  \right)  \label{part gauge fixed}
\end{equation}%
and has the following Lagrangian 
\begin{equation}
L=ic\bigg\{ \dint\nolimits_{\text{M}}\tilde{\Omega }\wedge \text{Tr}%
\big( A\wedge \partial _{\tau }A-2A_{\tau }F_{A}\big) +\Omega _{\tau
}\dint\nolimits_{\text{M}} CS(A)+\dint\nolimits_{\text{M}}d\tilde{\Omega }\wedge \text{Tr}\Big ( A_{\tau
}A+\left( \kappa _{\tau }A-\tilde{\kappa }A_{\tau }\right) i_{\mathcal{R}%
}\mathbb{A}\Big ) \bigg\}.  \label{Lagrangian 2}
\end{equation}
To verify the invariance of \eqref{part gauge fixed} under $\Omega$-shifts it is useful to notice that
\begin{equation}
0=i_{\mathcal{R}}\Big( d_{\mathbb{M}}\Omega \wedge \Omega \wedge \kappa \wedge \text{Tr}(s\mathbb{A}) \Big)=
-d_{\mathbb{M}}\Omega \wedge \Omega \wedge \text{Tr}(s\mathbb{A})+d_{\mathbb{M}}\Omega \wedge \Omega \wedge \kappa \text{Tr}(si_{\mathcal{R}}\mathbb{A}). \label{identity for T-duality}
\end{equation}

We restart the Dirac procedure by taking \eqref{part gauge fixed} as the new action functional. For the canonical momentum, we have now
\begin{equation}
P_{\tau}= 0, \text{ \ \ }P=ic\ \tilde{\Omega }\wedge A , 
\end{equation}%
or in components,
\begin{equation}
P_{z} =ic\big( \Omega _{\overline{z}} A_{\sigma }-\Omega _{\sigma } A_{\overline{z}} \big) , \text{ \ \ } P_{\overline{z}} =ic\big( \Omega _{\sigma } A_{z} -\Omega _{z} A_{\sigma } \big)
, \text{ \ \ } P_{\sigma } =ic\big( \Omega _{z} A_{\overline{z}} -\Omega _{\overline{z}} A_{z} \big) , 
\end{equation}%
leading to the existence of four primary constraints, given by
\begin{equation}
\begin{aligned}
P_{\tau}& \approx 0,\\
\phi_{z}&=P_{z} -ic\big( \Omega _{\overline{z}} A_{\sigma }-\Omega _{\sigma } A_{\overline{z}} \big) \approx 0, \\
\phi_{\overline{z}}&= P_{\overline{z}} -ic\big( \Omega _{\sigma } A_{z} -\Omega _{z} A_{\sigma } \big) \approx 0, \\ 
\phi_{\sigma}&=P_{\sigma } -ic\big( \Omega _{z} A_{\overline{z}} -\Omega _{\overline{z}} A_{z} \big)\approx 0.  \label{new prim cons}
\end{aligned}
\end{equation}%
The total Hamiltonian is now
\begin{equation}
H_{T}=H+\text{Tr}\left( u_{\tau },P_{\tau }+u_{z},\phi_{z}+u_{\overline{z}},\phi_{\overline{z}}+u_{\sigma},\phi_{\sigma}\right) _{(z,\sigma )}, \label{total Ham 2}
\end{equation}%
where
\begin{equation}
H=ic\bigg\{2 \dint\nolimits_{\text{M}}\tilde{\Omega }\wedge \text{Tr}%
\big( A_{\tau }F_{A}\big) -\Omega _{\tau
}\dint\nolimits_{\text{M}} CS(A)-\dint\nolimits_{\text{M}}d\tilde{\Omega }\wedge \text{Tr}\Big ( A_{\tau
}A+\left( \kappa _{\tau }A-\tilde{\kappa }A_{\tau }\right) i_{\mathcal{R}%
}\mathbb{A}\Big ) \bigg\}. \label{can Ham 2}
\end{equation}
We expect the first class constraint associated to the $\Omega$-shift symmetry to re-emerge as a particular linear combination of the four primary constraints \eqref{new prim cons}.

From \eqref{can Ham 2}, we get the functional derivatives
\begin{equation}
\begin{aligned}
\frac{\delta H}{\delta A_{\tau }} &=2ic\Big( \Omega _{z}F_{\overline{z}%
\sigma }+\Omega _{\sigma }F_{z\overline{z}}+\Omega _{%
\overline{z}}F_{\sigma z}-\Omega _{z\overline{z}}\left( A_{\sigma
}-\kappa _{\sigma }i_{\mathcal{R}}\mathbb{A}\right) \Big) ,\\
\frac{\delta H}{\delta A_{z}}& =2icQ_{\overline{z}\sigma }, \text{ \ \ }
\frac{\delta H}{\delta A_{\overline{z}}} =2icQ_{\sigma z},\text{ \ \ }\frac{%
\delta H}{\delta A_{\sigma }}=2ic\Big (Q_{z\overline{z}}+\Omega _{z\overline{z}%
}\left( A_{\tau }-\kappa _{\tau }i_{\mathcal{R}}\mathbb{A}\right) \Big),\label{new variations}
\end{aligned}
\end{equation}%
where this time we have defined
\begin{equation}
\begin{aligned}
Q &=-\tilde{\Omega }\wedge
d_{A}A_{\tau }-\Omega _{\tau }F_{A}.
\end{aligned}
\end{equation}

Similar as done before, we must verify the time preservation of the constraints \eqref{new prim cons} under the new time flow defined by the total Hamiltonian \eqref{total Ham 2}. From the variations \eqref{new variations}, we find
\begin{equation}
\big\{ \text{Tr}\left( \eta ,P_{\tau }\right) _{(z,\sigma )},H_{T}\big\}
=\gamma(\eta)\approx 0
\end{equation}%
and obtain a secondary constraint given by
\begin{equation}
\gamma(\eta)=-2ic\dint\nolimits_{\text{M}}\tilde{\Omega }\wedge \text{Tr}\left( \eta
F_{A}\right) +2ic\dint\nolimits_{\text{M}}d\tilde{\Omega }%
\wedge \text{Tr}\Big( \eta \left( A-\tilde{\kappa }i_{\mathcal{R}}%
\mathbb{A}\right) \Big), \label{sec cons}
\end{equation}  
which is nothing but \eqref{Secondary cons} with $\Phi=0$. We also find that
\begin{equation}
\begin{aligned}
\left\{ H_{T},\phi _{z}\right\}  &=2ic\Big( Q_{\overline{z}\sigma }+\left(
\Omega _{\overline{z}}u_{\sigma }-\Omega _{\sigma }u_{\overline{z}}\right)
\Big) , \\
\left\{ H_{T},\phi _{\overline{z}}\right\}  &=2ic\Big( Q_{\sigma z}+\left(
\Omega _{\sigma }u_{z}-\Omega _{z}u_{\sigma }\right) \Big) , \\
\left\{ H_{T},\phi _{\sigma }\right\}  &=2ic\Big( Q_{z\overline{z}}+\Omega_{z\overline{z}}\left(A_{\tau}-\kappa_{\tau}i_{\mathcal{R}}\mathbb{A}  \right)+\left(
\Omega _{z}u_{\overline{z}}-\Omega _{\overline{z}}u_{z}\right) \Big) ,
\end{aligned}
\end{equation}
are actually conditions over the Lagrange multipliers $u_{\sigma},u_{z}, u_{\overline{z}}$. Concerning the time preservation of the secondary constraint \eqref{sec cons}, a similar argument leading to \eqref{time-sec} holds, hence no new constraints are produced.

By introducing the constraint 2-form
\begin{equation}
\phi=\phi_{z}d\overline{z}\wedge d\sigma+\phi_{\overline{z}}d\sigma \wedge dz+\phi_{\sigma} dz \wedge d\overline{z},
\end{equation}
we recover the $\Omega$-shift symmetry generator
\begin{equation}
\mathcal{S}_{\Omega}(s)=\text{Tr}(s,\Omega_{\tau}P_{\tau}+\tilde{\Omega}\wedge \phi )_{(z,\sigma)},
\end{equation}
which obeys 
\begin{equation}
\{H_{T}, \mathcal{S}_{\Omega}(s) \}=\{ \gamma(\eta) ,\mathcal{S}_{\Omega}(s)\}=0.
\end{equation}

At this point, as follows from the Dirac procedure, it is necessary to classify the set of constraints found so far as first or second class constraints. However, we will do this only after implementing the step II mentioned above, see section \eqref{3}, which corresponds to taking the degenerate limit $\zeta\rightarrow 0$. The reason for this is that here we are mainly interested in recovering the conventional 4d CS theories from the Hamiltonian theory point of view and not in pursuing a thorough Hamiltonian analysis of the generalized 4d CS theory. 

In the $\zeta\rightarrow 0$ limit, we have that
\begin{equation}
\Omega_{\tau}=0,\text{ \ \ } \tilde{\Omega}=\omega=\Omega_{z}dz=\varphi dz,
\end{equation}
where the component $\Omega_{z}=\varphi$ is identified with the twist function of the underlying integrable field theory. We also supplement the limit with the condition \eqref{new bdry condition}. The action functional \eqref{part gauge fixed}, the Lagrangian \eqref{Lagrangian 2} and the canonical Hamiltonian \eqref{can Ham 2} are those of the 4d CS theory and are given, respectively, by \eqref{hol CS action} and
\begin{equation}
\begin{aligned}
L&= ic\dint\nolimits_{\text{M}}\omega \wedge \text{Tr}%
\big( A\wedge \partial _{\tau }A-2A_{\tau }F_{A}\big) +ic\dint\nolimits_{\text{M}}d{\omega }\wedge \text{Tr}\big ( A_{\tau
}A\big ) , \\
H&=2ic \dint\nolimits_{\text{M}}\omega \wedge \text{Tr}%
\big( A_{\tau }F_{A}\big)-ic\dint\nolimits_{\text{M}}d\omega \wedge \text{Tr}\big ( A_{\tau}A\big ) . \label{ham in the limit}
\end{aligned}
\end{equation}
Step II is then accomplished.

Concerning the Hamiltonian constraints, we have from \eqref{new prim cons} and \eqref{sec cons}, that
\begin{equation}
\begin{aligned}
P_{\tau} \approx 0,\text{ \ \ } \phi_{z}=P_{z} \approx 0, \text{ \ \ } \phi_{\overline{z}}= P_{\overline{z}} +ic\varphi A_{\sigma }  \approx 0, \text{ \ \ }\phi_{\sigma}=P_{\sigma } -ic \varphi A_{\overline{z}} \approx 0  
\end{aligned} \label{prim hol CS}
\end{equation}%
and
\begin{equation}
\gamma(\eta)=-2ic\dint\nolimits_{\text{M}}\omega \wedge \text{Tr}\left( \eta
F_{A}\right) +2ic\dint\nolimits_{\text{M}}d\omega %
\wedge \text{Tr}\big( \eta A \big ). \label{gamma hol CS}
\end{equation} 
The constraint $\phi_{z}\approx 0$, is actually an identity reflecting the fact that the field component $A_{z}$ completely decouples from the theory. The rest of the Hamiltonian analysis follows exactly the lines considered in \cite{Vicedo-PCM, me-PCM}, to which the reader is referred for further details. Thus, we will not repeat their results here, but instead gather some relevant facts to be used later. 

The constraint $P_{\tau}\approx 0$ is first class and can be gauged fixed by choosing a gauge fixing condition of the form
\begin{equation}
A_{\tau}=A_{\tau}(A_{\sigma}|_{\mathfrak{p}}), \label{gauge fix 1}
\end{equation}
i.e. the component $A_{\tau}$ is chosen to be a function of the gauge field component $A_{\sigma}$ evaluated at the set of poles $\mathfrak{p}$ of the twist 1-form $\omega$. The constraint $\phi_{z}$ together with the component $A_{z}$, can be ignored. The constraints $\phi_{\overline{z}}$ and $\phi_{\sigma}$ form a second class pair and \eqref{gamma hol CS} is a first class constraint, provided we restrict the gauge parameters to satisfy the condition $\eta|_{\mathfrak{p}}=0$, which is a possible way to cancel the obstruction \eqref{d-explicit}. The latter constraint, i.e. $\gamma(\eta)\approx 0$, then reduces to $F_{A}\approx 0$ and as a gauge fixing condition, we can choose
\begin{equation}
A_{\overline{z}}\approx 0, \label{gauge fix 2}
\end{equation}
together with the restriction $\partial_{\overline{z}}g=0$ over the gauge parameters.

The 4d CS theory is completely recovered but apparently it is not gauge invariant because the WZ-type term is now proportional to $\omega\wedge \chi(g)$, as is well known in the literature.   

\subsection{Recovering the lambda-PCM Lax connection}

Here, we quickly explore the implications of the condition \eqref{new bdry condition} in determining the analytic structure of the lambda-PCM Lax connection.

In the gauge $\Phi =0$, the equations of motion \eqref{full eom} reduce to%
\begin{equation}
\Omega \wedge F_{\mathbb{A}}=d_{\mathbb{M}}\Omega \wedge \left( \mathbb{A}%
-\kappa i_{\mathcal{R}}\mathbb{A}\right) \label{total eom 1}
\end{equation}
and \eqref{def of Phi} becomes a trivial identity. The gauge fixed action functional is \eqref{part gauge fixed} and in the $\zeta \rightarrow 0$ limit, it becomes \eqref{zeta 0 limit action}, i.e.
\begin{equation}
S =ic\dint\nolimits_{\mathbb{M}}\omega \wedge CS\left( 
\mathbb{A}\right)+ic\dint\nolimits_{\mathbb{M}}d\omega \wedge \kappa\wedge \text{Tr}\left( \mathbb{A}i_{\mathcal{R}}\mathbb{A}  \right).  \label{gauge action step 2}
\end{equation}
The eom \eqref{total eom 1} reduce to
\begin{equation}
\omega \wedge F_{\mathbb{A}}=d\omega \wedge \left( \mathbb{A}%
-\kappa i_{\mathcal{R}}\mathbb{A}\right). \label{total eom 2}
\end{equation}
The usual 4d Chern-Simons theory is recovered by imposing the condition \eqref{new bdry condition}, i.e.
\begin{equation}
\dint\nolimits_{\mathbb{M}}d\omega \wedge \kappa\wedge \text{Tr}\left( \mathbb{A}i_{\mathcal{R}}\mathbb{A}  \right)=0. \label{condition}
\end{equation}
In this case, the action and the eom are given, respectively, by
\begin{equation}
S =ic\dint\nolimits_{\mathbb{M}}\omega \wedge CS\left( 
\mathbb{A}\right),\text{ \ \ }\omega \wedge F_{\mathbb{A}}=d\omega \wedge  \mathbb{A}. \label{both}
\end{equation} 

Notice that, despite of the fact that we are imposing the condition \eqref{condition}, both expressions in \eqref{both} are invariant under the residual $\omega$-shift symmetry given by $^{\omega}\mathbb{A}=\mathbb{A}+s\omega$. This follows from the fact that $d\omega \wedge \omega= \underline{\pi}^{\ast}(d_{C}\omega_{C}\wedge \omega_{C})=0$, where $d_{C}$ is the exterior derivative on the base manifold $C$. Locally, we have $\omega=\varphi(z)dz$, thus the gauge field component $A_{z}dz$ decouples from the theory, a prominent characteristic of the action \eqref{1.1}.

In components, the eom in \eqref{both} are equivalent to the set of equations
\begin{equation}
\varphi F_{\overline{z}\mu }=\omega _{z\overline{z}}A_{\mu },\text{ \ \ }\varphi F_{\tau \sigma }=0, \label{three eom}
\end{equation}
where $\mu=\tau,\sigma$ and $d\omega=\omega_{z\overline{z}}dz\wedge d\overline{z}$. The lambda deformed PCM is specified by the twist function \cite{k-def}
\begin{equation}
\varphi (z)=b\frac{z^{2}-1}{z^{2}-a^{2}},\text{ \ \ }\omega =\varphi (z)dz,
\end{equation}%
where $a,b\in 
%TCIMACRO{\U{211d} }%
%BeginExpansion
\mathbb{R}
%EndExpansion
.$ The zeroes $\mathfrak{z}$ and the poles $\mathfrak{p}$ of $\omega$ on the chart $\mathcal{U}_{1}$ are located at $z=\pm 1$ and $z=z_{\pm }=\pm a$, respectively\footnote{There is an order 2 pole at $\infty$ covered by an analysis on the chart $\mathcal{U}_{0}$. We will not consider this pole here, due to the fact that the lambda deformed PCM Lax connection to be re-derived below, is known \cite{unifying} to satisfy all the analytic properties at $\infty$.} and all are real numbers.
In local coordinates around each pole in the chart $\mathcal{U}_{1}$, we have that
\begin{equation}
d\omega =i\pi b\frac{(z_{+}^{2}-1)}{2z_{+}}\Big( \delta _{zz_{+}}-\delta
_{zz_{-}}\Big) dz\wedge d\overline{z},
\end{equation}%
where we have used the expression \eqref{delta complex}. Then,
\begin{equation}
\omega _{z\overline{z}}=i\pi b\frac{(z_{+}^{2}-1)}{2z_{+}}\Big( \delta
_{zz_{+}}-\delta _{zz_{-}}\Big) .
\end{equation}%

The first equation in \eqref{three eom}, in the gauge $A_{\overline{z}}=0$ (cf. \eqref{gauge fix 2}), implies
\begin{equation}
\frac{(z^{2}-1)}{(z^{2}-z_{+}^{2})}\partial _{\overline{z}}A_{\mu }(z)=-%
\frac{(z_{+}^{2}-1)}{2z_{+}}\partial _{\overline{z}}\left( \frac{A_{\mu
}(z_{+})}{z-z_{+}}-\frac{A_{\mu }(z_{-})}{z-z_{-}}\right). 
\end{equation}%
Thus, we have%
\begin{equation}
(z^{2}-1)A_{\mu }(z)=H_{\mu }(z)+\frac{(z_{+}^{2}-1)}{2z_{+}}\Big(
(z-z_{+})A_{\mu }(z_{-})-(z-z_{-})A_{\mu }(z_{+})\Big),
\end{equation}%
with $H_{\mu }(z)$ holomorphic. Consistency, i.e. $A_{\mu }(z)|_{z=z_{\pm
}}=A_{\mu }(z_{\pm })$, fixes $H_{\mu }(z)$ and we end up with the result
\begin{equation}
A_{\mu }(z)=f_{+}(z)A_{\mu }(z_{-})+f_{-}(z)A_{\mu }(z_{+}),\text{ \ \ }%
f_{\pm }(z)=\pm \frac{(1-z_{+}^{2})}{2z_{+}}\frac{(z-z_{\pm})}{(z^{2}-1)}. \label{general lax}
\end{equation}
This connection interpolates between the set of poles $\mathfrak{p}$.

Now we solve the condition \eqref{condition}, which is equivalent to 
\begin{equation}
\dint\nolimits_{\mathcal{U}_{1}}d\omega \text{Tr}\left( A_{\tau
}^{2}-A_{\sigma }^{2}\right) =0,
\end{equation}%
or to
\begin{equation}
\text{Tr}\Big( \left( A_{\tau }(z_{+})^{2}-A_{\sigma }(z_{+})^{2}\right)
-\left( A_{\tau }(z_{-})^{2}-A_{\sigma }(z_{-})^{2}\right) \Big) =0. \label{cond}
\end{equation}%
This equation can be solved by taking the following linear combination (cf. \eqref{gauge fix 1})
\begin{equation}
A_{\tau }(z_{\pm })=p(z_{\pm })A_{\sigma }(z_{+})+q(z_{\pm })A_{\sigma
}(z_{-}).
\end{equation}%
Then, \eqref{cond} imply that
\begin{equation}
p(z_{+})^{2}-p(z_{-})^{2}=1,\text{ \ \ }q(z_{+})^{2}-q(z_{-})^{2}=-1,\text{
\ \ }p(z_{+})q(z_{+})=p(z_{-})q(z_{-}).
\end{equation}%
The solutions are
\begin{equation}
p(z_{\pm})=\frac{1}{2}s\left(z_{\pm}\pm z_{\pm}^{-1}\right),\text{ \ \ } q(z_{\pm})=\frac{1}{2}s^{\prime}\left(z_{\pm}\mp z_{\pm}^{-1}\right),
\end{equation}
where $s=\pm 1$ and $s^{\prime}=\pm 1$ are sign functions. Thus, \eqref{general lax} for $\mu=\tau$, becomes
\begin{equation}
A_{\tau }(z)=g_{+}(z)A_{\sigma }(z_{-})+g_{-}(z)A_{\sigma }(z_{+}), \label{atau solution}
\end{equation}%
where%
\begin{equation}
g_{+}(z)=-\frac{s^{\prime}
}{2}\left( 1-z_{+}^{2}\right) \frac{\left( z-z_{+}^{-1}\right) }{\left(
z^{2}-1\right) },\text{ \ \ }g_{-}(z)=-\frac{s}{2}\left( 1-z_{+}^{2}\right) 
\frac{\left( z-z_{-}^{-1}\right) }{\left( z^{2}-1\right) }.
\end{equation}

In the light-cone coordinates the Lax connection $\mathscr{L}_{\pm}(z)$, now identified with $A_{\pm}(z)$, has the following analytic structure
\begin{equation}
A_{\pm}(z)=\frac{1}{2}\Big( A_{\tau}(z)\pm A_{\sigma}(z)   \Big)=\frac{I_{\pm}}{1\pm z},\label{PCM Lax}
\end{equation}
for some currents $I_{\pm}$, i.e. the Lax connection has poles at the zeroes of the twist 1-form $\omega$. Using the solutions \eqref{general lax} with $\mu=\sigma$, \eqref{atau solution} and comparing with \eqref{PCM Lax}, we find that $s=s^{\prime}=-1$ and identify
\begin{equation}
I_{\pm }=\frac{(z_{+}-z_{+}^{-1})}{4}\Big( (1\mp z_{+})A_{\sigma }(z_{+})-(1\pm z_{+})A_{\sigma
}(z_{-})\Big) .
\end{equation}
The latter are nothing but the lambda deformed PCM currents \cite{Sfetsos,lambda-bos}. Finally, the last equation in \eqref{three eom} boils down to
\begin{equation}
\partial_{+}I_{-}+\partial_{-}I_{+}=0,\text{ \ \ }\partial_{+}I_{-}-\partial_{-}I_{+}+\left [ I_{+},I_{-}  \right ]=0,
\end{equation}
which are the lambda deformed PCM eom. They also coincide, formally, with the conventional PCM eom. 

It is interesting to notice the instrumental r\^ole played by the condition \eqref{condition} in deriving the Lax connection of the associated integrable field theory within our approach. We want to emphasize that the condition \eqref{condition} is new and never used in conventional 4d CS theories, where the Lax connection of the integrable model associated to the CS theory is constructed without making any reference to it. See for instance \cite{CY} for the original construction for integrable field theories with order and disorder surface defects and section \S 5 of \cite{Lacroix} for a quick review of the Lax pair/CS theory relation.

Finally, let us summarize the main results accomplished in this section. We have performed a thorough study of the generalized theory in the particular example $(\text{M}=S^{3}, \alpha)$, covering integrable theories of the PCM type. We constructed explicitly the contact form $\alpha$, applied the Hamiltonian analysis (implementing the steps I and II introduced in section \eqref{3}), found the important result \eqref{metric inner final} to be used later and constructed the PCM lambda deformed Lax connection.

In the next section we will consider a complementary approach that allows to introduce the quadratic action \eqref{mu mu 2} in a very intuitive and simple way. Then, in section \eqref{6} we move to the final part of the present work, where we present the main result \eqref{1.3}, which is valid for the PCM type models considered in this section.

\section{Quadratic action from a duality approach }\label{5}

This time, by using a duality approach, we recover the generalization of the 4d Chern-Simons theory introduced above in section \eqref{2}. The argument follows the same logic used in \cite{NA loc CS} but adapted now to the present case, where we have two shift transformations in $\mathcal{S}$. The original 4d CS theory needs to be modified first in an specific way in order for the theory to be dualized in a consistent manner. We also show how the quadratic action \eqref{Generalized 4CS action} emerges naturally as a dual model.

We start with the 4d CS theory on $\mathbb{M}=\Sigma\times C$, with action \eqref{1.1} 
\begin{equation}
S=ic\dint\nolimits_{\mathbb{M}}\omega\wedge CS(\mathbb{A}). \label{semi-hol CS}
\end{equation} 
The difficulty in dualizing \eqref{semi-hol CS} comes from the structure of the `twist' form $\omega$ and the 4d manifold $\mathbb{M}$, which turns the action ill-defined for implementing  a duality transformation via gauging. In what follows, we shall break the strategy for constructing the dual model in the correct way into two simple steps.

The first step consist in regularizing the theory. Instead of \eqref{semi-hol CS} we consider its $\Omega$-shift invariant extension \eqref{part gauge fixed}, which is defined now on the manifold $\mathbb{M}=\mathbb{R}\times \text{M}$, with a non-trivial circle bundle space M and with an action functional defined by
\begin{equation}
S(\mathbb{A}) =ic\dint\nolimits_{\mathbb{M}}\Omega \wedge CS\left( 
\mathbb{A}\right) +ic\dint\nolimits_{\mathbb{M}}d_{\mathbb{M}}\Omega \wedge \kappa\wedge \text{Tr}\left( \mathbb{A}i_{\mathcal{R}}\mathbb{A}  \right). \label{reg action}
\end{equation}%
A simpler way to deduce the action \eqref{reg action} without invoking the Hamiltonian formalism and so on, starts by replacing $\omega \rightarrow \Omega$ in \eqref{semi-hol CS} and follows by adding the necessary terms that make the new action invariant under $\Omega$-shifts. To see this, we simply use the result
\begin{equation}
\Omega\wedge CS\left(^{\Omega}\mathbb{A}\right)=\Omega\wedge CS(\mathbb{A})+d_{\mathbb{M}}\Omega \wedge \Omega \wedge \text{Tr}(s \mathbb{A}) 
\end{equation}
and the identity \eqref{identity for T-duality} in order to replace
\begin{equation}
d_{\mathbb{M}}\Omega \wedge \Omega \wedge \text{Tr}(s\mathbb{A})=-d_{\mathbb{M}}\Omega \wedge \kappa\wedge \Omega \text{Tr}(si_{\mathcal{R}}\mathbb{A}).
\end{equation}
Thus, after combining both expressions, using $s\Omega=\!^{\Omega}\mathbb{A}-\mathbb{A}$ and the fact that $i_{\mathcal{R}}(^{\Omega}\mathbb{A})=i_{\mathcal{R}}\mathbb{A}$, we quickly realize that the action \eqref{reg action} raises as the obvious $\Omega$-shift invariant extension of the action \eqref{semi-hol CS}. Notice that, it is also invariant under the rescalings \eqref{rescalings}. At this stage, it is interesting to compare \eqref{reg action} with \eqref{mu mu 2} for $\Phi=0$. Notice that $S(\mathbb{A}=s\Omega)=0$, thus the functional \eqref{reg action} descends to the quotient $\mathcal{A}/\mathcal{S}_{\Omega}$, where one component of $\mathbb{A}$ manifestly decouples. Furthermore, under finite gauge transformations, we find that
\begin{equation}
S(^{g}\mathbb{A}) =S(\mathbb{A})+ic\dint\nolimits_{\mathbb{M}}\Omega \wedge \chi(g)+ic\dint\nolimits_{\mathbb{M}}d_{\mathbb{M}}\Omega \wedge \kappa \wedge \text{Tr}\left(Y' \mathbb{J}\right),\label{A gauge}
\end{equation} 
where we have used \eqref{CS gauge change}, \eqref{relation} and defined
\begin{equation}
Y'= 2i_{\mathcal{R}}\mathbb{A}+i_{\mathcal{R}}\mathbb{J}. \label{Y prime}
\end{equation}
The last contribution on the rhs of \eqref{A gauge} is related to the obstruction \eqref{d}, while the second term is the WZ-type term found before.  

The second step consists of introducing the $\kappa$-shift symmetry. To do so, we consider a Lie algebra valued field $\Phi\in \Omega_{\mathbb{M}}^{0}\otimes \mathfrak{g}$, which transforms like
\begin{equation}
^{\kappa}\Phi=\Phi+s, 
\end{equation}  
for an arbitrary $s\in\Omega_{\mathbb{M}}^{0}\otimes \mathfrak{g}$ under $\kappa$-shifts and like $\Phi \rightarrow t^{-1}\Phi$ under the arbitrary rescalings \eqref{rescalings}. We also demand that it is invariant under $\Omega$-shifts, i.e. $^{\Omega}\Phi=\Phi$. The key idea now \cite{NA loc CS} is to notice that the combination $\mathbb{A}-\kappa \Phi$ is invariant under $\kappa$-shifts and rescalings. Thus, a double-shift invariant action functional is obtained from \eqref{reg action} after making the substitution $\mathbb{A}\rightarrow \mathbb{A}-\kappa \Phi$. We find that\footnote{Notice how the top-form $\Omega \wedge \kappa \wedge d_{\mathbb{M}} \kappa$ appears naturally from this point of view. This result actually motivated the definition of the inner product \eqref{inner product} introduced above.}
\begin{equation}
S(\mathbb{A}, \Phi) =S(\mathbb{A})+ic\dint\nolimits_{\mathbb{M}}\Omega \wedge \kappa \wedge d_{\mathbb{M}} \kappa \text{Tr}\left(\Phi^{2} \right)-2ic\dint\nolimits_{\mathbb{M}}\text{Tr}\Big( \Phi \left( \Omega \wedge
\kappa \wedge F_{\mathbb{A}}+d_{\mathbb{M}}\Omega \wedge \kappa \wedge 
\mathbb{A}\right) \Big) . \label{T-dual action}
\end{equation}
We have not specified the behavior of $\Phi$ under gauge transformations. However, in order to be consistent with the transformation of $\Phi$, as defined in \eqref{Phi and B}, under gauge transformations \eqref{on-shell Phi under gauge} we demand that 
\begin{equation}
^{g}\Phi=g^{-1}\left (\Phi+B(\mathbb{J})    \right )g.
\end{equation}
Then, we get
\begin{equation}
S(^{g}\mathbb{A},^{g}\Phi) =S(\mathbb{A},\Phi)+ic\dint\nolimits_{\mathbb{M}}\Omega \wedge \chi(g)+ic\dint\nolimits_{\mathbb{M}}d_{\mathbb{M}}\Omega \wedge \kappa \wedge \text{Tr}\left(W \mathbb{J}\right), \label{A,Phi gauge}
\end{equation}  
for some $W\in\Omega_{\mathbb{M}}^{0}\otimes \mathfrak{g}$. Again, there is a term related to \eqref{d} and a WZ-type contribution. Even if we set $^{g}\Phi=g^{-1}\Phi g$, we obtain the same type of expression with a change $W\rightarrow W'$ for another $W'\in\Omega_{\mathbb{M}}^{0}\otimes \mathfrak{g}$, whose explicit form is not relevant in what follows. 

Now we consider the duality manipulations. Using the $\kappa$-shift symmetry we can reach a gauge where $\Phi=0$, in this case we recover the regularized theory \eqref{reg action}. Alternatively, if we calculate the $\Phi$ field eom and put them back into the action \eqref{T-dual action}, we find a dual action that is classically equivalent to \eqref{Generalized 4CS action}. Indeed, the $\Phi$ eom is nothing but the first expression defined in \eqref{Phi and B}, i.e.
\begin{equation}
\Phi =\frac{\Omega \wedge \kappa \wedge F_{\mathbb{A}}+d_{\mathbb{M}%
}\Omega \wedge \kappa \wedge \mathbb{A}}{\Omega \wedge \kappa \wedge d_{%
\mathbb{M}}\kappa } \label{Phi eom}
\end{equation}
and fulfills all the required transformation properties, see for instance \eqref{1 rule} and \eqref{2 rule}. At this point we can see an advantage of the regularized theory, as the denominator in the expression right above never vanishes, making the variational problem for the quadratic field $\Phi$ well-defined over the manifold $\mathbb{M}$. Otherwise, we would have a contribution of the form $\omega\wedge d\alpha=0$, see \eqref{Clever}, in the second term on the rhs of \eqref{T-dual action} turning the theory linear in the field $\Phi$. The effective action obtained after replacing \eqref{Phi eom} in \eqref{T-dual action} gives the dual action
\begin{equation}
S_{\text{dual}}=ic(\mu,\mu), \label{last}
\end{equation}
which is precisely the $\mathcal{S}$-invariant quadratic action constructed in \eqref{Generalized 4CS action}. Thus, the actions \eqref{reg action} and \eqref{last} are dual to each other. Once in the form \eqref{last}, we can recover \eqref{reg action} if we gauge fix the $\kappa$-shift symmetry with the gauge fixing condition $\Phi=0$. The original theory \eqref{semi-hol CS} is finally recovered by taking the degenerate limit $\zeta\rightarrow 0$ and by imposing the boundary conditions \eqref{condition} on the gauge connection $\mathbb{A}$. This further clarifies the why of the two step strategy introduced before in section \eqref{3}. From \eqref{Phi eom}, we find that
\begin{equation}
\Phi(\mathbb{A}=s\Omega)=0, \text{ \ \ } \Phi(\mathbb{A}=s\kappa)=s \label{Phi on kappa and Omega}
\end{equation} 
and from this result we obtain $S(\mathbb{A}=s\Omega)_{\text{dual}}=0$ and $S(\mathbb{A}=s\kappa)_{\text{dual}}=0$. This confirms that \eqref{last} descends to the quotient $\overline{\mathcal{A}}$, where a second component of $\mathbb{A}$ manifestly decouples from the theory. The number of components of $\mathbb{A}$, as a Lie algebra valued 1-form is two, see \eqref{explicit A} and \eqref{Pi bundle}. Recall that we still have the action of the gauge group $\mathcal{G}$ on $\overline{\mathcal{A}}$ to be taken into account. 

The following diagram roughly summarizes our findings: \\
\begin{equation*}
\begin{array}{ccc}
S=\small\text{\{4d CS theory \eqref{1.1}\}} & \overset{\Omega \text{-shift
extension}}{\xrightarrow{\hspace*{3.0cm}}  } & S(\mathbb{A})=\small\text{\{regularized action \eqref{reg action}\}}
\\ 
&  &  \\ 
\; \;  \Bigg\uparrow
\begin{array}{c}
\footnotesize\text{step I: }\Phi _{\eqref{Phi eom}}=0  \\ 
\! \! \! \! \! \! \! \! \! \! \footnotesize\text{step II: } \zeta \rightarrow 0%
\end{array}
&  & \Bigg\downarrow \kappa \footnotesize\text{-shift extension} \\ 
&  &  \\ 
S_{\text{dual}}=\small\text{\{}\text{generalized 4d CS theory \eqref{last}\}}
& \overset{\Phi \text{-integration}}{\xleftarrow{\hspace*{2.4cm}} } & S(\mathbb{A},\Phi )=%
\small\text{\{}\text{extended action \eqref{T-dual action}\}}.%
\end{array}%
\end{equation*} 
The dual pair of action functionals is formed by \eqref{reg action} and \eqref{last}, i.e. $S(\mathbb{A})$ and $S_{\text{dual}}$ right above. Steps I and II are to be supplemented with a solution to the condition \eqref{new bdry condition}.  

In the previous sections we have studied systematically several classical aspects of the generalized 4d CS theory. Now, it is time to consider the path integral formulation of it. After all, as mentioned in the introduction, our main goal is to show that the usual 4d CS theories \eqref{1.1} can be embedded into a more general theory whose path integral formulation takes the canonical form \eqref{1.3}. This is precisely the topic of the next and last section.

\section{Path integral and non-Abelian localization}\label{6}

Here we comment on the second ingredient \eqref{main 2} involved in the formula of non-Abelian localization, i.e. the path integral symplectic measure. The first one, discussed extensively above, being the quadratic form \eqref{main 1} of the action functional. As announced, we will show that the path integral for the generalized theory, at least for the main example considered in section \eqref{4}, takes the form \eqref{1.3}. See \cite{NA loc CS}, for the original 3-dimensional CS theory formulation.

Consider the theory defined formally by the path integral%
\begin{equation}
Z=\mathcal{N}'\int\nolimits_{\mathcal{A}\times \mathcal{S}_{\kappa}} \mathcal{D}\mathbb{A}\mathcal{D}\Phi \; \text{exp}\left[\frac{i}{\hbar} S(\mathbb{A}, \Phi)\right],
\end{equation}
where the action in the exponential is given by \eqref{T-dual action}, $\mathcal{N}'$ is defined by
\begin{equation}
\mathcal{N}'=\mathcal{N}\times \frac{1}{ \text{Vol}(\mathcal{S}_{\kappa})\times \text{Vol}(\mathcal{S}_{\Omega})\times \text{Vol}(\mathcal{G})}
\end{equation}
and $\mathcal{N}$ is a normalization constant. As we have shown in \eqref{1 rule}, \eqref{2 rule}, \eqref{Phi on kappa and Omega}, the field $\Phi$ belongs to the orbit generated by the action of the $\kappa$-shift group $\mathcal{S}_{\kappa}$, while it is a fixed point under the action of the $\Omega$-shift group $\mathcal{S}_{\Omega}$. Then, the integral associated to the measure $\mathcal{D}\Phi$ is over $\mathcal{S}_{\kappa}$, while the integral associated to $\mathcal{D}\mathbb{A}$ is, as usual, over the whole space $\mathcal{A}$ of gauge connections. 

The translation-invariant measure $\mathcal{D} \Phi$ is defined independently of any
metric on $\mathbb{M}$ by the invariant, quadratic form%
\begin{equation}
\left( \Phi ,\Phi \right) _{\mathbb{M}}=-\int\nolimits_{\mathbb{M}}\Omega
\wedge \kappa \wedge d_{\mathbb{M}}\kappa \text{Tr}\left( \Phi ^{2}\right) .
\end{equation}
This quadratic form is, up to scale, used to formally define the volume of $\mathcal{S}_{\kappa}$. Similar expressions are used to define $\mathcal{S}_{\Omega}$ and $\mathcal{G}$, as anticipated before in \eqref{inner product} and \eqref{norm S}. This is to be complemented with the formal definition of the translation-invariant measure $\mathcal{D}\mathbb{A}$ induced by the norm \eqref{break metric} and its orthogonal decomposition between the spaces $\overline{\mathcal{A}}$ and $\mathcal{S}$.  

On the one hand, using the $\kappa$-shift symmetry, we can fix $\Phi=0$ trivially with unit Jacobian, and the resulting integral over $\mathcal{S}_{\kappa}$ produces a formal factor of $\text{Vol}(\mathcal{S}_{\kappa})$. Hence, the theory is equivalent to \eqref{reg action}, i.e. to the $\Omega$-shift invariant extension of the 4d CS theory \eqref{1.1}. The resulting action functional is valued in the quotient $\mathcal{A}/\mathcal{S}_{\Omega}$ and $\mathcal{D}\mathbb{A}$ integrates over $\mathcal{A}/\mathcal{S}_{\Omega}\times \mathcal{S}_{\Omega}$, where the integral over $\mathcal{S}_{\Omega}$ produces a formal factor or $\text{Vol}(\mathcal{S}_{\Omega})$. Then, in principle, we get  
\begin{equation}
Z=\mathcal{N}\times \frac{1}{\text{Vol}(\mathcal{G})}\int\nolimits_{\mathcal{A}/\mathcal{S}_{\Omega}} \mathcal{D}\mathbb{A} \; \text{exp}\left[\frac{i}{\hbar} S(\mathbb{A})\right].
\end{equation}

On the other hand, because of the field $\Phi$ appears only quadratically in the action \eqref{T-dual action}, we can perform the path integral over $\Phi$ directly. Integrating out $\Phi$, produces a contribution
\begin{equation}
I=\int\nolimits_{\mathcal{S}_{\kappa}} \mathcal{D}\Phi \; \text{exp}\left[-\frac{c}{\hbar} \int\nolimits_{\mathbb{M}}\Omega
\wedge \kappa \wedge d_{\mathbb{M}}\kappa \text{Tr}\left( \Phi ^{2}\right) \right].
\end{equation}
The resulting action functional is given by the dual quadratic expression \eqref{last} and is valued in $\overline{\mathcal{A}}$. Thus $\mathcal{D}\mathbb{A}$ integrates over $\overline{\mathcal{A}}\times \mathcal{S}$. Here, we make use of the results \eqref{break metric}, \eqref{metric inner final} related to the fact that the quotient space $\overline{\mathcal{A}}$ is symplectic and equipped with a K\"ahler metric in order to write the measure along $\overline{\mathcal{A}}$ in the form (see \eqref{main 2})
\begin{equation}
\mathcal{D}\mathbb{A}|_{\overline{\mathcal{A}}}=\text{exp}\; \hat{\underline{\Omega}}. \label{symp measure}
\end{equation}
As we showed above, this measure is to be taken over elements $\Pi({\mathbb{A}})\in \overline{\mathcal{A}}$ of the form \eqref{A in quotient}.   

An important consequence of the fact that the metric on $\overline{\mathcal{A}}$ is K\"ahler is that the Riemannian measure $\mathcal{D}\mathbb{A}$ on $\overline{\mathcal{A}}$ is actually the same as
the symplectic measure defined by $\hat{\underline{\Omega}}$. Indeed, if $X$ is a symplectic manifold of dimension $2n$ with
symplectic form $\hat{\Omega}$, then the symplectic measure on $X$ is given by the top-form $\hat{\Omega}^{n}/n!$.
This measure can be represented by the expression $\text{exp}\; \hat{\Omega}$, where we implicitly pick out from the series expansion of the exponential the term which is of top degree on $X$. Consequently, because of the Riemannian and the symplectic
measures on $\overline{\mathcal{A}}$ agree, we can formally replace $\mathcal{D}\mathbb{A}$ over $\overline{\mathcal{A}}$ in the path integral by the expression \eqref{symp measure} above and write instead 
\begin{equation}
Z(\epsilon)= \mathcal{N}\times I \times \frac{1}{\text{Vol}(\mathcal{G})}  \dint\nolimits_{\overline{\mathcal{A}}}\ \text{exp} \left[ \hat{\underline{\Omega}}-\frac{1}{2\epsilon}(\mu,\mu)  \right ], \label{final path integral}
\end{equation}
where we have used $\text{Vol}(\mathcal{S})=\text{Vol}(\mathcal{S}_{\kappa})\times \text{Vol}(\mathcal{S}_{\Omega})$ and defined $\epsilon=\hbar/2c$. This integral takes the canonical form \eqref{1.3} with $X=\overline{\mathcal{A}}$, as required by the non-Abelian localization method. The normalization constant $\mathcal{N}$ can be adjusted to an specific valued if needed. 

The main consequence of an expression like \eqref{final path integral} is that it suggests an interesting relationship between the quantum integrable structure of the 4d CS theory and the geometry of the symplectic quotient space $\overline{\mathcal{A}}$. In principle, the 4d CS theories and their associated integrable models and field theories could be explored via standard localization techniques.  

\section{Concluding remarks}\label{7}

In this concluding section, we make some comments, provide further explanations concerning the results presented along the text and touch on some topics we judge interesting to be considered in the future.

Clearly, the pre-symplectic form \eqref{pre-symplectic} plays a crucial r\^ole in the construction of the path integral \eqref{1.3}, as it specifies the moment map $\mu$ used to define the quadratic action $S\sim(\mu,\mu)$, as well as the path integral symplectic measure $e^{\underline{\Omega}}$ over the quotient space $\overline{\mathcal{A}}$. Thus, some comments on what inspired its definition are in order.

Consider the original pre-symplectic form defined in \cite{NA loc CS}, which in the present notation takes the form
\begin{equation}
\hat{\Omega}=-\frac{1}{2}\dint\nolimits_{\text{M}}
\alpha \wedge \text{Tr}\left( \hat{\delta }\mathbb{A}\wedge\hat{\delta }\mathbb{A}\right). \label{Witten symp}
\end{equation}%
The circle fibers, correspond to the integral curves of the Reeb vector field $R$ satisfying the normalization condition $\alpha(R)=1$. After introducing the time direction, we extend M to $\mathbb{M}=\mathbb{R}\times \text{M}$. Locally, the manifold $\mathbb{M}$ looks like $\mathbb{M}=\Sigma \times C$ and the light-cone tangent vectors to the Minkowskian cylinder $\Sigma$, are given by $\partial_{\pm}\sim \partial_{\tau}\pm \partial_{\sigma}$. It is then natural to extend them, respectively, to their global counterparts \eqref{time+Reeb}, \eqref{time-Reeb} 
\begin{equation}
\mathcal{R}\sim \frac{1}{\alpha_{\tau}}\partial_{\tau}+R,\text{ \ \ }\mathcal{R}'\sim \frac{1}{\alpha_{\tau}}\partial_{\tau}-R
\end{equation}
and to introduce two 1-forms $\kappa$ and $\Omega '$ such that $\kappa(\mathcal{R})=1$ and $\Omega '(\mathcal{R}')=1$, see \eqref{solutions}, \eqref{Omega prime}. Because of $R$ defines vertical and horizontal directions in the tangent space of the total space M, we can add any horizontal 1-form $\omega$ with no $d\tau$ term to $\kappa$ or to $\Omega '$, without spoiling the normalization conditions and this is because $i_{\mathcal{R}}\omega=i_{\mathcal{R}'}\omega=0$. We choose to add it to $\Omega '$ and define $\omega$ as the pull-back, by the projection map $\underline{\pi}$, of the twist 1-form $\omega_{C}$ on $C$ that specifies the associated integrable field theory. There is some room to introduce an arbitrary parameter, which we call $\zeta$. Then, a natural generalization of \eqref{Witten symp} to four dimensions, in which we include the time direction, is given by 
\begin{equation}
\hat{\Omega}=-\frac{1}{2}\dint\nolimits_{\mathbb{M}}\Omega \wedge
\kappa \wedge \text{Tr}\left( \hat{\delta }\mathbb{A}\wedge\hat{\delta }\mathbb{A}\right). \label{sy}
\end{equation}%

Now we show how the 1-forms $\Omega$, $\kappa$ of \eqref{solutions} used in \eqref{sy} are constructed. Let us start with the interpolating expressions%
\begin{equation}
\kappa =s\alpha _{\tau }d\tau +(1-s)\alpha ,\text{ \ \ }\Omega =\omega
+\zeta ^{\prime }\left( (1-s)d\tau -\frac{s}{\alpha _{\tau }}\alpha \right) ,
\end{equation}%
and%
\begin{equation}
\mathcal{R}=\frac{1}{2}\left( \frac{1}{s\alpha _{\tau }}\partial _{\tau }+%
\frac{1}{1-s}R\right) ,\text{ \ \ \ }\mathcal{R}^{\prime }=\frac{1}{2\zeta
^{\prime }}\left( \frac{1}{1-s}\partial _{\tau }-\frac{\alpha _{\tau }}{s}%
R\right) ,
\end{equation}%
where $s\in (0,1)$ and $\zeta ^{\prime }\in 
%TCIMACRO{\U{211d} }%
%BeginExpansion
\mathbb{R}
%EndExpansion
$. They satisfy the normalization conditions $i_{\mathcal{R}}\kappa =i_{%
\mathcal{R}^{\prime }}\Omega =1.$ By demanding that $\Omega \wedge \kappa
\wedge d_{\mathbb{M}}\kappa =\zeta d\tau \wedge \alpha \wedge d\alpha $, we
find%
\begin{equation}
\zeta ^{\prime }=\frac{\zeta }{(1-s)\left[ s^{2}+(1-s)^{2}\right] }.
\end{equation}%
However, the conditions $i_{\mathcal{R}}\Omega =i_{\mathcal{R}^{\prime
}}\kappa =0$ require that $s$ take the specific value $s=1/2$, corresponding to the `light-cone' solutions introduced above in \eqref{solutions}, \eqref{solutions 2}. 
It is also possible to make the changes $(\alpha,R)\rightarrow (-\alpha,-R)$ in all formulae.

The path integral \eqref{1.3} is specified by the symplectic data associated to $\hat{\Omega}$ and the Hamiltonian action of $\mathcal{H}$ on $\overline{\mathcal{A}}$. Also notice that the generalized 4d CS theory does not require a twist 1-form $\omega$ to be well-defined. Actually, for $\omega=0$, we have that $\Omega \wedge \kappa=\Omega_{\tau} d\tau \wedge \alpha$ and this case can be seen as the simplest canonical 4d extension of \eqref{Witten symp}. Then, the generalized theory allows, in principle, to embed any 4d CS theory, regardless of the analytic structure of $\omega_{C}$, into a quantization framework based on the non-Abelian localization method. In particular, it may offer an approach for quantizing non-ultralocal integrable field theories \cite{Maillet} from a more geometric perspective. These type of theories all have twist 1-forms $\omega_{C}$ with zeroes \cite{Vicedo-PCM}, like the PCM type models considered above, and quantization is problematic because of the non-ultralocality prevents a straightforward use of techniques coming from the quantum inverse scattering method. Furthermore, as argued heuristically in \cite{CWY1}, the zeroes of $\omega_{C}$ corresponds to points where $\hbar \rightarrow \infty$, hence an approach based on localization may provide a 4d CS theories description even in this regime.  

Let us notice that the combined shift symmetries in $\mathcal{S}$ dictate the very form of the quadratic action \eqref{last}. Thus, a natural question to be asked is what is the r\^ole played by the vector field $\mathcal{R}'$ and how its associated moment map modifies the quadratic action.\\
Let us start with the complete induced vector field on $\mathcal{A}$, cf. \eqref{total vector field}, which is given by
\begin{equation}
V(\tilde{p},\eta ,a)=d_{\mathbb{A}}\eta +p\pounds _{\mathcal{R}}%
\mathbb{A+}p^{\prime }\pounds _{\mathcal{R}^{\prime }}\mathbb{A}, \label{full vector field}
\end{equation}%
where\footnote{Alternatively, we may introduce a four entry notation $(p',p,\eta,a)$. } $\tilde{p}=p+p^{\prime }$. For the moment map $\mu$ associated to $V(p',0 ,0)$, we find that%
\begin{equation}
\left\langle \mu ,(p^{\prime },0,0)\right\rangle =\frac{p^{\prime }}{2}%
\dint\nolimits_{\mathbb{M}}\Omega \wedge \kappa \wedge \text{Tr}\left( 
\pounds _{\mathcal{R}^{\prime }}\mathbb{A\wedge A}\right), 
\end{equation}%
which descends to the quotient space $\overline{\mathcal{A}}$. We also find the Poisson bracket, cf. \eqref{R with F},
\begin{equation}
\left\{ \left\langle \mu ,(p^{\prime },0,0)\right\rangle ,\left\langle \mu
,(0,\lambda ,0)\right\rangle \right\} =\left\langle \mu ,(0,-p^{\prime }%
\pounds _{\mathcal{R}^{\prime }}\lambda ,0)\right\rangle.
\end{equation}%
Thus, the bracket \eqref{total bracket} is replaced by
\begin{equation}
\Big[ (\tilde{p},\eta ,a),(\tilde{q},\lambda ,b)\Big] =\Big( 0,[\eta,\lambda]-p%
\pounds _{\mathcal{R}}\lambda-p'\pounds_{\mathcal{R}'}\lambda +q\pounds _{\mathcal{R}}\eta+q'\pounds_{\mathcal{R}'}\eta ,c(\eta ,\lambda
)\Big) ,\label{extended bracket}
\end{equation}%
where $\tilde{q}=q+q^{\prime }$. Now, we extend the inner product \eqref{inner product} to%
\begin{equation}
\big( \left( \tilde{p},\eta ,a\right) ,\left( \tilde{q},\lambda
,b\right) \big) =-\int\nolimits_{\mathbb{M}}\Omega \wedge \kappa \wedge
d_{\mathbb{M}}\kappa \text{Tr}\left( \eta \lambda \right) -\tilde{p}b-%
\tilde{q}a.\label{extended inner product}
\end{equation}%
The final step is to verify if the inner product just defined is invariant, which is equivalent to having, cf. \eqref{inv},
\begin{equation}
\Big( \left[ \left( \tilde{p},\eta ,a\right) ,\left( \tilde{q}%
,\lambda ,b\right) \right] ,(\tilde{r},\phi ,c)\Big) =\Big(
\left( \tilde{p},\eta ,a\right) ,\left[ \left( \tilde{q},\lambda
,b\right) ,(\tilde{r},\phi ,c)\right] \Big) ,
\end{equation}%
where $\tilde{r}=r+r^{\prime }$. This conditions boils down to
\begin{equation}
\begin{aligned}
\tilde{r}d(\eta,\lambda)+r'\int\nolimits_{\mathbb{M}}\Omega \wedge \kappa \wedge
d_{\mathbb{M}}\kappa \text{Tr}&\Big ( \eta \big ( \pounds_{\mathcal{R}}\lambda-\pounds_{\mathcal{R}'}\lambda   \big) \Big )\\
&=\tilde{p}d(\lambda,\phi)+p'\int\nolimits_{\mathbb{M}}\Omega \wedge \kappa \wedge
d_{\mathbb{M}}\kappa \text{Tr}\Big ( \lambda \big ( \pounds_{\mathcal{R}}\phi-\pounds_{\mathcal{R}'}\phi   \big) \Big ). 
\end{aligned}\label{Inv 2}
\end{equation} 
We solve this by taking $d(\ast,\ast)=0$ and $p'=q'=r'=0$. Thus, the vector field \eqref{full vector field}, the bracket \eqref{extended bracket} and the inner product \eqref{extended inner product} reduce to the ones considered before. As a consequence, the quadratic action remains unaltered. The vector fields $\mathcal{R}$ and $\mathcal{R}'$ have different uses in the formulation of the generalized 4d CS theory, at least as implied by the solutions to \eqref{Inv 2} chosen in the present paper. It would be interesting to consider other possible solutions and their implications.

For integrable field theories on (semi)-symmetric spaces, the formulation presented here requires to consider non-trivial $S^{1}$ bundles over the base space $C=\mathbb{CP}^{1}/\mathbb{Z}_{4}$. Coset spaces of the form $\mathbb{CP}^{1}/ \mathbb{Z}_{T}$ were first considered in \cite{dihedral} on an approach devised to reformulate $\mathbb{Z}_{T}$-graded coset $\sigma$-model as dihedral affine Gaudin models. The particular case $T=4$, was also studied in \cite{me def coset}, where the symmetric-space $\lambda$-model exchange algebra was recovered from the point of view of the conventional 4d CS theory. It is then desirable to study the generalized 4d CS theory on non-trivial circle bundles over spaces of this type, due to their relation to important non-ultralocal integrable field theories like the $\sigma$-models on (semi)-symmetric spaces and their integrable deformations \cite{eta-def bos,eta-def fer,Rivelles-Hector,lambda-bos,lambda-fer,Hybrid lambda,PS lambda} too. We expect to consider this in the near future. 

The generalized theory presents a behavior, under the action of finite gauge transformations, that is similar to the conventional 4d CS theory. Thus, the last couple of terms in \eqref{gauge change} must be properly handled first in order for the expression \eqref{final path integral} to make perfect sense at the quantum level. In this work we have adopted the strategy of imposing restrictions over the gauge elements $g\in\mathcal{G}$ in order to cancel both contributions, making the generalize theory gauge invariant. We do not know if this approach is the only way to do it or if there is some gauge group structure that can be exploited instead. We expect to consider this subtle issue in a more systematic way elsewhere. 

The present construction relies on having a compact direction in the 2-dimensional space-time $\Sigma$, which we chose to be a cylinder $\Sigma=\mathbb{R} \times S^{1}$. An interesting problem would be to consider instead a strip $\Sigma=\mathbb{R} \times [-L,L]$. This case would cover integrable field theories defined on the real line or a finite segment, depending on the choice of the parameter $L$.  \
It is a well-known fact that Chern-Simons theories and WZW models are closely related and that WZW models with open string boundary conditions require an specific set of D-brane configurations on group manifold or subsets of it. For example and just to name a few, D-branes are considered in \cite{Schomerus},\cite{NA Kinks},\cite{Driezen} in the context of WZW models, symmetric space sine-Gordon theories and lambda deformed integrable field theories, respectively. How our generalized 4-dimensional Chern-Simons theory is related to integrable models defined on a segment is not clear at this moment and remains as an open problem, but a sensible starting point could be to first explore the CS/WZW relation for open string configurations.

\section*{Acknowledgements}

The author thanks the referee for valuable comments and suggestions.

\end{document}